\documentclass[twocolumn]{aastex631}
\usepackage{sidecap}
\usepackage{amsmath}
\usepackage{ulem}
\usepackage{tabularx}
\DeclareUnicodeCharacter{0301}{\'{e}}

\usepackage{graphicx}
\usepackage{dcolumn}
\usepackage{bm}
\usepackage{longtable}
\usepackage{ulem}
\usepackage{float}
\begin{document}

\preprint{APS/123-QED}

\title{Anomalously fast core and \\ envelope rotation in red giants}

\author{Siddharth Dhanpal}
\email{dhanpal.siddharth@gmail.com}
\affiliation{Department of Astronomy and Astrophysics, Tata Institute of Fundamental Research, Mumbai, 400005, India}
\affiliation{Scientific Computing Department, Rutherford Appleton Laboratory \\
Science and Technology Facilities Council, Didcot, United Kingdom}
%Scientific Computing Depar, Tata Institute of Fundamental Research, Mumbai, 400005, India}

\author{Othman Benomar}
\affiliation{ Center for Space Science, NYUAD Institute, New York University Abu Dhabi, PO Box 129188, Abu Dhabi, UAE}
\affiliation{Division of Solar and Plasma Astrophysics, NAOJ, Mitaka, Tokyo, Japan}

\author{Shravan Hanasoge}
\affiliation{Department of Astronomy and Astrophysics, Tata Institute of Fundamental Research, Mumbai, 400005, India}
\affiliation{ Center for Space Science, NYUAD Institute, New York University Abu Dhabi, PO Box 129188, Abu Dhabi, UAE}

\author{Jim Fuller}
\affiliation{TAPIR, California Institute of Technology, Pasadena, CA 91125, USA}



\begin{abstract}
Red giants undergo dramatic and complex structural transformations as they evolve. Angular momentum is transported between the core and envelope during this epoch, a poorly understood process. Here, we infer envelope and core rotation rates from Kepler observations of $\sim$1517 red giants. While many measurements are consistent with the existing studies, our investigation reveals systematic changes in the envelope-to-core rotation ratio and we report the discovery of anomalies such as clump stars with rapidly rotating cores, and red giants with envelopes rotating faster than their cores. We propose binary interactions as a possible mechanism by which some of these cores and envelopes are spun up. These results pose challenges to current theoretical expectations and can have major implications for compact remnants born from stellar cores.
\end{abstract}

%\maketitle

%\tableofcontents

\section{Introduction}

Rotation exerts a profound influence on a star's internal structure, properties, and evolutionary processes. Following H-exhaustion in the main sequence phase, the stellar core undergoes contraction and spins up, while the envelope expands and spins down due to the local conservation of angular momentum. This radial differential rotation gives rise to various magneto-hydrodynamical instabilities, which significantly contribute to the transport of angular momentum and chemical elements within the star \citep{fuller:19}. Characterising the internal stellar rotation rates is thus a problem of major importance. In this regard, asteroseismology is an indispensable tool with which to precisely measure rotation rates in both the cores and envelopes of red giants \citep{beck:12,mosser:12,mosser:18,goupil:13}.

During the evolution of a star from its main-sequence phase to the red-giant stage, the density of its helium core increases \citep{montalban:13}, thereby allowing for coupling between core gravity modes and envelope pressure modes, leading to the appearance of {\it mixed modes} \citep{bedding:11}. Consequently, red giants exhibit a complicated oscillation spectrum, encompassing radial modes, quadrupolar pressure modes and dipolar mixed modes, allowing us to gain insight into the core as well as the envelope.

Rotation splits non-radial modes of degree $\ell$ into 2$\ell$+1 multiplets. Since pressure waves corresponding to quadrupolar modes are primarily sensitive to properties of the stellar envelope, the splitting is linearly proportional to the rotation rate of the envelope. On the other hand, the dipolar mixed-modes are sensitive to the physical processes within the entire star. This leads to the splitting of mixed mode being contingent on rates of the core and envelope, as well as the extents of the \textit{p}- and \textit{g}- characters within the mixed mode \citep{goupil:13}. The rotational splitting of an $\ell=1$ mode may be paramaterized by the following equation,
 \begin{equation}
    \mathrm{\delta\nu_{\rm rot}} = \frac{1}{2}\frac{\Omega_{\rm core}}{2\pi}\zeta(\nu) + \frac{\Omega_{\rm env}}{2\pi} [1-\zeta(\nu)],
    \label{eq:rot_mm}
\end{equation}
where $\Omega_{\rm core}$ and $\Omega_{\rm env}$ correspond to the average rotation rates of the core and envelope, respectively, and $\zeta(\nu)$ relates to the function that computes the mixing fraction in the mode. Dialing up $\zeta$ from 0 to 1 changes the mixed mode from being \textit{p}-dominated to \textit{g}-dominated, thereby resulting in a change in the rotational splitting.

In this article, we explore the problem of characterizing internal rotation in red giant stars. Specifically, we focus on two key rotational parameters: the core rotation rate ($\Omega_{\rm core}$), and the envelope rotation rate ($\Omega_{\rm env}$). These quantities are critical for probing the transport of angular momentum in different evolutionary phases. We have devised a neural network architecture to infer seismic parameters, such as rotation rates and inclination angles, directly from the oscillation power spectra of red giants in a single step. The output parameters reflect the underlying stellar structure and rotational dynamics, and the resulting probability distributions are in good agreement with those obtained by Monte Carlo Markov Chain (MCMC) analyses \citep{benomar:09,handberg:11,corsaro:14}. 

The remainder of the paper is organized as follows. In Section \ref{Methods}, we describe the training datasets and the neural network architecture used to infer seismic parameters and rotational rates. Section \ref{Results on Kepler data} presents the results of our analysis applied to a population of red giant observations. In Section \ref{Stellar Models}, we construct theoretical stellar models to interpret the observed rotation rates, incorporating both single-star evolution and binary interaction scenarios. Section \ref{Additional Observational Context} expands on the observational context by examining correlations with surface abundances, binarity indicators from GAIA, and other stellar activity proxies. Finally, in Section \ref{Conclusions}, we present our conclusions.

\begin{figure*}[!ht]
\centering
\includegraphics[width=\linewidth]{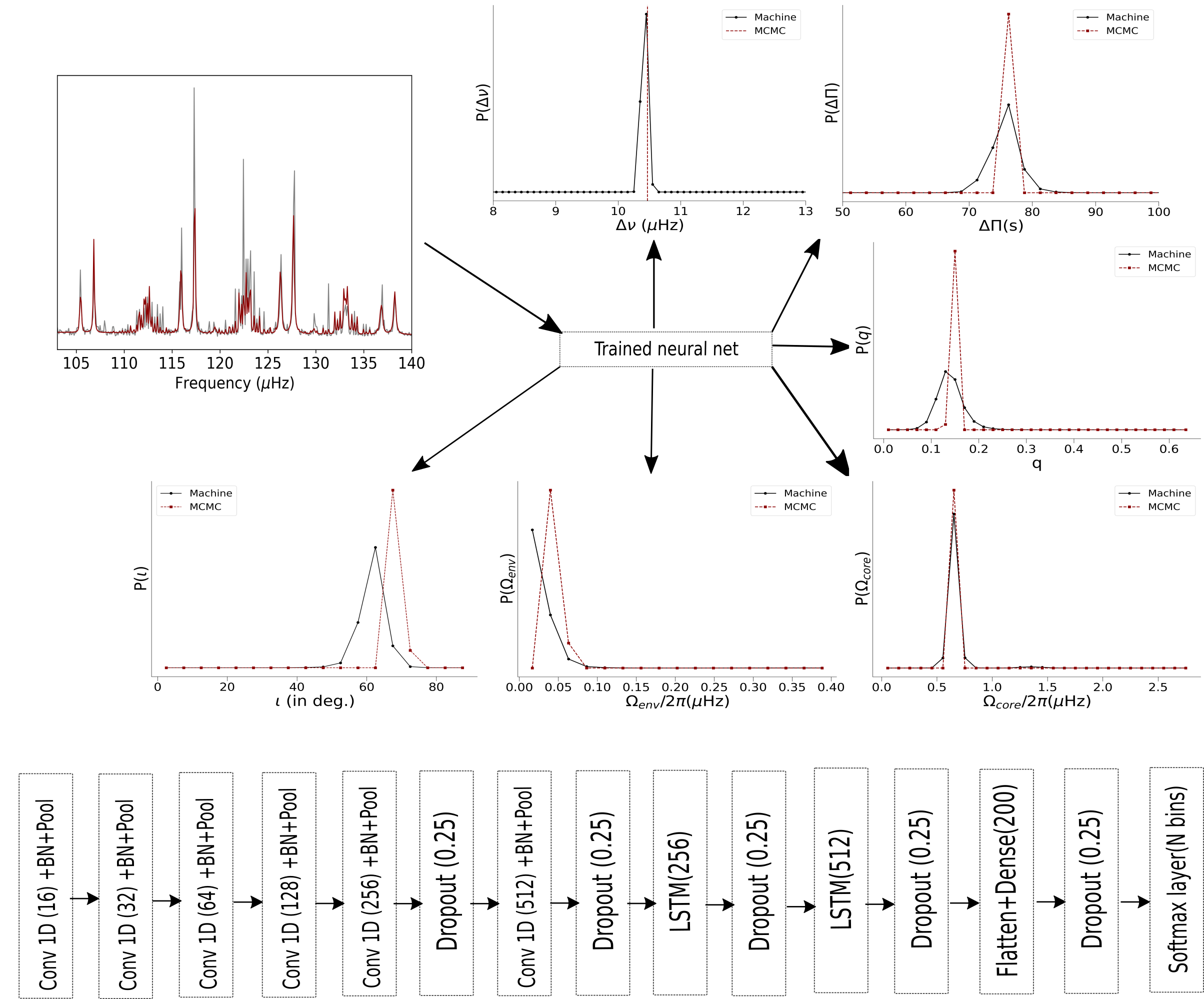}
\caption{Diagrammatic representation of the neural network applied to analyze the oscillation spectrum KIC 6191190. The trained neural network takes as input the normalized power spectrum and outputs the probability distributions of various seismic parameters, such as the large-frequency separation ($\Delta \nu$), large-period separation ($\Delta \Pi$), coupling constant ($q$), core rotation ($\Omega_{core}/2\pi$), envelope rotation ($\Omega_{env}/2\pi$), and inclination angle ($\iota$). To obtain the actual marginal distributions, MCMC was carried out to fit said spectrum, followed by a comparison with the distributions generated by the neural network. The agreement between the best-fit model and the data is seen to be satisfactory. The network's distributions, depicted in gray, were juxtaposed against the posterior distributions obtained from MCMC in red. The choice of bin sizes for representing MCMC posterior distributions was aligned with the bin sizes of the neural network to ensure an effective comparison.}
\label{fig:fig1_machine}
\end{figure*}

\section{Methods}\label{Methods}

For the ensemble study of Kepler stars, we utilized neural networks, while for conducting in-depth analysis of stars of interest identified by the neural network, we applied Markov Chain Monte Carlo (MCMC).

The efficacy of a neural network hinges mainly on two factors: (a) the training data and (b) the network design. Our training data, comprising solely of synthetic spectra, offers two key benefits. Firstly, it allows for the production of an abundant number of samples, unrestricted by the limitations that come with a finite number of observational samples. Secondly, it permits the avoidance of biases that may arise from human intervention, such as the labeling of peaks in the data. However, the primary challenge of using synthetics lies in the construction of a realistic spectra.

\subsection{Training Data}

We employ the asymptotic theory of oscillations to model power spectra of red giants, as described previously \citep{garcia:19,aerts:10}. This theory integrates the fundamental physical characteristics of red giants such as their structure, compositional gradient, and rotation. The detailed modeling procedures implemented in our simulations may be found in Appendix A of \cite{dhanpal:22} and \cite{dhanpal:23}. In these papers, we demonstrate that our simulations are of sufficiently high quality, realistic, and encompass a sufficiently wide range of parameter space. 

Asymptotic theory postulates that dipole mixed modes in solar-like stars are given by the equation \citep{mosser:15,farnir:21,lindsay:22,ong:23}
\begin{equation}
    \tan \bigg[ \pi \frac{\nu-\nu_p}{\Delta\nu} \bigg] = q \tan \bigg[ \frac{\pi}{\Delta \Pi} \left(\frac{1}{\nu}-\frac{1}{\nu_g}\right) \bigg] .\hspace{0.2cm} \label{eq:mixedmodes}
\end{equation}

The expression for pure \textit{p} modes ($\nu_p$) in this equation is given by $\frac{\nu_{p;n,\ell}}{\Delta \nu} = n_p + \frac{\ell}{2}+ \varepsilon_p(\Delta \nu) - d_{0\ell}(\Delta \nu) + \frac{\alpha_{\ell}}{2}\left(n_p-\frac{\nu_{max}}{\Delta \nu}\right)^2$. Here, $\Delta \nu$ represents the mean-frequency separation between two successive radial modes, $\varepsilon_p(\Delta \nu)$ denotes the \textit{offset parameter}, $d_{0\ell}$ corresponds to the \textit{small-frequency separation}, and $\alpha_{\ell}$ represents the degree-dependent gradient, given by $\alpha_{\ell} = \left(d\log \Delta \nu/dn\right)_{\ell}$. Additionally, the frequency $\nu_g$ can be expressed as $\nu_{g}^{-1} = (n+\varepsilon_{g})\Delta \Pi$, where $\Delta \Pi$ is the asymptotic period spacing and $\epsilon_g$ represents the offset parameter associated with the g modes. 

The impact of rotation on mixed modes is shown in equation~\ref{eq:rot_mm}. Although all $\ell=2$ modes are formally mixed \citep{benomar:13}, those near the $\ell=0$ ridge predominantly propagate as envelope-trapped $p$ modes or $p$ dominated mixed modes. Their rotational kernels are highly sensitive to the outer layers, making $\ell=2$ mode splittings approximate measures of $\Omega_{\rm env}/2\pi$. While $\ell=1$ mixed modes contain information on both the core and envelope, $\ell=2$ modes serve as the primary source for estimating the envelope rotation rate \citep{ahlborn:20}. However, since $\ell=2$ modes may retain some sensitivity to deeper regions, the inferred rotation rates represent a weighted average over the envelope with a minor contribution from the core \citep{ahlborn:20}. This limitation should be considered when interpreting the envelope rotation trends presented in this paper.

In low spatial-resolution observations, the observed amplitude of a stellar oscillation mode \( f_{n,\ell,m} = A Y_\ell^m(\theta,\phi) \) is expressed as
\[
a_{n,\ell,m} = r_{\ell,m}(\iota) V(\ell) A,
\]
where \( V(\ell) \) is the mode visibility and \( r_{\ell,m}(\iota) \) is the relative amplitude that depends on the inclination angle \( \iota \) \citep{gizon:03,ballot:06}. The visibility function, which depends on the limb-darkening profile of the star and the observational method, decreases with increasing degree of spherical harmonic \( \ell \). As a result, asteroseismic observations predominantly detect modes with \( \ell = 0, 1, \) and \( 2 \), with \( \ell = 1 \) generally having the highest visibility. Modes with \( \ell = 3 \) or higher are rarely observed due to their low visibilities.

For red giants, typical visibility values are \( V(0) \simeq 1 \), \( V(1) \in [1.2, 1.75] \), \( V(2) \in [0.2, 0.8] \), and \( V(3) \in [0, 0.1] \) \citep{mosser:12}. The relative amplitude is given by
\begin{equation}
%\[
    r_{\ell,m}\left(\iota\right) = \left[Y_{\ell,m}(\iota,0)\right]^2
    \label{eq:eq_inc}
\end{equation}
where $Y_{\ell,m}(\theta,\phi)$ is a spherical harmonic.

While generating the training dataset, $\ell=3$ modes were included using representative visibility values. However, because of their intrinsically low amplitudes, they are nearly undetectable in most red giants and do not significantly interfere with nearby $\ell=1$ components \citep{aerts:10,mosser:11,benomar:15}. Similarly, $\ell=2$ mixed modes, though formally present, are effectively p-dominated modes \citep{basu:20}. Consequently, their omission in the MCMC analysis and negligible impact on the neural network's performance are consistent with observational limitations and do not affect the primary rotation inferences reported in this work.

We create datasets using a simulator code\footnote{The version used for this paper is available in the {\it Siddarth2023} branch.} \citep{othman_benomar_2023_8296459}, capable of generating synthetic spectra across a wide range of parameters. We have incorporated the frequencies given by asymptotic theory into this simulator. The heights and widths of these peaks are modeled using a variety of templates. 

Specifically, the simulator takes as input a set of global seismic parameters (randomly generated in our case) and produces a corresponding spectrum. To ensure realistic simulations, these spectra also incorporate random noise drawn from a $\chi^2$ distribution with 2 degrees of freedom. To train the machine to distinguish features from noise, we generated multiple realizations of noise for each set of parameters. Using the aforementioned simulator, we constructed two extensive training datasets comprising 1 million high-frequency and 4 million low-frequency samples of red-giant and clump oscillators. A red giant was classified as ``low frequency" if its $\Delta \nu<9\,\mu$Hz, while those with $\Delta \nu\ge9\,\mu$Hz were classified as ``high frequency". This classification was chosen heuristically, considering that a substantial number of He-burning clump stars and a number of H-burning stars show $\Delta \nu < 9.5\,\mu$Hz \citep{mosser:14}. The distributions of both datasets can be found in Table \ref{tab:tab_dataset}. We employed a uniform random distribution to sample most parameters in the training dataset, except for three specific ones: (a) the inclination angle, which was sampled through an isotropic distribution ($P(\iota) \propto \sin\iota$), and (b) both the observation time and signal-to-noise ratio (SNR), which were sampled using distributions derived from Kepler red giants. In creating these datasets, all parameters were treated as free variables.

\begin{table*}[!ht]
\centering
\begin{tabular}{|l|l|l|}
\hline
Parameter & High-frequency red giants & Low-frequency red giants\\
{} & {} & and Red clumps \\
\hline
Range of $\Delta \nu$ & 9-19$\,\mu$Hz & 1-9$\,\mu$Hz\\
\hline
Range of $\nu_{\mathrm{max}}$ & 108-286$\,\mu$Hz & 5-108$\,\mu$Hz\\
\hline
Range of $\Delta \Pi$ & 40-500$\,$s & 40-500$\,$s\\
\hline
Range of $q$ & 0-0.5 & 0-0.65\\
\hline
Range of $\epsilon_p$ & 0-1  & 0-1 \\
\hline
Range of $\epsilon_g$ & 0-1  & 0-1\\
\hline
Range of $\alpha_{\ell}$ & 0.0-0.008  & 0.0-0.008 \\
\hline
Range of Core & 0.005-2.8  & 0.005-2.8 \\
rotation ( in$\,\mu$Hz) & {} & {} \\
\hline
Range of Envelope & 0.005-0.4  & 0.005-0.4 \\
rotation ( in$\,\mu$Hz) & {} & {}  \\
\hline
Range of inclination $\iota$ ( in deg) & 0-90  & 0-90\\
\hline
Frequency range used  & 0-283$\,\mu$Hz  &  0-283$\,\mu$Hz \\
for ML training & {} & {}  \\
\hline
Range of Observation & 9-1460 days & 9-1460 days \\
time & {} & {} \\
\hline
SNR distribution & 8-155 & 8-155\\
\hline
\end{tabular}
\caption{\label{tab:tab_dataset}  The two columns represent parameter spaces in different evolutionary stages of giant stars. The chosen range of parameters is listed in this table, which was used to generate different synthetic datasets. The parameters were carefully selected to cover the full range of published results on \textit{Kepler} data \citep{mosser:15,vrard:16,mosser:17} and ensure that our synthetic data sets are consistent with existing observations.}
\end{table*}

\subsection{Machine learning algorithm}\label{Machine learning algorithm}
Having constructed a suitable training dataset, we utilized a convolutional neural network to infer seismic parameters from the power spectra. The architecture of this neural network comprises six CNN layers, two LSTM cells, and one feed-forward dense layer in sequential order. To downsample the features, we employed a combination of a max-pool and average-pooling layers. Finally, this network connects to six outputs which allow for the inference of the parameters of large-frequency separation $\Delta \nu$, large-period separation $\Delta \Pi$, coupling factor $q$, inclination angle $\iota$, core rotation $\Omega_{\mathrm{core}}/2\pi$, and envelope rotation $\Omega_{\mathrm{env}}/2\pi$.

To infer the seismic parameters, regression is typically used since they are real numbers. However, we have reformulated this as a classification problem by discretizing the parameters into ordinal categories. This requires choosing a suitable bin size that provides both good parameter resolution and sufficient examples per bin for effective training. In our study, we have used bin sizes of 0.1$\,\mu$Hz for $\Delta \nu$, 2.5$\,$s over the range 40-150$\,$s  and 7$\,$s  in the range of 150-500$\,$s  for $\Delta \Pi$, 0.02 for $q$, 0.025$\,\mu$Hz for $\Omega_{env}/2\pi$, 0.095$\,\mu$Hz for $\Omega_{core}/2\pi$ and 5$^\circ$ for $\iota$. The bin sizes were chosen so that the precision of the neural network estimates would match the uncertainty of the classical methods.

We minimize the cross-entropy loss to train the neural network. Since we aim to infer six parameters simultaneously, the total loss is the summation of cross-entropy losses for all parameters, as given by,
\begin{equation} \mathrm{Loss}=\sum_{\mathrm{y} \in \{\Delta \nu, \Delta \Pi, \mathrm{q}, \Omega_{\mathrm{core}}/2\pi, \Omega_{\mathrm{env}}/2\pi, \iota\}} -\log(p(y_{\mathrm{true}})),
\end{equation}
where $\mathrm{y}_{\mathrm{true}}$ represents the true category of the parameter $\mathrm{y}$ used to generate the input power spectrum X and $p(y_{\mathrm{true}})$ denotes the output probability given by the softmax function in the bin $\mathrm{y}_{\mathrm{true}}$. We utilized a total of 5 million (X,$\mathrm{y}_{\mathrm{true}}$) sets for training the neural network. To minimize the loss function, we employed the Adam optimizer and trained the network for 20 epochs. 

After processing an input power spectrum, the trained network produces probability distributions for various parameters, as illustrated in figure \ref{fig:fig1_machine} \footnote{Since we do not directly deduce the rotation rate ratio, we calculate the uncertainties as shown in Figure \ref{fig:fig3_splitting} under the assumption of independent distributions for envelope and core rotation. However, it should be noted that the actual uncertainty related to the ratio might differ from this estimation.}. These distributions are Bayesian \citep{richard:91} and are represented as probabilities in each bin for each parameter. It is important to note that the sum of probabilities in all bins for each parameter is equal to one. The highest value over the output probabilities of each parameter is referred to as confidence and is denoted by $p_{max}$. For N bins, the value of $p_{max}$ can range from 1/N to 1, depending on the confidence of the network prediction. The deduced value, indicated as $\mathrm{y}_{pred}$, corresponds to the median of the probability distribution associated with parameter y.

In order to investigate rotation rates, we carefully choose stars that have well-constrained seismic parameters related to their internal structure. We also impose a minimum threshold on the inclination angle for the selected stars. If the network is capable of accurately determining all the rotational components, it indicates that the aforementioned conditions have been met. The criteria used to select the samples are as follow: (a) $p_{max}(\Delta \nu)>0.3$, (b) $p_{max}(q)>0.15$ and $q_{pred}>0.05$, (c) $p_{max}(\Delta \Pi)>0.2$, (d) $p_{max}(\Omega_{\mathrm{core}}/2\pi)>0.2$, (e) $p_{max}(\Omega_{\mathrm{env}}/2\pi)>0.2$, and (f) $\iota>45^{\circ}$. Additionally, these criteria ensure that the rotation rate inferences have a sufficient level of confidence, which we term ``confident".

\begin{figure*}[!ht]
\centering
\includegraphics[width=0.9\linewidth]{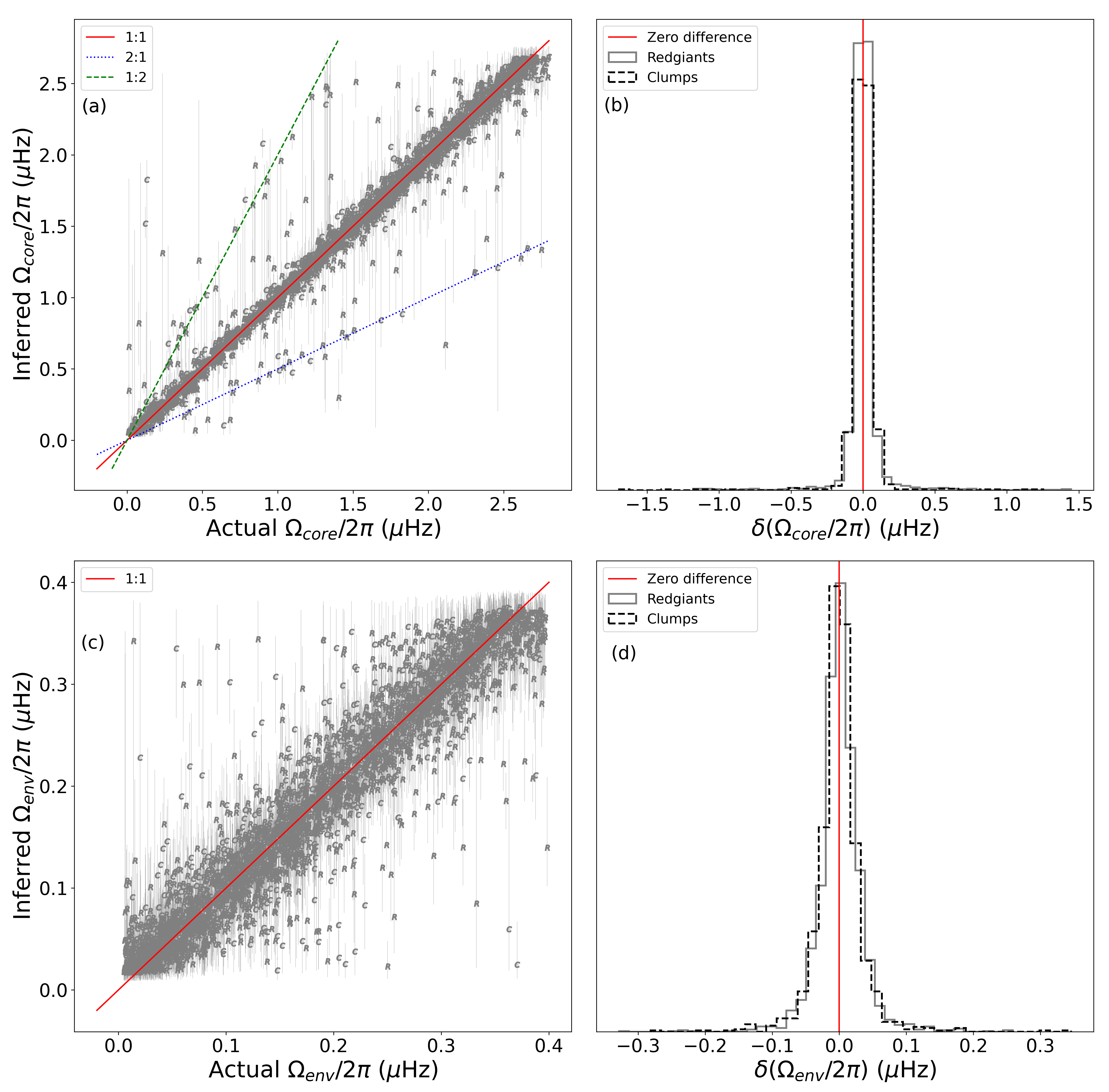}
\caption{Results on synthetic spectra. Shown in panels (a) and (c) are the inferred values of $\Omega_{\mathrm{core}}/2\pi$ and $\Omega_{\mathrm{env}}/2\pi$ plotted against the actual injected values for red giants (R) and clumps (C). Each point is associated with grey lines representing 1-$\sigma$ uncertainties. In panel (a), the red-solid, blue-dotted, and green-dashed lines indicate the 1:1, 1:2, and 2:1 ratios between actual and inferred values, respectively. Some 96\% of the inferences lie around the 1:1 ratio, and the remaining 4\% of inferences are distributed equally around the 1:2 and 2:1 lines. Panels (c) and (d) depict the distributions of the differences between inferred and actual values for red giants (grey-solid line) and clump stars (black-dashed line). The results indicate that core rotation may be inferred to within 0.1$\,\mu$Hz and envelope rotation to within 0.05$\,\mu$Hz of injected values for 90\% of the stars.}
\label{fig:synthetics_results}
\end{figure*}

We present the performance of our network on 5000 unseen examples that meet the specified criterion in figure \ref{fig:synthetics_results}. Our results indicate that 90\% of the inferred core rotation values are within 0.1$\,\mu$Hz of the actual values, and over 95\% of the core-rotation values are near the 1:1 ratio of the actual values. Moreover, we observe that 90\% of the inferred envelope-rotation values are within 0.05$\mu$Hz of the actual values. These findings demonstrate that our network accurately estimates rotation rates from synthetic data.

The neural network is trained on synthetic data based on specific physical assumptions, including a defined range of seismic parameters, rotation profiles, mode visibility, and noise characteristics. As a result, it performs reliably only for stars that fall within this simulated parameter space. For stars that differ from these assumptions, such as those with unusual chemical composition, strong magnetic fields, or unexpected internal rotation, the predictions may be biased or inaccurate. We have reported inferences only where the machine-learning model is confident. %We have reported inferences only for cases where the network is confident. 
However, some biases may still persist, as neural network inferences are not directly explainable in the same way as traditional methods.

\begin{figure*}[!ht]
\centering
\includegraphics[width=\linewidth]{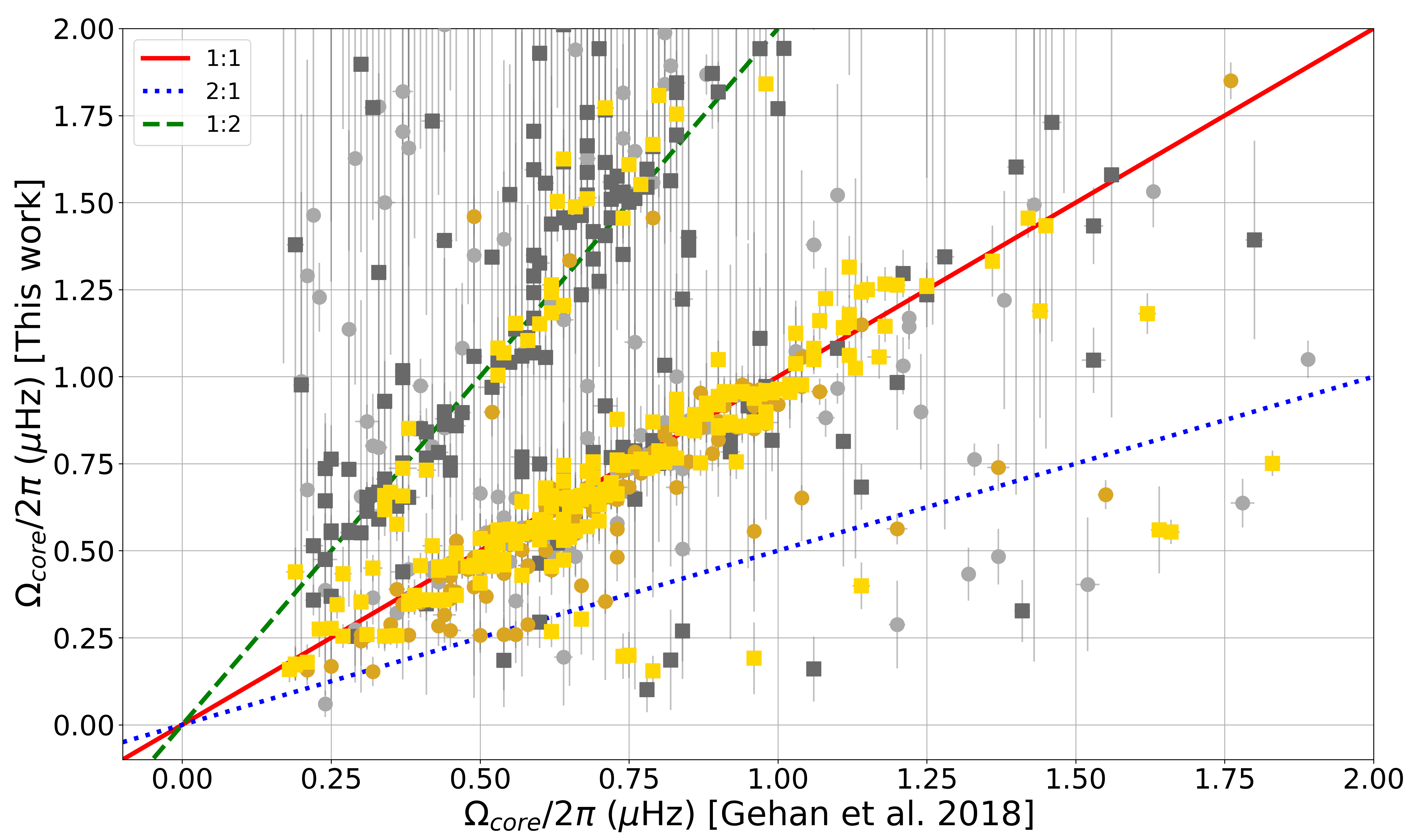}
\caption{Comparison of core rotation rate measurements from this work with those of \citet{gehan:18} for 842 Kepler red giants. Confident neural‑network measurements are highlighted in yellow; the solid red line marks one‑to‑one agreement, while blue dotted and green dashed lines denote 2:1 and 1:2 ratios, respectively. Dark‑yellow circles indicate stars with three rotational components detected by \citet{gehan:18}; light‑yellow squares show stars with two components. Grey bars give 16–84 percent uncertainties.}
\label{fig:rot_study_gehan_latest}
\end{figure*}

\begin{figure*}[!ht]
\centering
\includegraphics[width=\linewidth]{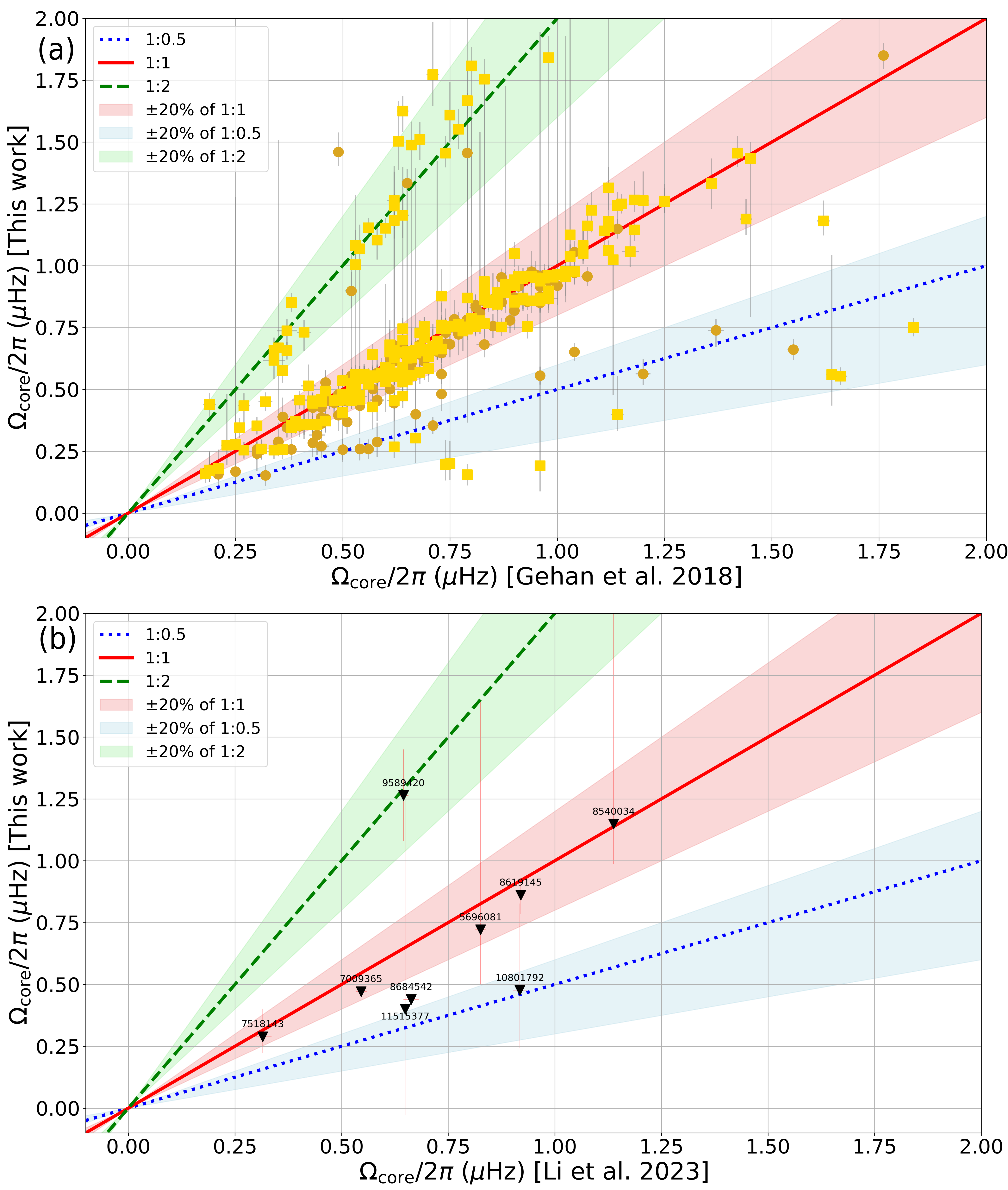}
\caption{(a) Same as Figure~\ref{fig:rot_study_gehan_latest}, but only displaying measurements deemed confident by the network. The red, green, and blue lines indicate 1:1, 1:2, and 1:0.5 ratios, respectively. Shaded bands around each reference line show $\pm\,$20\% bounds. Based on this criterion, 252 stars lie within the 1:1 band, 42 within the 1:2 band, and 24 within the 1:0.5 band.(b) Comparison of core rotation rates for stars with known magnetic fields, as reported by \cite{li:23}. The numbers around each point represent the KIC IDs of the stars.}

\label{fig:rot_study_conf_only_mag}
\end{figure*}

\subsection{Markov Chain Monte Carlo}

To obtain samples from the underlying parameter distributions, we employ Markov Chain Monte Carlo (MCMC) sampling. Unlike other gradient-descent methods such as Maximum Likelihood Estimation (MLE) or Maximum A Posteriori (MAP), MCMC is resilient to the presence of local maxima, allowing it to consistently provide accurate parameter inferences. In our approach, we sample the two-degrees-of-freedom $\chi^2$-likelihood distribution by utilizing the power spectral data of the star $\textbf{y}$ and the oscillation spectrum model $M(\nu,\Theta)$. The methodology we follow for fitting the model to the data and obtaining the underlying parameter distribution is described in \cite{benomar:09}.

The model $M(\nu,\Theta)$ comprises the oscillation spectrum $S(\nu,\Theta_{S})$ and the noise profile $N(\nu,\Theta_{N})$. The oscillation spectrum is computed based on asymptotic theory, with $\Theta_{S}$ representing the seismic parameters such as $\Delta \nu$ and $\Delta \Pi$. On the other hand, the noise profile is obtained by combining two Harvey profiles and white noise, and $\Theta_{N}$ corresponds to the parameters of these different profiles.

The oscillation power spectrum is a summation of Lorentzians, characterized by heights denoted as $H(n,\ell,m)$ and centered around frequencies $\nu(n,\ell,m)$, each with corresponding widths $\Gamma(n,\ell,m)$, such that
\begin{equation}
S(\nu,\Theta_{S})=\sum_{n} \sum_{\ell=0}^{\ell=3} \sum_{m=-\ell}^{m=\ell} \frac{H(n,\ell,m)}{1+4({\frac{\nu - \nu(n,\ell,m)}{\Gamma(n,\ell,m)}})^2} \, .
\end{equation}
The heights in the model are given by $ H(n,\ell,m) = r_{\ell,m}(\iota)V(\ell)A(n)$, where $V(\ell)$ signifies the mode visibility, $A(n)$ varies with radial order, and $r_{\ell,m}(\iota)$ denotes the relative amplitude of the mode, which is dependent on the inclination angle $\iota$. The visibility function is influenced by both the limb-darkening function, which is variable based on the star's type, and the employed measurement technique.

In order to estimate the seismic parameters from the available data, we employ a Bayesian formalism, which allows us to obtain the posterior distribution of the parameters given the observations, denoted as $\pi(\Theta|y,M,I)$. The posterior distribution $\pi(\Theta|y,M,I) \propto \mathcal{L}(y|\Theta,M,I) \pi(\Theta|M,I)$ where $\mathcal{L}(y|\Theta,M,I)$ is the likelihood function, $\pi(\Theta|M,I)$ is the prior distribution. In addition to establishing limits for visually distinguishable $\ell=0,2$ frequencies specific to individual stars, we have implemented uniform priors: 40 - 500$\,$s for $\Delta \Pi$, 0 - 1.0 for $q$, 0-2$\,\mu$Hz for $\Omega_{core}/2\pi$, 0 - 0.3$\,\mu$Hz for $\Omega_{env}/2\pi$, and an isotropic prior for the inclination angle. The likelihood function $\mathcal{L}(y_{i}|\Theta,M_{i},I)$ follows a 2-degree-of-freedom $\chi^2$ distribution due to the observed power spectrum statistics. Specifically, the likelihood function is given by $\mathcal{L}(y_{i}|\Theta,M_{i},I) = \frac{1}{M_{i}} \exp\left(-\frac{y_i}{M_i}\right)$, where $y_i$ represents the observed data at frequency $i$, and $M_i$ corresponds to the model prediction at the same frequency. To sample the posterior distribution, we employ the Metropolis-Hastings algorithm \citep{metropolis:53,hastings:70}, which allows for effective exploration of the parameter space and generation of representative samples from the posterior distribution.

While both the machine-learning model and the MCMC fits use the same prior distribution for seismic parameters, the two methods remain independent. The neural network extracts parameters by identifying patterns from the power spectra, whereas the MCMC approach estimates them through direct likelihood maximization. The shared prior ensures consistency in the assumed parameter space and does not imply methodological coupling or inference bias.

\section{Results on Kepler data}\label{Results on Kepler data}

The Kepler space telescope \citep{borucki:04, borucki:10} has observed 197,906 stars \citep{mathur:17} and detected stellar pulsations in tens of thousands, enabling large-scale asteroseismic studies across various stellar populations. We utilized long-cadence (29.4-minute) light-curve data from \textit{Kepler}, accessed via the \texttt{Lightkurve} software \citep{lightkurve:18}. For each target, all available quarters were downloaded, stitched together to produce continuous light curves, and processed to remove NaN values and outliers. The light curves were then normalized to parts-per-million (ppm) units and transformed into power spectral densities with a maximum frequency cutoff of 283$\,\mu$Hz to encompass the frequency range of long-cadence red giants. The sample of approximately 21000 red giants was selected based on previous catalogs and studies by \citet{hekker:10,stello:13,pinsonneault:14,mathur:16,yu:18,pinsonneault:18,elsworth:19,gaulme:20,yu:20,benbakoura:21,mosser:15,vrard:16,mosser:17,hon:19} and \citet{dhanpal:22}.

The neural network was applied to analyze this full sample. However, rotation-rate estimates may be unreliable for stars with low signal-to-noise ratios (SNR), weak coupling constants, or small inclination angles. As a result, we have opted to curate a subset of 1517 stars from the larger pool of $\sim$21000 Kepler red giants. Only those stars where the machine was able to confidently infer all seismic parameters and rotation rates, and which possessed a minimum inclination angle of 45$^\circ$, were retained in this ensemble analysis. We show these measurements in Table \ref{tab:tab_ml_catalog}.

\subsection{Validation against previous studies}\label{Validation against previous studies}

To assess reliability, our core rotation rate measurements were compared with existing studies \citep{gehan:18} as shown in Figure \ref{fig:rot_study_gehan_latest}. Among the 426 stars that meet the selection criteria, as shown in Figure \ref{fig:rot_study_conf_only_mag}(a), we found that 59.2\% (252 stars) of our measurements are within 20\% of the 1:1 ratio with the published values. Of the remaining stars, 9.9\% (42 stars) fall near the 2:1 green dashed line, and 5.6\% (24 stars) lie near the 1:2 blue dotted line, using the same $\pm\,$20\% slope threshold. The remaining 25.3\% of stars do not fall within any of these defined proximity zones. The uncertainties associated with the network predictions are examined through various tests detailed in the Appendix \ref{Analysis of uncertainties}.

These measurements are inherently challenging because rotationally split modes of different radial orders can become entangled when the rotational splitting exceeds half of the mixed-mode spacing. Under these conditions, labeling the rotationally split modes (m = -1, 1) and the m = 0 peak associated with a specific radial order becomes complex, which can lead to inaccuracies in mode identification. For instance, a spectrum with rotational splitting equal to one-quarter of the mixed-mode spacing at a high inclination angle (90$^\circ$) may appear very similar to a case with rotational splitting equal to one-third of the mixed-mode spacing at a lower inclination angle (55$^\circ$) \citep{gehan:18}. Consequently, discrepancies with prior studies or ground-truth values frequently arise at rotational-to-mixed-mode spacing ratios of 1:2 or 2:1. Most points around the 1:2 line were detected with only two rotational components by the authors \citep{gehan:18}, and most points around the 2:1 distribution were found to have all three components. This pattern suggests either that the network may have omitted or erroneously detected a component in some cases or that previous studies \citep{gehan:18} may have missed or misidentified an additional component.

It is crucial to highlight that previous methodologies \citep{gehan:18}, while generally effective in determining rotation rates, lack consideration for amplitudes. In earlier methods, the impact of the inclination angle is incorporated solely by adjusting the observable number of components that fit the data. However, the amplitudes of individual components vary with inclination, while the power summed over all three components remains constant, as shown in equation \ref{eq:eq_inc}. Our training data and synthetic models incorporate this amplitude variation, enhancing the precision of rotation rate determination. This inclusion also provides an explanation for the discrepancies observed in approximately 29\% of the stars analyzed using earlier methodologies. These results demonstrate that confident rotation rate measurements are reliable in most cases.

To assess the robustness of our rotation inferences in the presence of strong internal magnetic fields, we compared our core rotation estimates with values reported by \citet{li:23} for a sample of stars exhibiting magnetic signatures. As shown in Figure~\ref{fig:rot_study_conf_only_mag}(b), five out of nine stars fall near the 1:1 line, indicating that the neural network predictions remain consistent with the published values for a subset of magnetically active red giants. Although internal magnetic fields can induce asymmetries in mode splittings \citep{li:22,deheuvels:23,bugnet:21,mathis:21}, this comparison suggests that the network's core rotation estimates are not systematically biased in these cases.

\begin{figure*}[!ht]
\centering
\includegraphics[width=\linewidth]{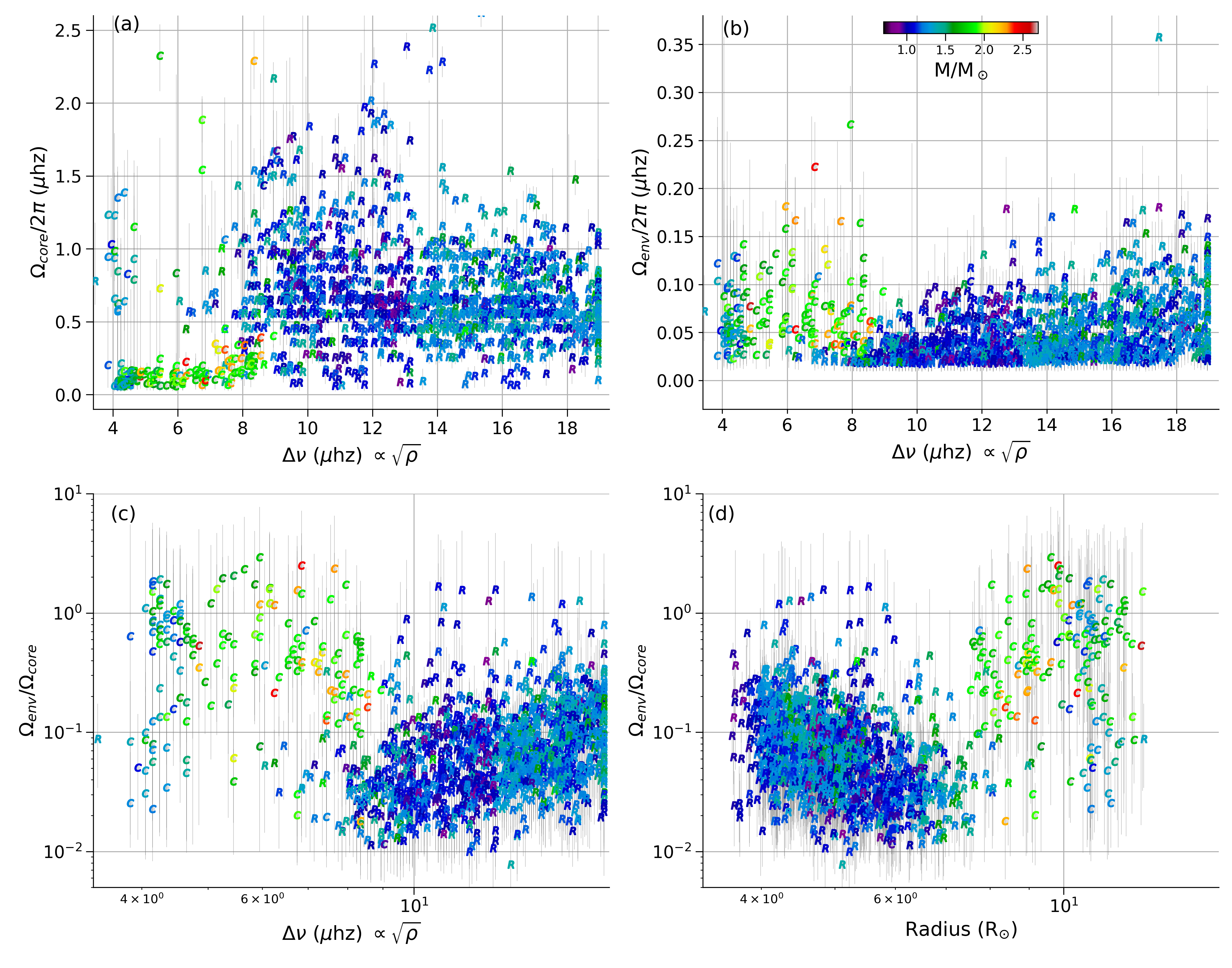}
\caption{ Evolution of core and envelope rotation. Plotted are (a) core rotation as a function of large frequency separation ($\Delta \nu$); (b) envelope rotation as a function of $\Delta \nu$; (c) envelope-to-core rotation ratio versus $\Delta \nu$; and (d) envelope-to-core rotation ratio versus stellar radius. Red giants and clump stars are denoted by “R” and “C,” respectively. Error bars for panels (a) and (b) are shown in grey.}
\label{fig:fig2_rot_trend}
\end{figure*}

\begin{figure*}[!ht]
\centering
\includegraphics[width=\linewidth]{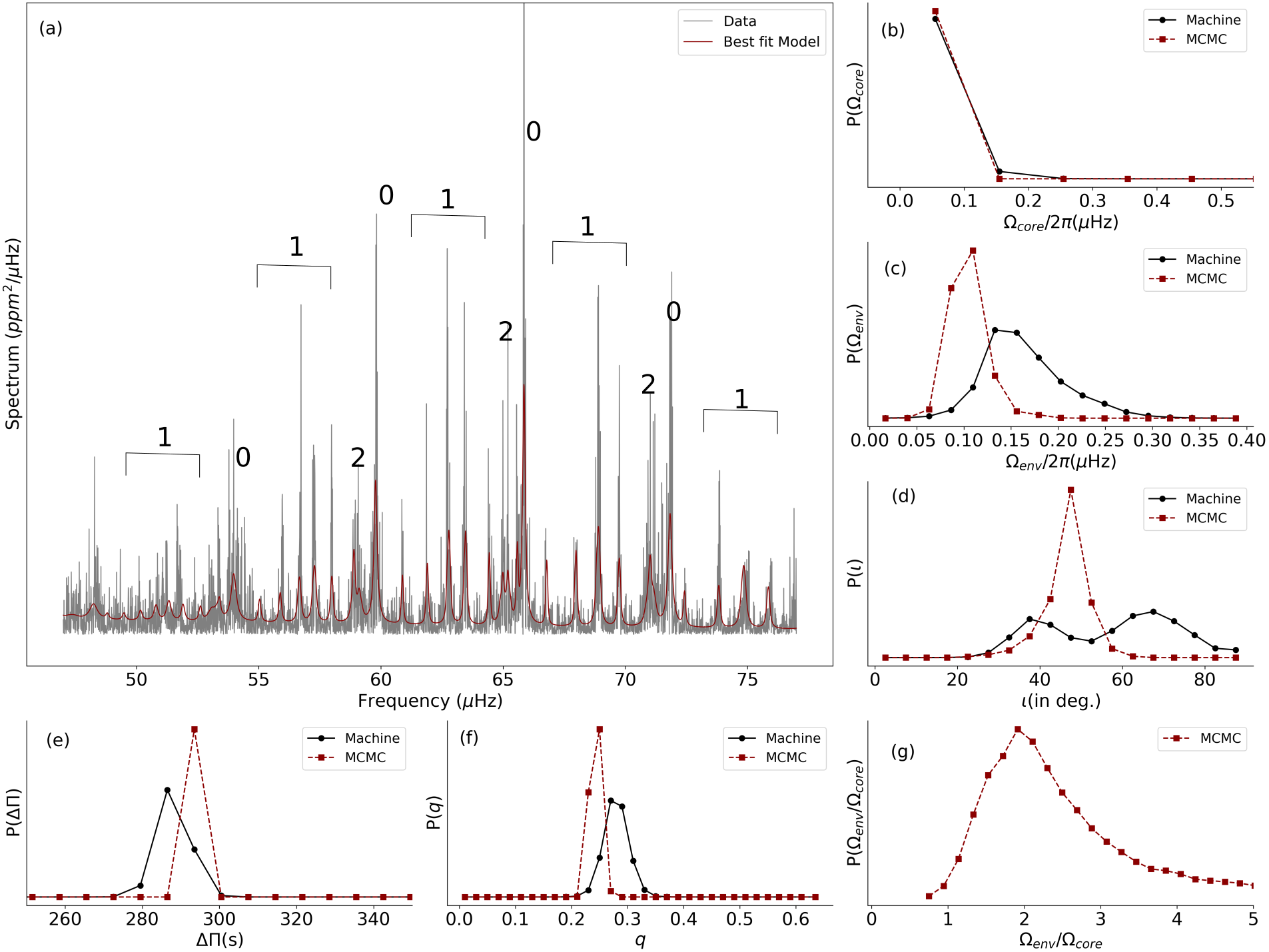}
\caption{MCMC fit to KIC 11615944. The best-fit model is shown in comparison with the observations in panel (a). Panels (b), (c), (d), (e), (f), and (g) show the posterior distributions of $\Omega_{core}$, $\Omega_{env}$, $\iota$, $\Delta \Pi$, $q$ and $\Omega_{env}/\Omega_{core}$, respectively, in red-dashed lines. In addition to these distributions, each panel from (a)-(f) displays the inferences obtained from the neural network. Panel (a) demonstrates a satisfactory alignment between the MCMC fit and the data. Furthermore, panels (b)-(f) exhibit a consistent agreement between the posterior distributions derived from MCMC and the inferences made by the neural network. Notably, panel (g) exhibits the posterior distribution of $\Omega_{\text{env}}/\Omega_{\text{core}}$, indicating a ratio greater than 1. Additional details of the best-fit model are presented in figure \ref{fig:detailed_fit_11615944}.}
\label{fig:extended_fig_2_fit_11615944}
\end{figure*}

\subsection{Evolution of rotation rates in red giants}

As solar-like stars undergo evolution, it is well-established that stellar density decreases and as a consequence, the large frequency separation $\Delta \nu$ decreases \citep{stello:09}. Therefore, we employ $\Delta \nu$ as a proxy for evolutionary progression, which, for stars of similar mass, may also reflect ages, and investigate the rotation rates as a function of this parameter. Red clump stars, in the phase of core helium burning, exhibit a significant period separation $\Delta \Pi$ ($>$150 s), which we utilize to distinguish between red giants and clump stars \citep{bedding:11,mosser:14}.

Figure~\ref{fig:fig2_rot_trend} shows the evolution of the core and envelope rotation as functions of seismic parameters. In panel (a), core rotation rates show a pronounced decline from the red giant branch to the clump phase, consistent with angular-momentum redistribution during evolution. A few clump stars exhibit unexpectedly high core rotation. Panel (b) highlights a narrowing in the envelope rotation rate dispersion during the red-giant phase, followed by a broader spread in the clump phase. Panels (c) and (d) illustrate a sharp difference in the envelope-to-core rotation ratio between red giants and clump stars. The data suggest that some red-giant and clump stars exhibit envelope rotation rates surpassing their core rotation rates, indicating potential deviations from standard angular-momentum transport models.

\subsubsection{Evolution of core rotation}

As shown in Figure \ref{fig:fig2_rot_trend}(a), during the course of their evolution from $\Delta \nu$ of $\sim$19$\,\mu$Hz to $\sim$13$\,\mu$Hz, red giants demonstrate a nearly uniform distribution of core rotation rates, between 0-1.5$\,\mu$Hz. The spread of rotation rates increases to 0-2$\,\mu$Hz in red giants with $8\,\mu$Hz$\lesssim\Delta \nu\lesssim12\,\mu$Hz. In red clump stars with $\Delta \nu \lesssim 8\,\mu$Hz, the core rotation rate drops below 0.25$\,\mu$Hz. However, we also observe some anomalously fast rotating cores in both the clump and red-giant phases.

\subsubsection{Evolution of envelope rotation}

Figure \ref{fig:fig2_rot_trend}(b) reveals that as the red giants progress from $\Delta \nu$ of $\sim$19\,$\mu$Hz to $\sim$8$\,\mu$Hz, the envelope rotation rate dispersion diminishes from 0-0.15$\,\mu$Hz to 0-0.03$\,\mu$Hz. In red clump stars with $\Delta \nu \lesssim 8\,\mu$Hz, the range of the envelope rotation rates escalates to 0-0.2$\,\mu$Hz. The higher envelope rotation rates in clump stars is likely related to the correspondingly larger average masses, as seen in Figure \ref{fig:fig2_rot_trend}.

\subsubsection{Evolution of rotation-rate ratio}

As illustrated in figures \ref{fig:fig2_rot_trend}(c,d), the evolutionary progression of red giants, characterized by an increase in their radii from $\sim$3\,$\mathrm{R_{\odot}}$ to $\sim$8$\,\mathrm{R_{\odot}}$ and a decrease in $\Delta \nu$ from $\sim$19$\mu$Hz to $\sim$8$\,\mu$Hz, coincides with a sharp decline in the distribution of the envelope-to-core rotation rate ratio from $\sim$[0-0.6] to $\sim$[0-0.08]. Subsequently, in the red-clump phase, the expansion of the stellar core causes a corresponding increase in the envelope-to-core rotation ratio to $\sim$[0.01-4] while the radii of these stars grow from $\sim$8$\,\mathrm{R_{\odot}}$ to $\sim$10.5$\,\mathrm{R_{\odot}}$. 

\subsection{Anomalous rotators}

Figure~\ref{fig:fig2_rot_trend} reveals a group of red giants and clump stars in which their envelopes rotate faster than their cores. In addition to these cases, several clump stars exhibit unusually rapid core rotation. Together, these stars represent a class of anomalous rotators.

We have sought to ensure that the values of seismic parameters related to structure for anomalous rotators align with the distributions established in the literature as presented in Appendix \ref{Parameter set of Anomalous stars}. Furthermore, we study some of the anomalous stars with envelopes rotating faster than the cores, as well as the rapidly rotating cores, through the use of Markov Chain Monte Carlo (MCMC) algorithm. This investigation helps us corroborate the finds of  unexpected rotational phenomena.

To illustrate the anomalous cases, we provide an exemplary instance from each category of these peculiar stars. KIC 11615944 is a clump star with an envelope that rotates faster than its core, KIC 11546972 is a red giant branch star with an envelope that rotates faster than its core, and KIC 66955665 is a clump star with a rapidly rotating core. 

\begin{figure*}[!ht]
\centering
\includegraphics[width=\linewidth]{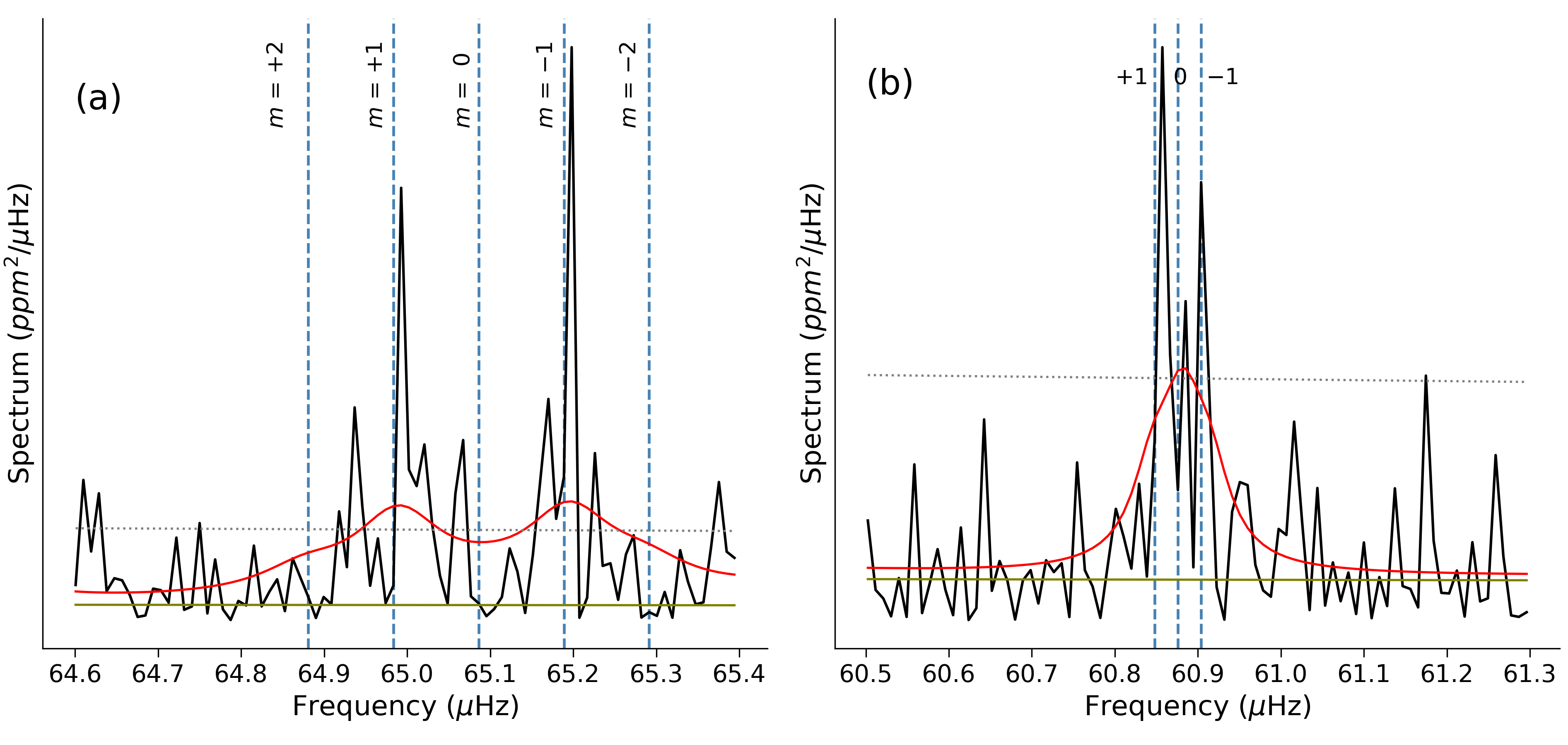}
\caption{Comparison between splittings of an $\ell=2$ pressure-dominated mode and a gravity-dominated $\ell=1$ mixed mode in KIC 11615944. In panel (a), the $\ell=2$ pressure-dominated mode is shown, while panel (b) depicts the gravity-dominated $\ell=1$ mixed mode. The red line depicts the best-fit model within the corresponding frequency range. The olive and grey-dotted lines indicate the noise baseline and 95\% confidence threshold relative to the baseline, respectively. Additionally, the blue-dashed vertical lines mark the frequency locations for reference.}
\label{fig:fig3_splitting}
\end{figure*}

\begin{figure*}[!ht]
\centering
\includegraphics[width=\linewidth]{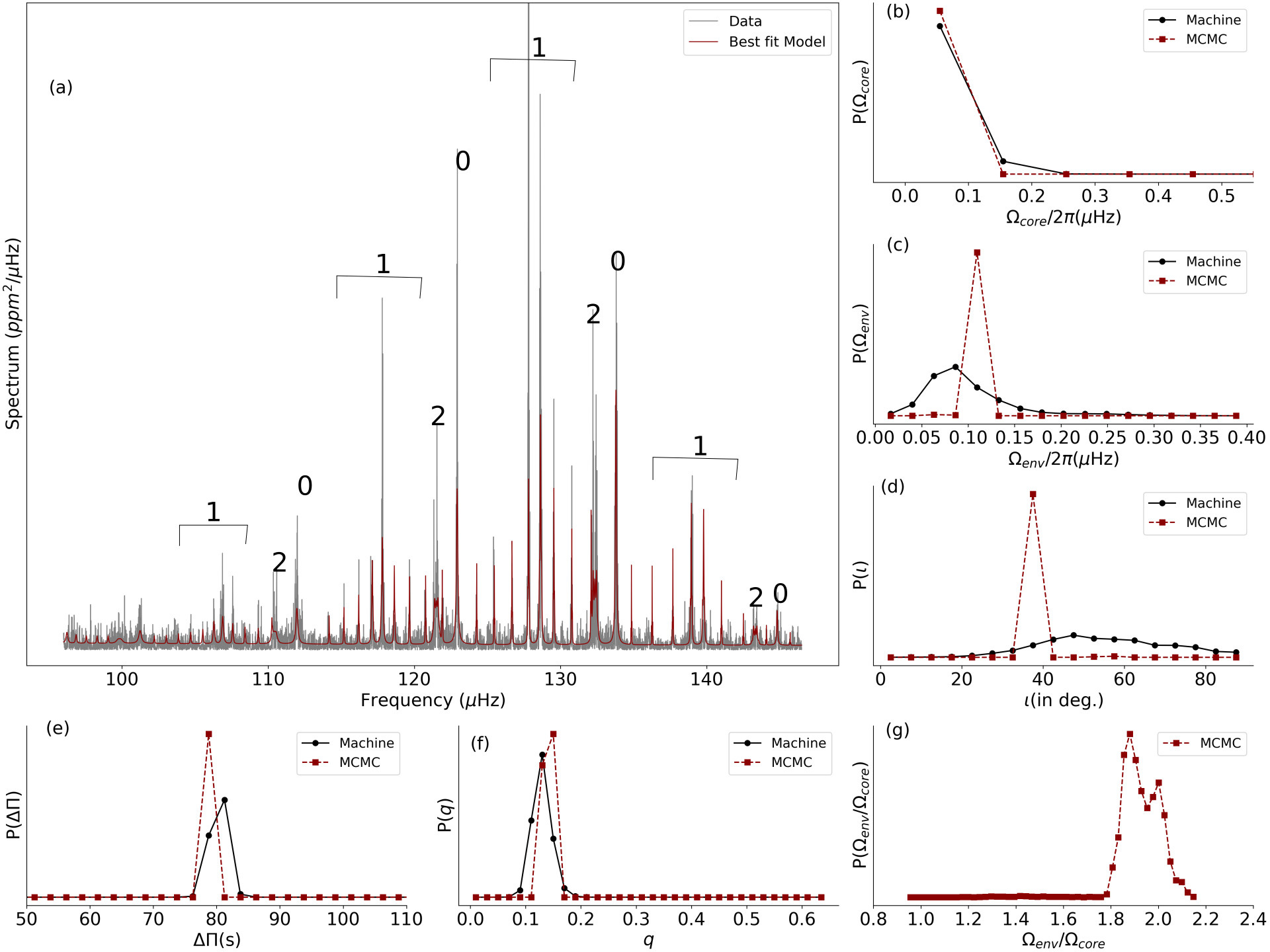}
\caption{MCMC fit to KIC 11546972. The best-fit model is shown in comparison with the data in panel (a). Panels (b), (c), (d), (e), (f), and (g) show in red dashed lines the posterior distributions of $\Omega_{core}$, $\Omega_{env}$, $\iota$, $\Delta \Pi$, $q$ and $\Omega_{env}/\Omega_{core}$, respectively. In addition to these distributions, each panel from (a)-(f) displays the inferences obtained by the neural network. Panel (a) exhibits a good agreement between the MCMC fit and the data. Additionally, panels (b)-(f) exhibit a consistent agreement between the posterior distributions derived from MCMC and the inferences made by the neural network. Notably, panel (g) exhibits a posterior distribution of $\Omega_{\text{env}}/\Omega_{\text{core}}$, indicating a ratio greater than 1. Additional details of the best-fit model are presented in figure \ref{fig:detailed_fit_11546972}.}
\label{fig:extended_fig_2_fit_11546972}
\end{figure*}

\begin{figure*}[!ht]
\centering
\includegraphics[width=\linewidth]{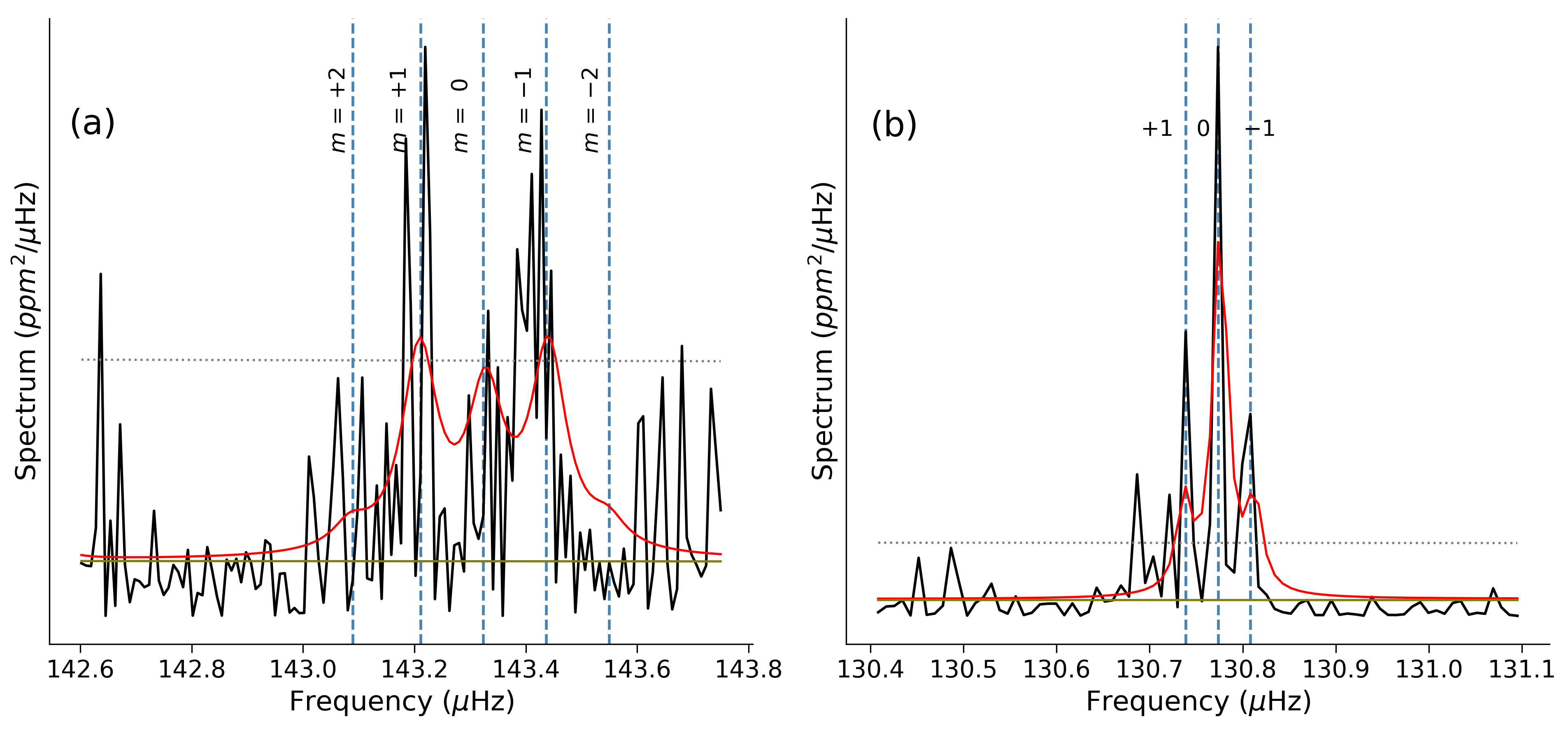}
\caption{Comparison between the splittings of an $\ell=2$ pressure dominated mode and a gravity-dominated mixed mode in KIC 11546972. In panel (a), the $\ell=2$ pressure dominated mode is shown, while panel (b) depicts the gravity-dominated mixed mode. The red line depicts the best-fit model within the corresponding frequency range. The olive line and grey-dotted line indicate the noise baseline and 90\% confidence threshold relative to the baseline, respectively. Additionally, the blue-dashed vertical lines depict the frequency locations for reference.}
\label{fig:extended_fig_1_splitting}
\end{figure*}

For the first of these (KIC 11615944), the neural network determines that $\Omega_{\mathrm{env}}/2\pi = 0.14^{+0.05}_{-0.03}\,\mu$Hz (figure \ref{fig:extended_fig_2_fit_11615944}(c)), and  $\Omega_{\mathrm{core}}/2\pi = 0.03^{+0.02}_{-0.02}\,\mu$Hz (figure \ref{fig:extended_fig_2_fit_11615944}(b)). Further analysis through MCMC reveals that $\Omega_{\mathrm{env }}/2\pi = 0.10^{+0.02}_{-0.02}\,\mu$Hz (figure \ref{fig:extended_fig_2_fit_11615944}(b)), $\Omega_{\mathrm{core}}/2\pi = 0.05^{+0.02}_{-0.02}\,\mu$Hz (figure \ref{fig:extended_fig_2_fit_11615944}(c)), and $\Omega_{\mathrm{env}}/\Omega_{\mathrm{core}}=2.30^{+1.75}_{-0.78}$ (figure \ref{fig:extended_fig_2_fit_11615944}(f)). The MCMC fit aligns well with the data, as depicted in figure \ref{fig:extended_fig_2_fit_11615944}(a), and all the parameter inferences from the neural network agree with the MCMC posterior distributions to within a 1-$\sigma$ margin, as shown in the various panels of figure \ref{fig:extended_fig_2_fit_11615944}.

Figure \ref{fig:fig3_splitting} displays the splittings of a $g-$dominated mixed mode and an $\ell=2$ $p$-dominated mode in KIC 11615944. Visual inspection reveals that all the azimuthal numbers of the $\ell=2$ mode, ranging from $m=-2$ to $m=+2$ in figure \ref{fig:fig3_splitting}(a), are uniformly separated by an approximate value of 0.12$\,\mu$Hz, which is consistent with the fitted value of $\Omega_{\mathrm{env}}/2\pi$. Moreover, in figure \ref{fig:fig3_splitting}(b), the rotational components of a $g$-dominated mixed-mode exhibit a separation of $\sim0.025\,\mu$Hz which is in agreement with the fitted value of $\Omega_{\mathrm{core}}/2\pi$. In light of these lines of evidence, it may be concluded that the envelope rotates faster than the core. 

The neural network inference for red giant KIC 11546972 is $\Omega_{\mathrm{core}}/2\pi=0.06^{+0.04}_{-0.04}\,\mu$Hz (figure \ref{fig:extended_fig_2_fit_11546972}(b)), and $\Omega_{\mathrm{env}}/2\pi=0.09^{+0.05}_{-0.03}\,\mu$Hz (figure \ref{fig:extended_fig_2_fit_11546972}(b)). These values indicate that the envelope may be rotating faster than the core. In-depth analysis using MCMC reveals that $\Omega_{\mathrm{core}}/2\pi=0.06^{+0.01}_{-0.01}\,\mu$Hz (figure \ref{fig:extended_fig_2_fit_11546972}(b)), $\Omega_{\mathrm{env}}/2\pi=0.11^{+0.01}_{-0.01}\,\mu$Hz (figure \ref{fig:extended_fig_2_fit_11546972}(c)), and $\Omega_{\mathrm{env}}/\Omega_{\mathrm{core}}=1.90^{+0.08}_{-0.06}$ (figure \ref{fig:extended_fig_2_fit_11546972}(f)). The MCMC fit shows good alignment with the data, as illustrated in figure \ref{fig:extended_fig_2_fit_11546972}(a), and all the parameter inferences from the neural network are in agreement with MCMC distributions within a 1-$\sigma$ margin, as indicated in the various panels of figure \ref{fig:extended_fig_2_fit_11546972}.

In Figure \ref{fig:extended_fig_1_splitting}, we observe the splittings of a $g$-dominated mixed mode and an $\ell=2$ $p$-dominated mode in KIC 11546972. The uniform separation of all azimuthal modes (ranging from $m=-2$ to $m=+2$) is approximately 0.11$\,\mu$Hz, as seen in Figure \ref{fig:extended_fig_1_splitting}(a), which is consistent with the fitted value of $\Omega_{\mathrm{env}}/2\pi$. Additionally, figure \ref{fig:extended_fig_1_splitting}(b) shows a $g$-dominated mixed mode whose components (ranging from $m=-1$ to $m=+1$) are spaced apart by approximately 0.03$\,\mu$Hz, consistent with $\Omega_{\mathrm{core}}/2\pi$. These observations suggest strongly that the envelope rotates at a faster rate than the core.

\begin{figure*}[!ht]
\centering
\includegraphics[width=\linewidth]{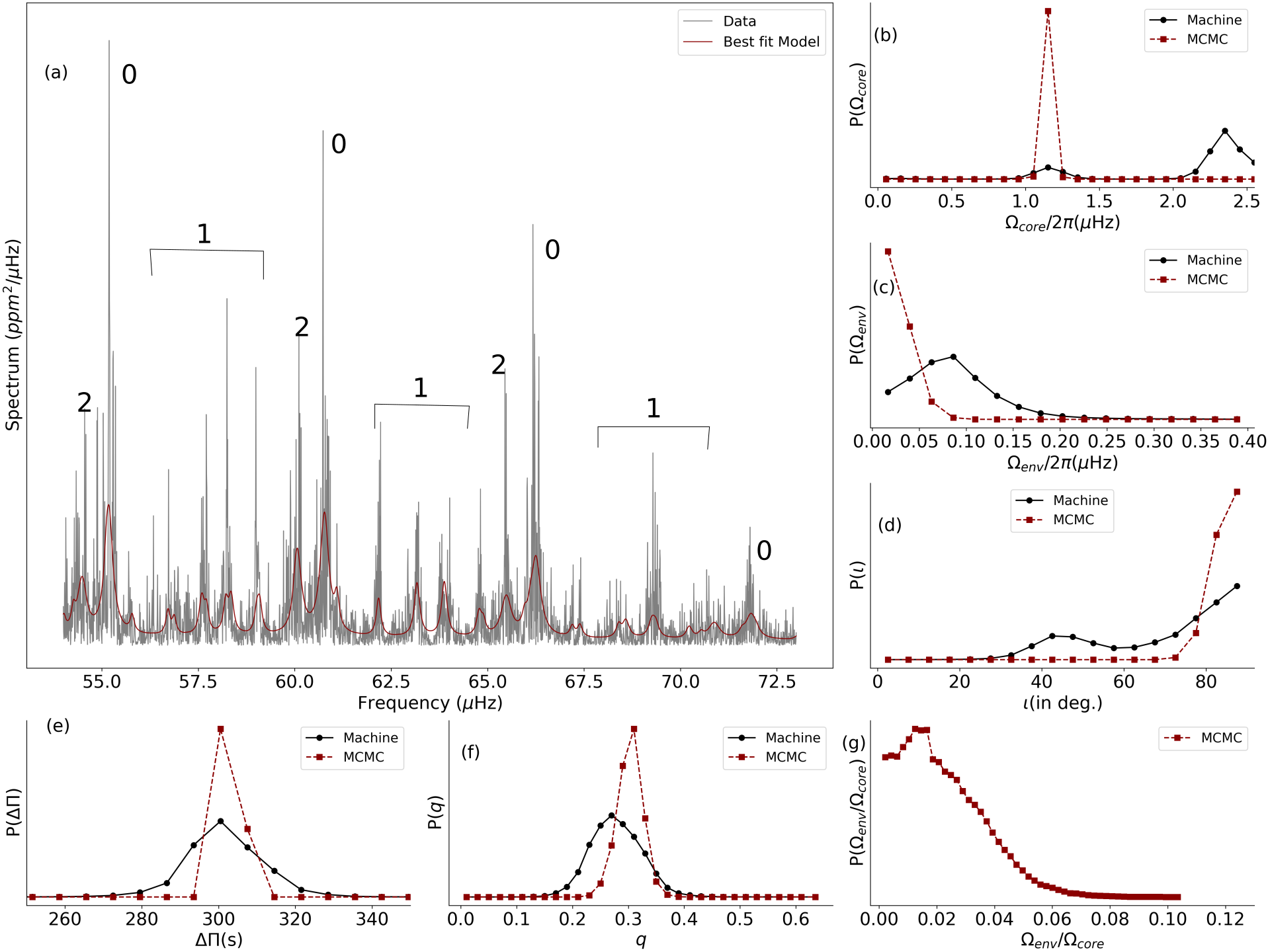}
\caption{MCMC fit to KIC 6695665. The best-fit model is shown in comparison with the observations in panel (a). Panels (b), (c), (d), (e), and (f) show in red-dashed lines the posterior distributions of $\Omega_{core}$, $\Omega_{env}$, $\Delta \Pi$, $q$, and $\iota$, respectively. In addition to these distributions, each panel displays the inferences obtained from the neural network. Panel (a) illustrates the concordance between the MCMC fit and the observed data. In addition, panel (b) reveals that the posterior distribution of $\Omega_{\text{core}}/2\pi$ obtained through MCMC aligns with the first peak of the neural network inference, which is characterized by a bimodal distribution. Additional details of the best-fit model are presented in figure \ref{fig:detailed_fit_6695665}.}
\label{fig:extended_fig_2_fit_6695665}
\end{figure*}

In regard to the clump star KIC 6695665, the neural-network inference of the core-rotation rate depicts a bimodal distribution with two distinct peaks located around 1.2$\,\mu$Hz and 2.3$\,\mu$Hz. In either case, this star's rotation rate greatly exceeds that of typical clump cores, by a factor of either 10 to 20 relative to the median clump rotation rate of 0.1$\,\mu$Hz. MCMC analysis further reveals that $\Omega_{\mathrm{core}}/2\pi=1.16^{+0.03}_{-0.03}\,\mu$Hz (figure \ref{fig:extended_fig_2_fit_6695665}(b)), in agreement with the machine's inference of 1.2$\,\mu$Hz. As this star exhibits fast rotation, distinguishing individual azimuthal components visually is challenging. However, the MCMC fit illustrated in figure \ref{fig:extended_fig_2_fit_6695665}(a) demonstrates good alignment with the data. Additional details of the best-fit models of the three rapid rotators are presented in appendix \ref{Analysis of best-fit models}.

\begin{figure*}[!ht]
\centering
\includegraphics[width=\linewidth]{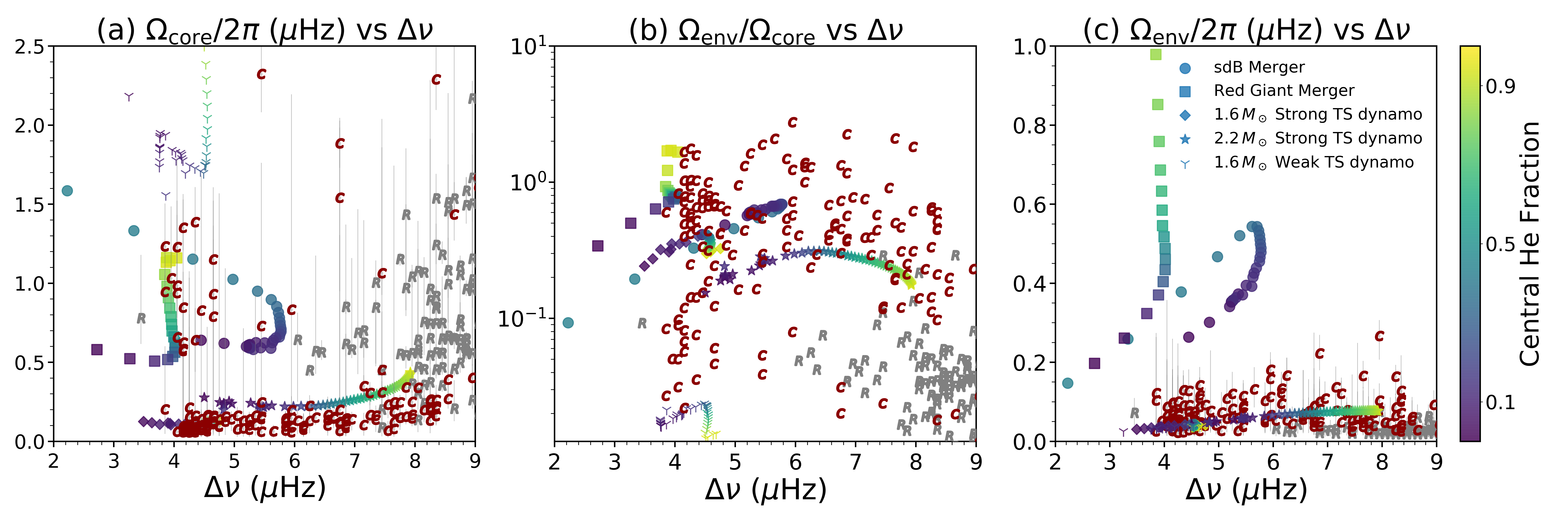}
\caption{Comparison between the inferred rotation rates from the neural network (red Cs and gray Rs) and several theoretical models (indicated in the legend). In panel (a), the red-giant and sub-dwarf binary merger models potentially account for the population of rapid clump rotators. Panel (b) reveals that the red-giant merger model also produces envelopes that rotate faster than the stellar cores. However, panel (c) demonstrates that envelope rotation rates of the merger models are too fast compared to observed clump stars, including those with rapidly rotating cores and super-rotating envelopes.}
\label{fig:fig4_theory}
\end{figure*}

\begin{figure*}[!ht]
\centering
\includegraphics[width=0.9\linewidth]{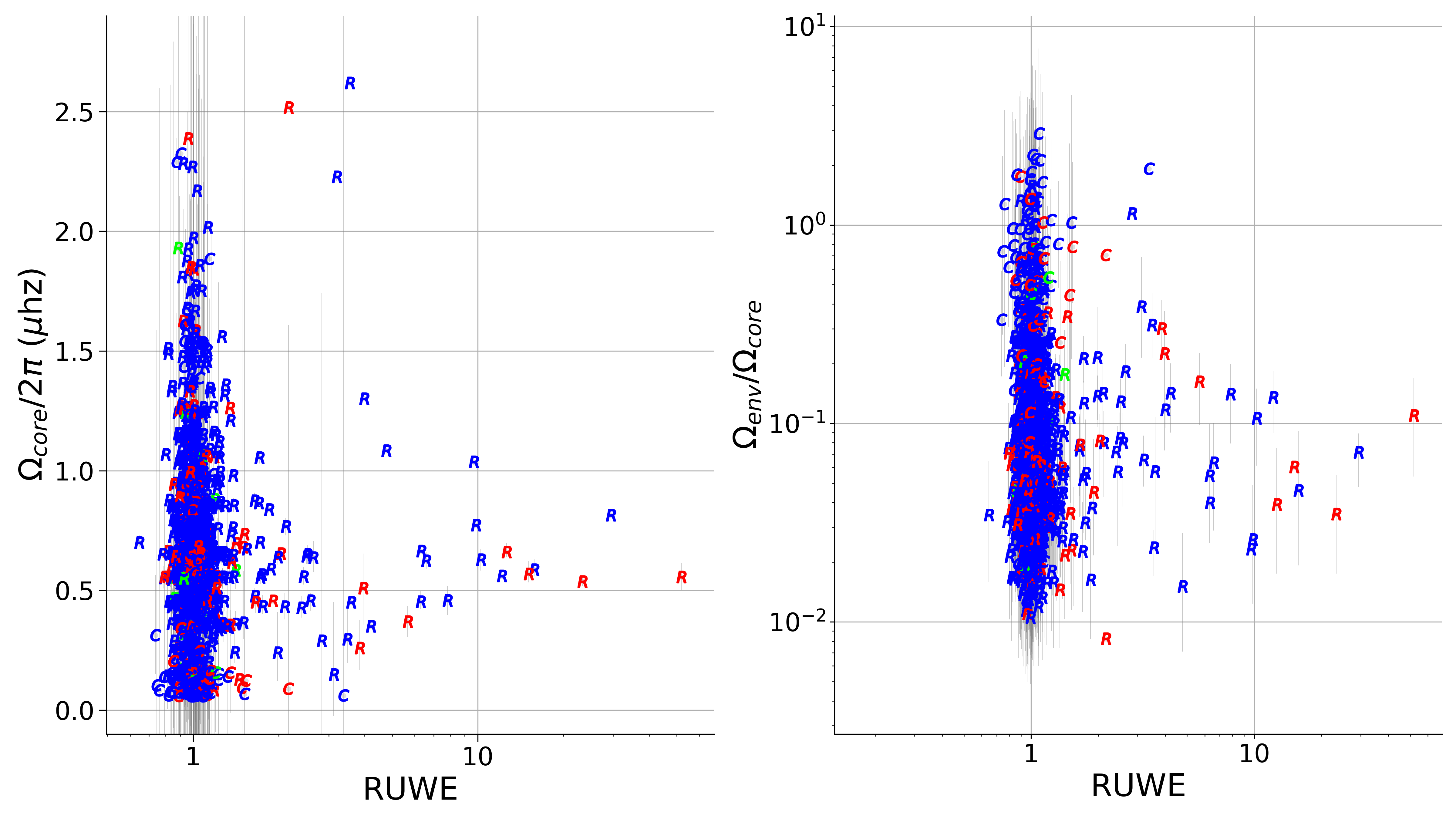}
\caption{Variation of core rotation (left) and rotation rate ratio (right) with RUWE (Re-normalized unit-weight error) parameter obtained from GAIA. RUWE significantly greater than 1 can potentially indicate non-single systems. The color indicates the number of resolved companions associated with the corresponding system (blue - 1 resolved companion, red - 2 resolved companions and green - 3 resolved companions). The symbols R and C denote red giant and clump star, respectively.}
\label{fig:ruwe_rotation}
\end{figure*}

\begin{figure*}[!ht]
\centering
\includegraphics[width=0.9\linewidth]{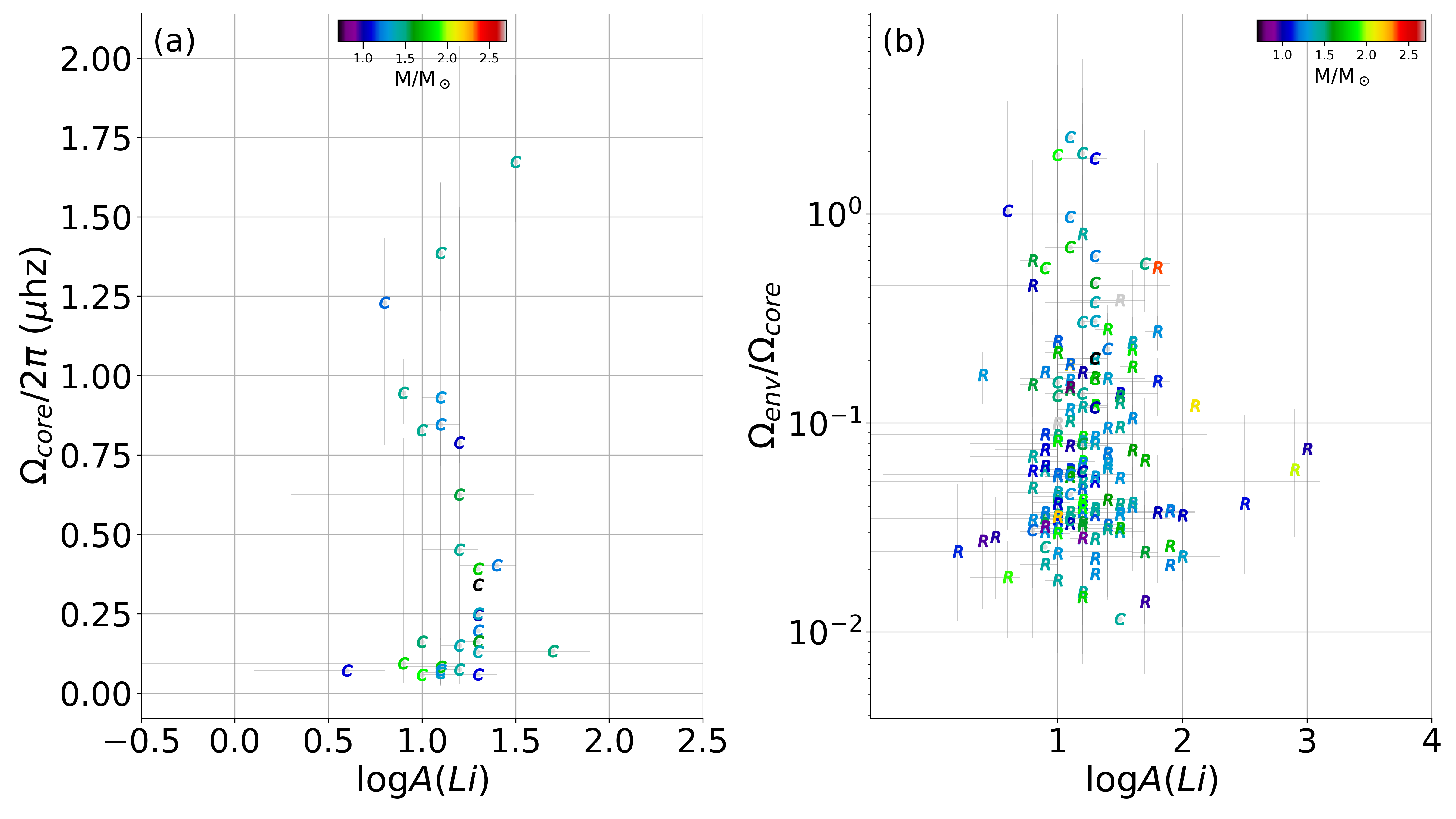}
\caption{(a) Core rotation rate vs. Lithium abundance distribution in 29 red clump stars. This plot focuses exclusively on clump stars since we identify anomalous core rotators mainly in the clump phase. (b) Rotation rate ratio vs. Lithium abundance distribution in 166 stars. These plots indicate that anomalous rotators do not exhibit any significant Lithium-abundance anomalies.}
\label{fig:li_rotation}
\end{figure*}

\begin{figure*}[!ht]
\centering
\includegraphics[width=0.9\linewidth]{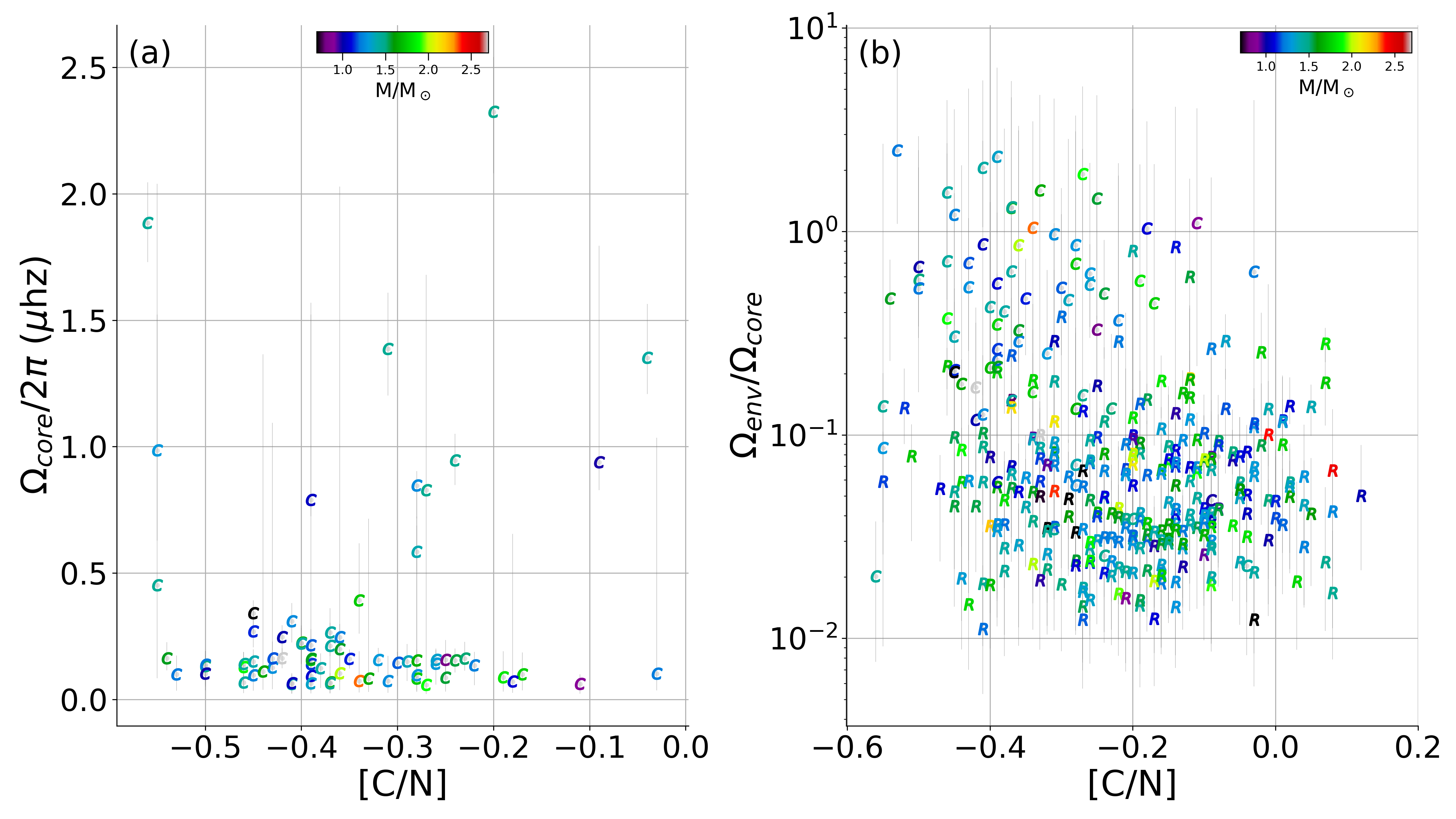}
\caption{(a) Core rotation rate vs. [C/N] abundance distribution in 72 red clump stars. (b) Rotation-rate ratio vs. [C/N] abundance distribution in 358 stars. These plots indicate that anomalous rotators do not exhibit unusual [C/N].}
\label{fig:carbon_rotation}
\end{figure*}

\begin{figure*}[!ht]
\centering
\includegraphics[width=0.9\linewidth]{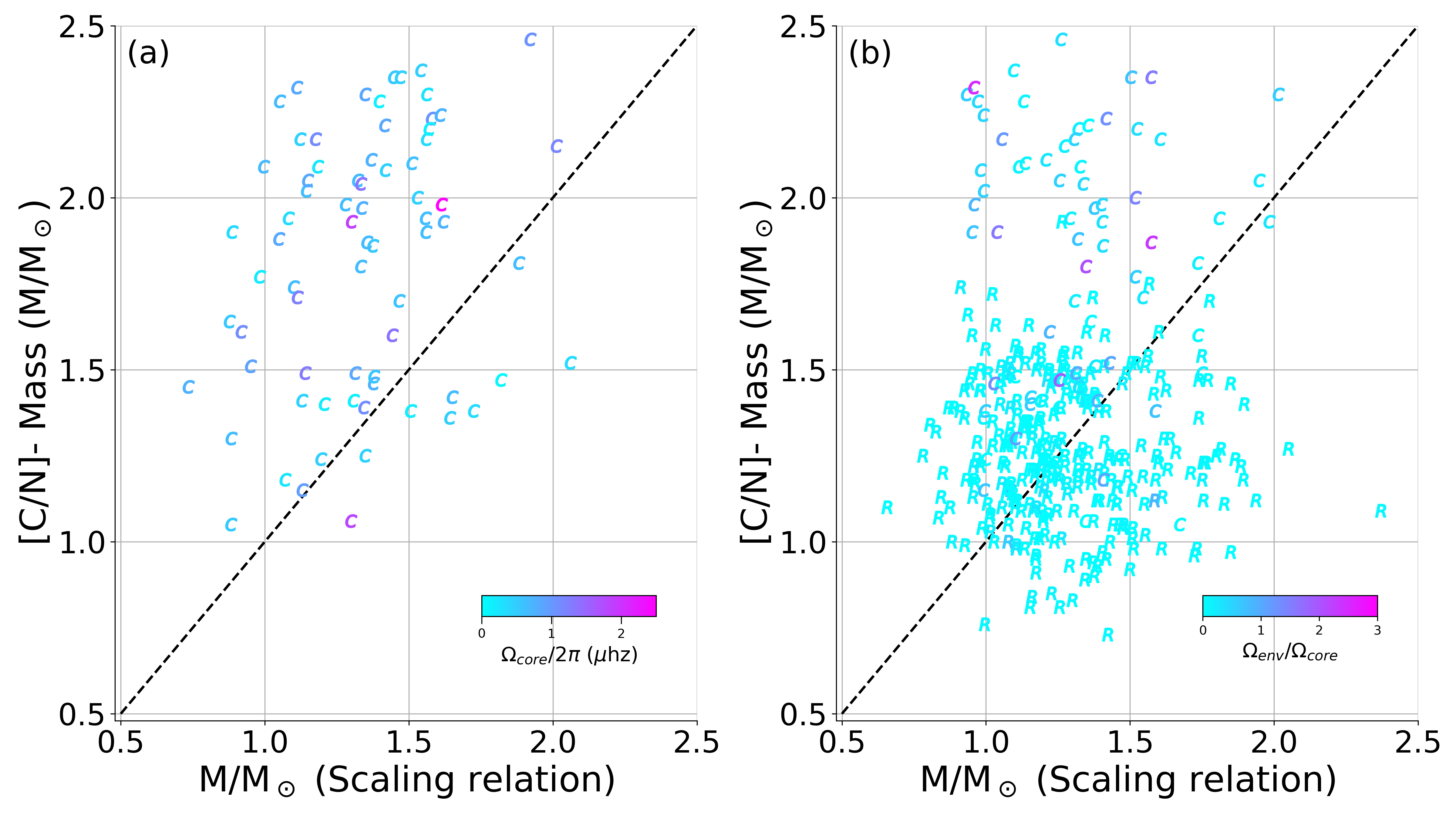}
\caption{(a) Predicted mass using [C/N] abundance \cite{martig:16} vs. estimated mass using scaling relations in 72 clump stars, where the colors of the points indicate core rotation rates. (b) Predicted mass \citep{martig:16} vs. estimated mass using scaling relations in 358 stars, where the colors indicate the rotation-rate ratios in this case. Predictions are generally overestimated, and anomalous rotators fall within the usual distribution.}
\label{fig:mass_cn_comparison}
\end{figure*}

\section{Stellar Models}\label{Stellar Models}

Most core and envelope rotation rates measured here are similar to prior measurements \citep{mosser:12,gehan:18}. These measured rotation rates are slower by a factor of ten or more than predicted by most angular‑momentum‑transport prescriptions in stellar‑evolution codes \citep{cantiello:14}, but typically remain within a factor of $\sim$2 of models that incorporate strong angular‑momentum transport through a modified version of the Tayler–Spruit dynamo \citep{fuller:19,spruit:02}, as shown in Figure~\ref{fig:fig4_theory}.

To predict core and envelope rotation rates, we built single‑star models with the MESA stellar‑evolution code \citep{paxton:11}. These models, taken directly from \cite{fuller:19}, include magnetic angular momentum transport via both the original Tayler–Spruit dynamo \citep{spruit:02} and a modified, more efficient version \citep{fuller:19}. The latter approximately reproduces the rotation rates of ordinary RGB and red‑clump stars, though significant discrepancies persist, especially for subgiant stars \citep{eggenberger:17}.

However, a newly identified population of clump stars with rapidly rotating cores ($\Omega_{\rm core}/2\pi\sim1\,\mu{\rm Hz}$) rotates roughly ten times faster than the modified Tayler–Spruit models \citep{fuller:19} predict and lies closer to the predictions of the original dynamo. These stars appear to form a separate population of predominantly low‑mass clump stars, suggesting a distinct formation pathway.

One possibility is that these stars experience weaker internal angular‑momentum transport, allowing their cores to spin up as they contract during the red‑giant phase. Figure~\ref{fig:fig4_theory} shows that models implementing the original Tayler–Spruit prescription produce core rotation rates roughly twenty times larger than those with more efficient transport \citep{fuller:19}, only slightly faster than the observed rapid rotators. Recent simulations of the Tayler instability reveal a bifurcation of magnetic‑dynamo action \citep{barrere:23}, implying that some stars may host weak dynamos, with weak angular‑momentum transport and rapidly rotating cores, while others harbor strong dynamos that enforce slow core rotation.

Binary interactions offer an alternative explanation. Both tidal spin‑up and merger scenarios can in principle create stars with the observed core‑rotation rates, yet such models usually also produce rapidly rotating envelopes that we do not observe. This discrepancy favors the weak transport interpretation for the rapid core population.

To explore binary effects directly, we constructed two MESA binary models that include angular‑momentum transport via the Tayler–Spruit dynamo (following the slightly modified prescription of \citealp{fuller:22}) and magnetic braking via the Rappaport prescription \citep{rappaport:83}.

The first binary model mimics tidal spin-up or merger with a low-mass companion near the tip of the red giant branch. We evolve a $1.2 \, \mathrm{M_\odot}$ star to the tip of the red giant branch, and then spin up the convective envelope to a rotation frequency of $\Omega_{\rm env}/2 \pi \simeq 1.5 \times 10^{-2} \, \mu {\rm Hz}$, corresponding to an equatorial rotation rate of $v_{\rm rot} \simeq 10 \, {\rm km}/{\rm s}$. This rotation rate is orders of magnitude faster than normal stars at the tip of the red giant branch, and roughly one-third of the breakup rotation rate. After the ignition of helium burning, this model contracts onto the clump where its envelope rotation rate increases to $\Omega_{\rm env}/2 \pi \approx 1.5 \, \mu {\rm Hz}$ ($v_{\rm rot} \approx 100 \, {\rm km/s}$), close to its breakup rotation rate. While the surface spins down due to magnetic braking, core-envelope coupling spins the core up, reaching a maximum rotation rate of $\Omega_{\rm core}/2 \pi \approx \, 1.2 \mu {\rm Hz}$. As the surface continues spinning down due to magnetic braking, the core does so as well, decreasing to $\Omega_{\rm core}/2 \pi \approx \, 0.5 \mu {\rm Hz}$ by the end of core helium burning. 

The second binary model mimics the merger of a sdB (hot subdwarf B) star \citep{heber:16} and a low-mass M-dwarf in a tight binary system. We create a $0.46 \, \mathrm{M_\odot}$ sdB star and evolve it to a central helium mass fraction of $X_{\rm He} \approx 0.5$, i.e., half-way through the core helium burning stage. We spin the star up to $\Omega/2 \pi = 1.6 \times 10^{-4} \, {\rm Hz}$, mimicking tidal synchronization by a companion in a $\approx \! 2$-hour orbit. We then add $\approx 0.3 \, \mathrm{M_\odot}$ of solar composition material to the surface of the sdB model, mimicking the tidal disruption and accretion of the M-dwarf star. We set the specific angular momentum of the accreted material to $j = 10^{17} \, {\rm cm}^2/{\rm s}$. Following the merger, the star resembles an ordinary clump star in structure, but with initial envelope and core rotation rates of $\Omega_{\rm env}/2 \pi \approx 0.5\,\mu {\rm Hz}$ and $\Omega_{\rm core}/2 \pi \sim 1\,\mu {\rm Hz}$. The star spins down via magnetic braking, with core and envelope rotation rates of $\Omega_{\rm env}/2 \pi \approx 0.3\,\mu {\rm Hz}$ and $\Omega_{\rm core}/2 \pi \sim 0.6\,\mu {\rm Hz}$ by the end of core helium burning.

Other binary-evolution pathways could yield different outcomes. A merger between a red‑giant core and a helium white dwarf might spin up the core while leaving the envelope slow; conversely, merging with a main‑sequence star near the base of the RGB might spin up the envelope while preserving a slow core, potentially explaining stars with super‑rotating envelopes. Engulfment of a $\sim10\,\mathrm{M}_{\rm Jup}$ planet near the RGB tip would inject enough angular momentum to produce an envelope rotation rate of $\Omega_{\rm env}/2\pi\sim1\,\mu{\rm Hz}$, comparable to the models in Figure~\ref{fig:fig4_theory}. After contraction onto the clump, subsequent angular momentum redistribution would likely spin up the core. More sophisticated binary models, incorporating varied prescriptions for core-envelope coupling, magnetic braking, and abundance tracking, are needed to test these ideas.

\section{Additional Observational Context}\label{Additional Observational Context}

We explored the connection between our sample and different metal abundances using Li measurements for 166 stars obtained from \cite{ding:24} and [C/N] measurements for 358 stars obtained from \cite{martig:16}. Figures \ref{fig:li_rotation} and \ref{fig:carbon_rotation} show that the anomalous rotators do not appear to have unusual Li or [C/N] abundances. Several previous studies have found correlations between Li and surface rotation (e.g., \cite{sneden:22,rolo:24,sayeed:24}, see also \cite{tayar:23} and references therein). However, these correlations are very weak, with many rapidly rotating stars showing normal Li abundances, and many Li-rich giants with normal rotation rates. The correlation could be even weaker in our asteroseismic sample, since rapidly rotating stars often have low-amplitude stellar oscillations that make it difficult to perform asteroseismic measurements (e.g., \cite{gaulme:14}). Compiling a larger sample of stars with asteroseismic constraints and Li abundance measurements may allow for a weak correlation to be uncovered.

Surface abundances of Carbon and Nitrogen have been used to estimate the masses and ages of red giants \citep{martig:16}. Binary mergers may have masses higher than estimates obtained using birth [C/N] abundances, though this hypothesis requires a large sample for verification. We investigated this relationship in 358 stars, as shown in Figure \ref{fig:mass_cn_comparison}. The predicted masses from \cite{martig:16} tend to be overestimates compared to the values derived from the scaling relations\footnote{We used the temperatures published in \cite{martig:16} to compute masses using scaling relations for this analysis}, and the anomalous rotators do not appear as outliers. It is possible that a larger sample of stars with asteroseismic constraints and [C/N] abundance measurements may reveal a weak correlation.

We also explored the connection between our rotation rate measurements and stellar binarity leveraging Renormalized Unit Weight Error (RUWE) values derived from GAIA data \citep{berger:20} as depicted in Figure \ref{fig:ruwe_rotation}. Although large RUWE ($>1.4$) suggests potential non-single systems, no significant correlation between internal rotation rates and RUWE was observed. However, if fast rotators were to arise from stellar mergers, they would not be expected to show large RUWE values due to close binary companions.

These stellar mergers may exhibit strong magnetic fields \citep{schneider:16,schneider:19}, Ca II H\&K emission \citep{deMedeiros:99}, or emit significant X-rays \citep{soker:07}. However, as stars evolve, their magnetic activity diminishes due to spin-down processes \citep{skumanich:64,skumanich:72}. Additionally, it has been observed that only 8\% of red giants show photometric rotational modulations caused by star spots \citep{gaulme:20}. Furthermore, no significant differences in Ca II H\&K emission have been observed between merger remnants and single stars \citep{gehan:22}. Consequently, detecting any significant activity in these anomalous stars may be challenging.

\section{Conclusions}\label{Conclusions}

We have devised and applied a convolutional neural network to power spectra of approximately 21000 Kepler red giants. This approach yielded 1517 reliable core and envelope rotation measurements, significantly expanding the sample size of stars with asteroseismic rotation estimates. In addition to accurately reproducing the well-established core rotation distribution \citep{mosser:12,gehan:18,mosser:18}, our investigation yields three findings. First, the ratio of envelope-to-core rotation undergoes a series of systematic changes during the clump phase, highlighting the radial transport of angular momentum (cf. Figure~\ref{fig:fig2_rot_trend}). Second, some red clump and red giant stars exhibit (anomalous) envelopes that rotate faster than their cores (cf. Figure~\ref{fig:fig2_rot_trend}(c)). Third, a group of clump stars hosts cores rotating at rates much higher than theoretically expected (cf. Figure~\ref{fig:fig2_rot_trend}(a)).

To evaluate the reliability of the neural-network predictions, we compared core rotation rates with values obtained from previous asteroseismic studies. For 426 stars that met our selection criteria, 59.2\% of the inferred core rotation rates were consistent with the published values within $\pm\,$20\% of the 1:1 ratio \citep{gehan:18}. An additional 9.9\% and 5.6\% of stars were found near the 2:1 and 1:2 deviation lines, respectively, using the same relative tolerance. The remaining stars fell outside of these proximity zones. Unlike earlier approaches that ignore inclination‑dependent amplitudes, our training set incorporates these effects, reducing misclassifications. A comparison with nine magnetically active stars from \citet{li:23} shows that five lie close to the 1:1 line, indicating that strong internal magnetic fields do not systematically bias the network.

To interpret these findings, we constructed MESA single‑star and binary models. Most core and envelope rotation rates are slower by an order of magnitude than predicted by standard angular‑momentum transport prescriptions, but typically lie within a factor of $\approx$2 of models employing strong Tayler–Spruit transport. MESA models with enhanced angular‑momentum transport reproduce most normal RGB and clump rotation rates but fail to capture anomalous rotators.

A plausible explanation is that weaker internal angular-momentum transport allowed the cores of these stars to spin up as they contracted during the red-giant phase, leading to the observed rapid core rotation. An alternative scenario involves binary interactions, where both tidal spin-up and mergers have the potential to generate stars with rapid core rotation or super-rotating envelopes. However, such models typically also result in rapidly rotating envelopes, which are not observed in the data. These findings support weak angular momentum transport models over binary models to create stars with rapid core rotation.

Chemical–abundance measurements and binarity indicators do not correlate with the anomalous rotators. Lithium and [C/N] measurements for 166 and 358 stars, respectively, show no systematic offsets relative to the broader sample (Figures~\ref{fig:li_rotation} and \ref{fig:carbon_rotation}); any Li‑rotation correlation appears weaker than in purely photometric samples. Likewise, RUWE values from Gaia (Figure~\ref{fig:ruwe_rotation}) exhibit no clear trend with internal rotation, consistent with the expectation that merger remnants need not retain close companions. Although these observations cannot exclude all binary pathways, they provide no compelling evidence that binarity dominates the anomalous‑rotator population. A larger sample with simultaneous asteroseismic and abundance constraints will be required to quantify any residual binary signature.

If the rapid core rotation persists into later evolutionary phases, it could produce rapidly rotating white dwarfs, as apparently indicated from asteroseismic rotation measurements of some white dwarfs \citep{hermes:17}. In massive stars, this could allow for rapidly spinning neutron stars or black holes to be produced upon core-collapse, allowing for the production of energetic supernovae and gamma-ray bursts. This highlights the broader astrophysical significance of understanding angular momentum evolution across stellar lifetimes.  Future work will refine the rotation inferences by incorporating explicit magnetic asymmetries into the training framework, expanding the sample with upcoming TESS and PLATO observations, and implementing more sophisticated angular momentum transport models.

\newpage

\section*{Acknowledgements}
S.D. acknowledges SERB, DST, Government of India, CII and Intel Technology India Pvt. Ltd. for the Prime minister's fellowship and facilitating research. All computations are performed on Intel\textsuperscript{\textregistered} Xeon\textsuperscript{\textregistered} Platinum 8280 CPU. We thank Dhiraj D. Kalamkar, Intel Technology India Pvt. Ltd. for suggestions which helped optimize the neural-network training. This paper includes data collected by the \textit{Kepler} mission and obtained from the MAST data archive at the Space Telescope Science Institute (STScI). Funding for the \textit{Kepler} mission is provided by the NASA Science Mission Directorate. STScI is operated by the Association of Universities for Research in Astronomy, Inc., under NASA contract NAS 5–26555. This research made use of Lightkurve, a Python package for \textit{Kepler} and TESS data analysis (Lightkurve Collaboration, 2018). 

\textbf{Funding:} This research was supported in part by a generous donation (from the Murty Trust) aimed at enabling advances in astrophysics through the use of machine learning. Murty Trust, an initiative of the Murty Foundation, is a not-for-profit organisation dedicated to the preservation and celebration of culture, science, and knowledge systems born out of India. The Murty Trust is headed by Mrs. Sudha Murty and Mr. Rohan Murty.

\textbf{Author contributions}: SMH, SD and OB conceived the idea. OB has developed the power spectra simulations. SD has built the training dataset, developed the neural network and analysed Kepler data, with numerous inputs from OB and SMH. OB and SD have worked on the MCMC analysis. JF has developed the theoretical models. All authors reviewed the manuscript.

\textbf{Competing interests:} The authors declare no competing interests.

\textbf{Software}: pytorch, lightkurve, MESA

\appendix
\twocolumngrid

\begin{table*}[!ht]
\centering
\begin{tabular}{|l|l|l|l|}
\hline
KIC & $\Omega_{env}/2\pi$ ($\mu$Hz) & $\Omega_{surf}/2\pi$ ($\mu$Hz) & $\Omega_{surf}/\Omega_{env}$\\
\hline
3758731 & $0.05^{+0.16}_{-0.03}$ & $0.10^{+0.00}_{-0.00}$ & $\sim$1.9 \\
\hline
6776494 & $0.06^{+0.02}_{-0.02}$ & $1.65^{+0.11}_{-0.10}$ & $\sim$26.6 \\
\hline
7604896 & $0.02^{+0.02}_{-0.01}$ & $0.13^{+0.02}_{-0.01}$ & $\sim$6.4 \\
\hline
7903173 & $0.05^{+0.03}_{-0.02}$ & $0.56^{+0.05}_{-0.05}$ & $\sim$12.5 \\
\hline
8648338 & $0.06^{+0.07}_{-0.03}$ & $0.24^{+0.02}_{-0.02}$ & $\sim$4.2 \\
\hline
10000151 & $0.16^{+0.04}_{-0.03}$ & $0.11^{+0.01}_{-0.01}$ & $\sim$0.7 \\
\hline
10281161 & $0.04^{+0.18}_{-0.03}$ & $0.11^{+0.01}_{-0.00}$ & $\sim$2.6 \\
\hline
11295851 & $0.04^{+0.02}_{-0.02}$ & $0.07^{+0.00}_{-0.00}$ & $\sim$1.9 \\
\hline
11709205 & $0.03^{+0.02}_{-0.02}$ & $0.20^{+0.02}_{-0.01}$ & $\sim$6.4 \\
\hline
\end{tabular}
\caption{\label{tab:tab_surf_env}Rotation rates of surface and envelope in few red giant stars. Surface rotation rates are measured using star spots. Envelope rotation rates are measured using this neural network.}
\end{table*}

\begin{figure*}[!ht]
\centering
\includegraphics[width=\linewidth]{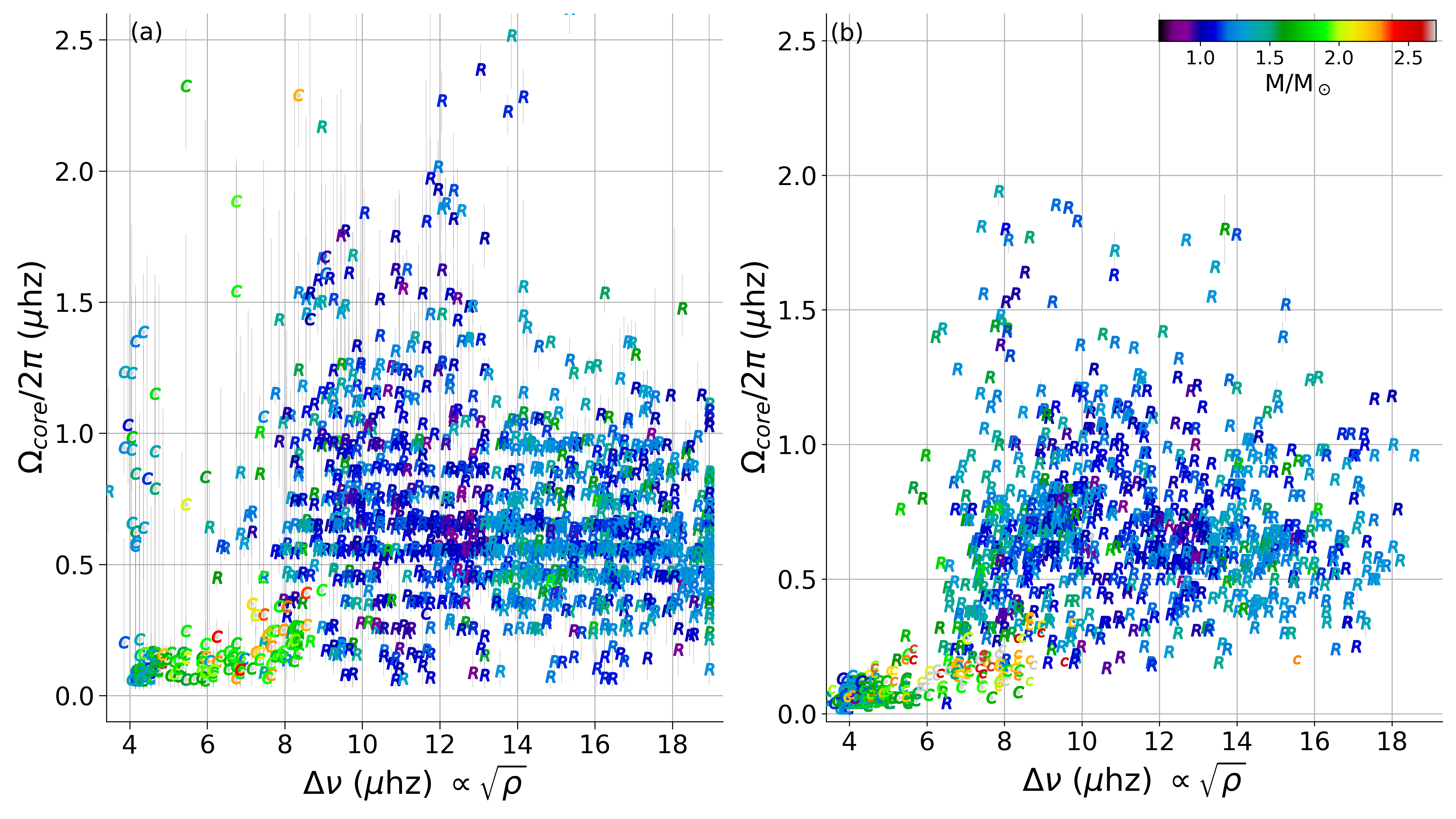}
\caption{Comparison between core rotation-rate trends derived from the neural network (1471 stars plotted in panel (a)) and those from established catalogs (1190 stars shown in panel (b); \citep{mosser:17,gehan:18,tayar:19}). The rotation rates are plotted as a function of the large-frequency separation ($\Delta \nu$). Notably, stars displaying anomalous characteristics, such as fast-rotating clump cores, situated within the parameter range $\Delta \nu<6\,\mu$Hz and $\Omega_{\text{core}}/2\pi>0.5\,\mu$Hz, are absent from existing catalogs.}
\label{fig:rot_compare_core_trend}
\end{figure*}

\begin{figure*}[!ht]
\centering
\includegraphics[width=\linewidth]{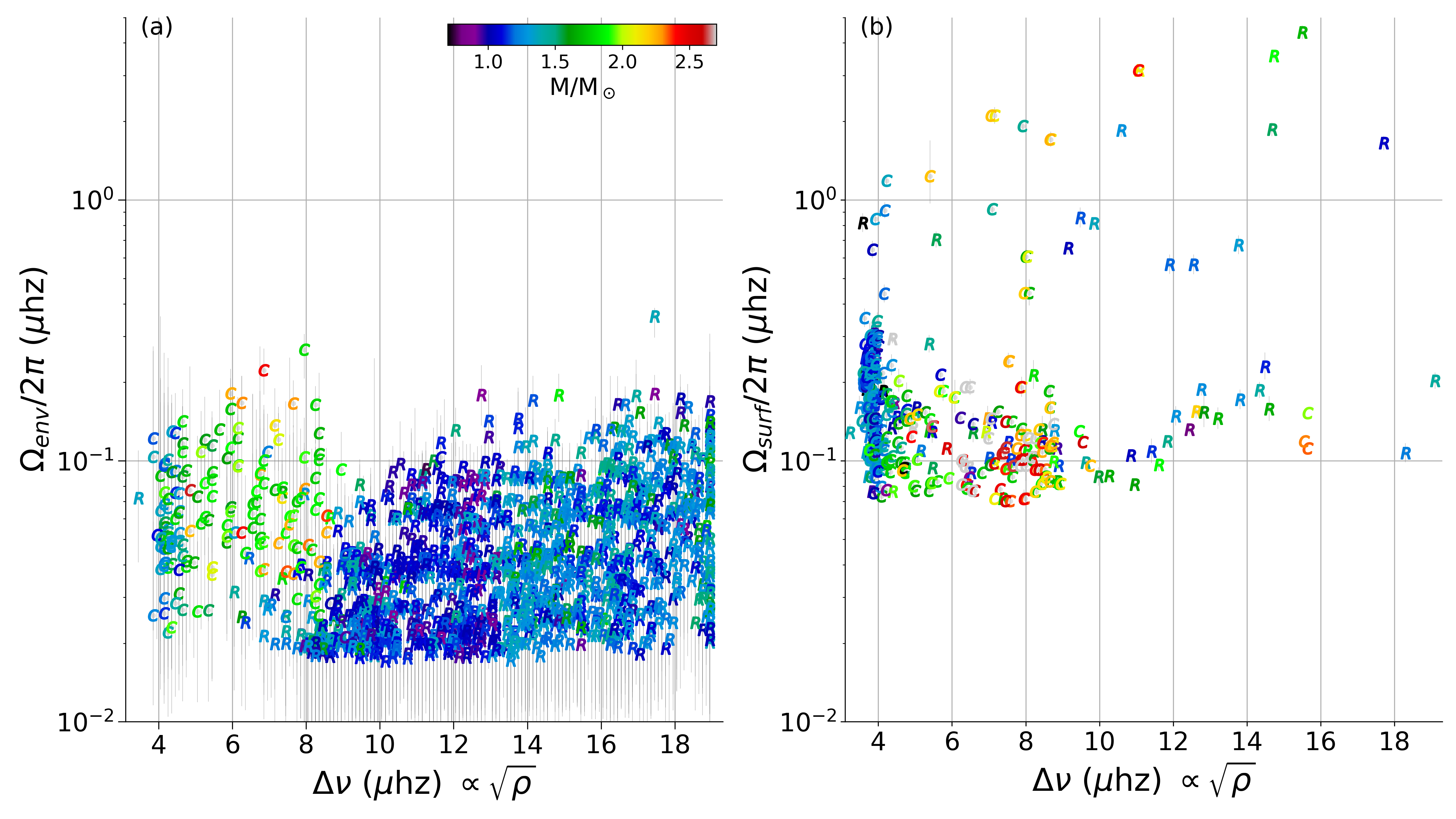}
\caption{Comparison of envelope rotation-rate trends derived from the neural network (1471 stars, shown in panel (a)) and those available in established catalogs (361 stars displayed in panel (b); \citep{ceillier:17,tayar:19}). The rotation rates are plotted as a function of $\Delta \nu$. 
\label{fig:rot_compare_env_trend}}
\end{figure*}

\begin{figure*}[!ht]
\centering
\includegraphics[width=\linewidth]{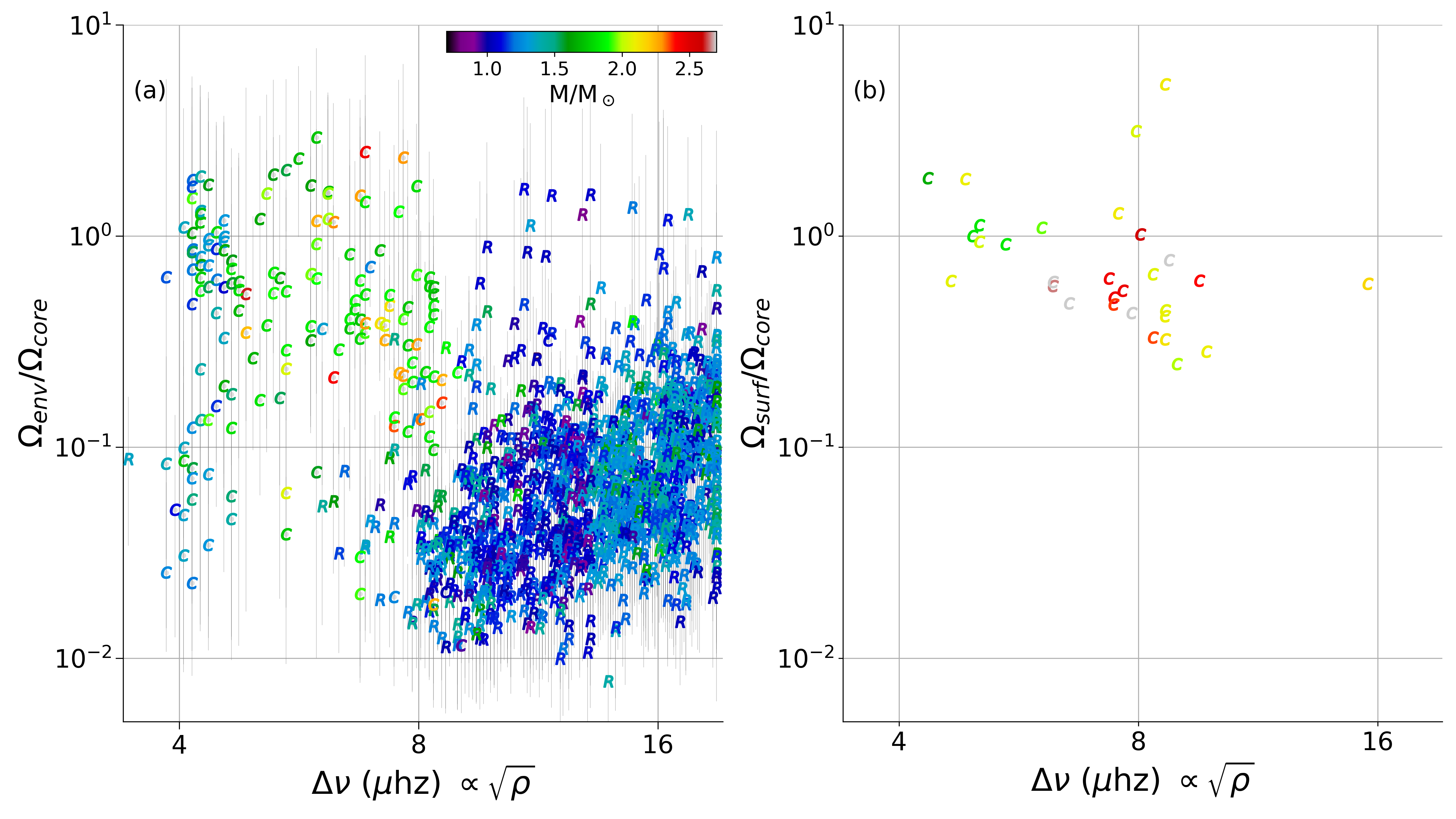}
\caption{Comparison of rotation-rate-ratio trends from a neural network (1471 stars, plotted in panel (a)) and an existing catalog (33 stars shown in panel (b); \citep{tayar:19}). Rotation rates are plotted as a function of $\Delta \nu$.}
\label{fig:rot_compare_ratio_trend}
\end{figure*}

\section{Comparison with existing observations}

In this section, we present a comparative analysis of trends in rotation rates derived here and those available in existing catalogs. The comparison is shown in Figures \ref{fig:rot_compare_core_trend},\ref{fig:rot_compare_env_trend},\ref{fig:rot_compare_ratio_trend}. Among the previous catalogs of core rotation rate \citep{mosser:17,gehan:18,tayar:19} and the current catalog in Table \ref{tab:tab_ml_catalog}, there are 449 stars in common. In figure \ref{fig:rot_compare_core_trend}, we focus on the evolution of core rotation rates. Contrasting Figures \ref{fig:rot_compare_core_trend}(a) and (b), we see that the fast-rotating clump cores, located in the parameter space $\Delta \nu<6\,\mu$Hz and $\Omega_{\text{core}}/2\pi>0.5\,\mu$Hz, are absent in the existing catalogs \citep{mosser:17,gehan:18,tayar:19}. Despite a comparable number of measurements on a statistical basis, earlier methodologies have been unable to identify these anomalous stars.

Comparisons between envelope-rotation and rotation-ratio trends are not possible because existing catalogs that employ asteroseismology have much smaller sample sizes. For instance, we use results from studies that measure surface rotation through star-spot variability \citep{ceillier:17,tayar:19}. The comparison between our envelope-rotation inferences with surface-rotation measurements from existing catalogs is illustrated in Figures \ref{fig:rot_compare_env_trend} and \ref{fig:rot_compare_ratio_trend}. However, those catalogs may be biased towards stars with rapidly rotating envelopes, enhancing the magnetic dynamo and associated spot modulation signal, whereas seismic catalogs are biased towards stars with slowly rotating envelopes and lower magnetic activity \citep{chaplin:11, gaulme:14}. Additionally, although \cite{ceillier:17} primarily interpret their surface-rotation samples in the framework of single-star evolution, they report that a fraction of rapidly rotating stars may result from binary interactions or mergers. Thus, potential contamination from binary-interaction products must be considered when interpreting differences between surface and seismic envelope-rotation measurements.

Starspot-inferred rotation rates are approximately 40 times faster than our envelope rotation measurements on an average. We illustrate the envelope rotation of nine stars in comparison to their respective surface rotation in Table \ref{tab:tab_surf_env}. In six of these cases, the surface rotation is at least 2.5 times faster, revealing a consistently larger surface rotation than envelope rotation. This observed trend is also corroborated in a subset of subgiants \citep{deheuvels:14}. 

The seismic power spectrum exhibits symmetric rotational splittings in non-radial modes, especially $\ell=2$ modes, providing a unique and reliable tool for measuring envelope rotation. The measurement of envelope rotation using non-radial modes is an outcome of the weighted average of the rotation rate across the entire envelope, rather than being solely influenced by the surface rotation \citep{deheuvels:14}. Seismic waves probe layers down to $r/R_{\star} \sim0.1$, whereas surface rotation measurements are restricted to the photosphere and above. Consequently, stars with rapid surface rotation in these catalogs may not truly qualify as anomalous stars such as the ones found using our method.

\begin{figure*}[!ht]
\centering
\includegraphics[width=\linewidth]{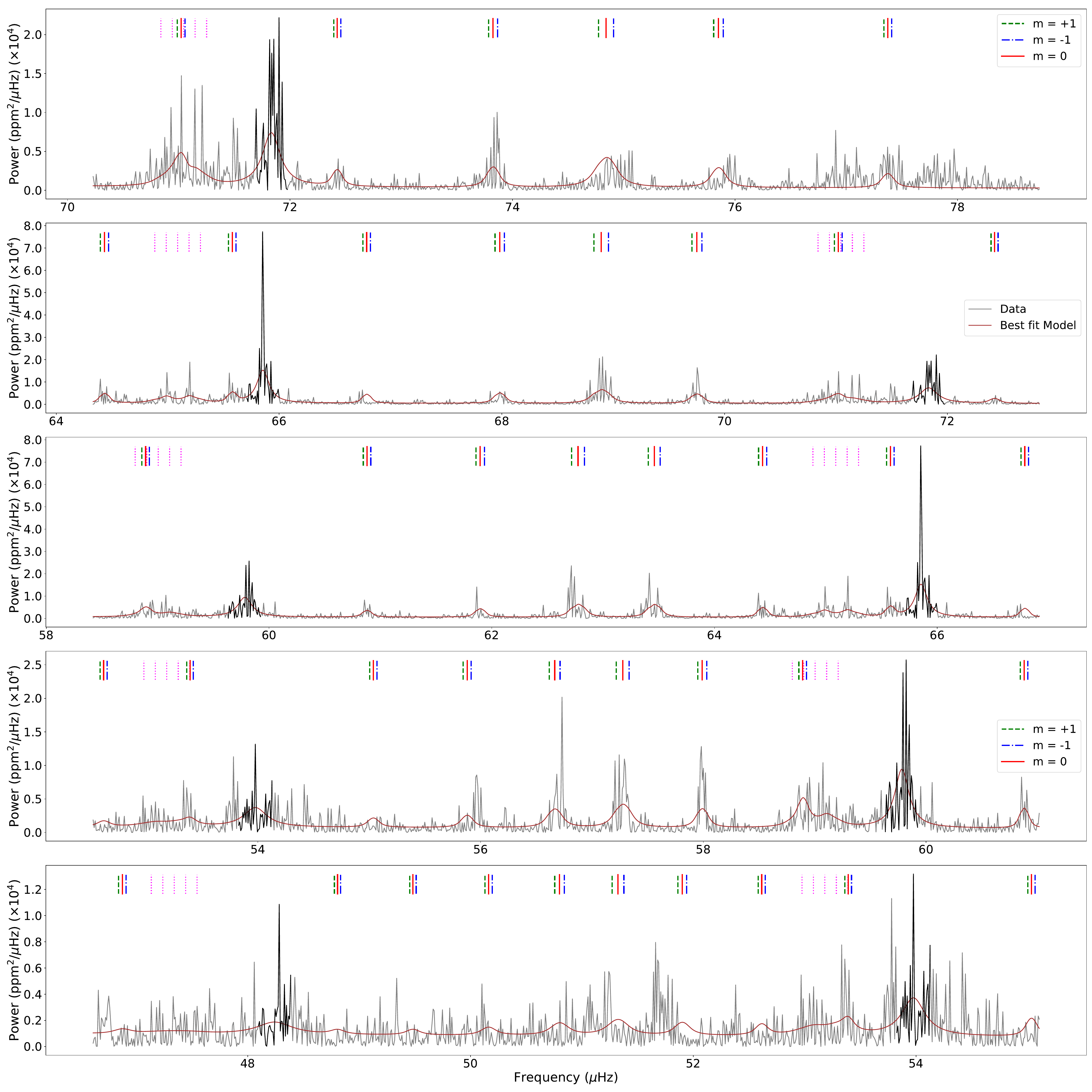}
\caption{A comprehensive comparison of the best-fit model (depicted in brown) and the observed power (depicted in gray) for the star KIC 11615944. Each panel in the analysis corresponds to a different frequency range. The dark segments in the data indicate $\ell=0$ modes identified through alignment with the best-fit model. Additionally, five dotted magenta lines sequentially depict $\ell=2$ modes with azimuthal components ranging from $m=+2$ to $m=-2$. The various constituents of $\ell=1$ modes obtained in the best-fit model are labeled using different lines as specified in the legends. Peaks in the data align with model predictions in a majority of locations with high amplitudes ($\gtrsim10^4$ ppm$^2/\mu$Hz).}
\label{fig:detailed_fit_11615944}
\end{figure*}

\begin{figure*}[!ht]
\centering
\includegraphics[width=0.7\linewidth]{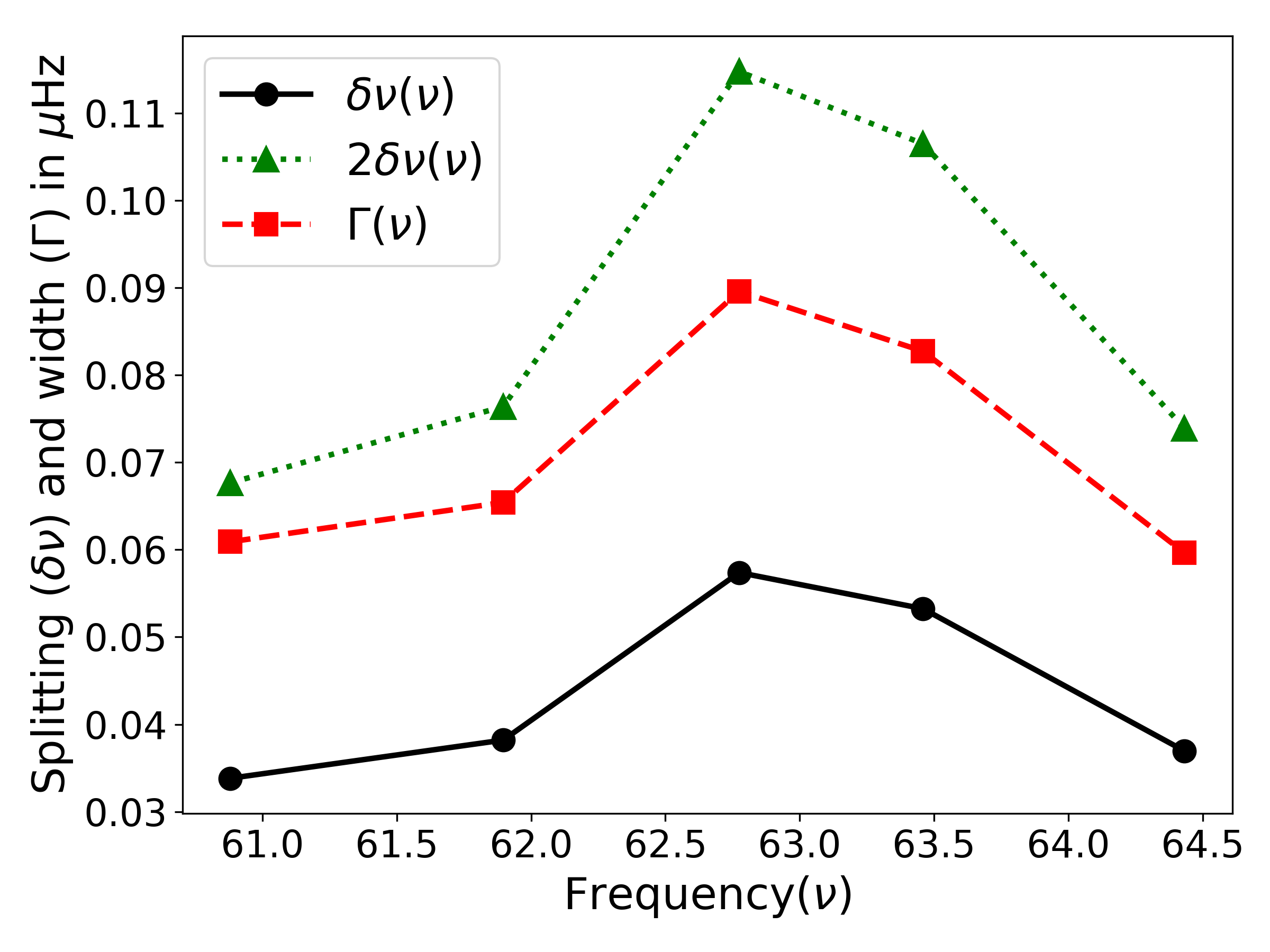}
\caption{Splitting and width as a function of $\ell=1$ mode frequencies in the range 60.5-64.5$\,\mu$Hz, as obtained in the MCMC fit for the star KIC 11615944 as detailed in Figure \ref{fig:detailed_fit_11615944}. It can be observed that 2$\,\delta \nu$ is always greater than the width of the mode, showing that rotationally split modes with m=$\pm1$ are not within the central m=0 mode.}
\label{fig:freq_mm_splitting_width}
\end{figure*}

\begin{figure*}[!ht]
\centering
\includegraphics[width=\linewidth]{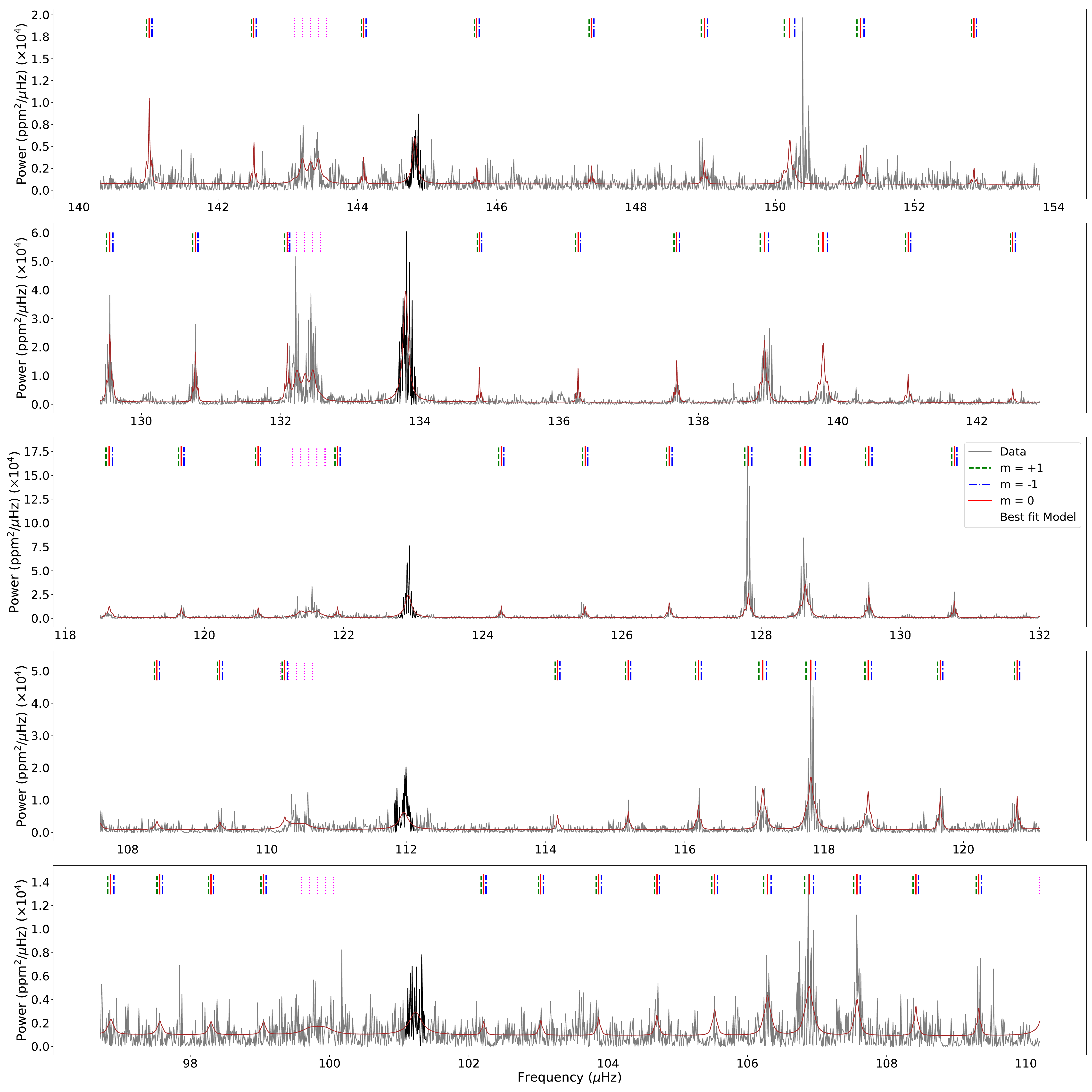}
\caption{Same as Figure \ref{fig:detailed_fit_11615944}, but for KIC 11546972.}
\label{fig:detailed_fit_11546972}
\end{figure*}

\begin{figure*}[!ht]
\centering
\includegraphics[width=\linewidth]{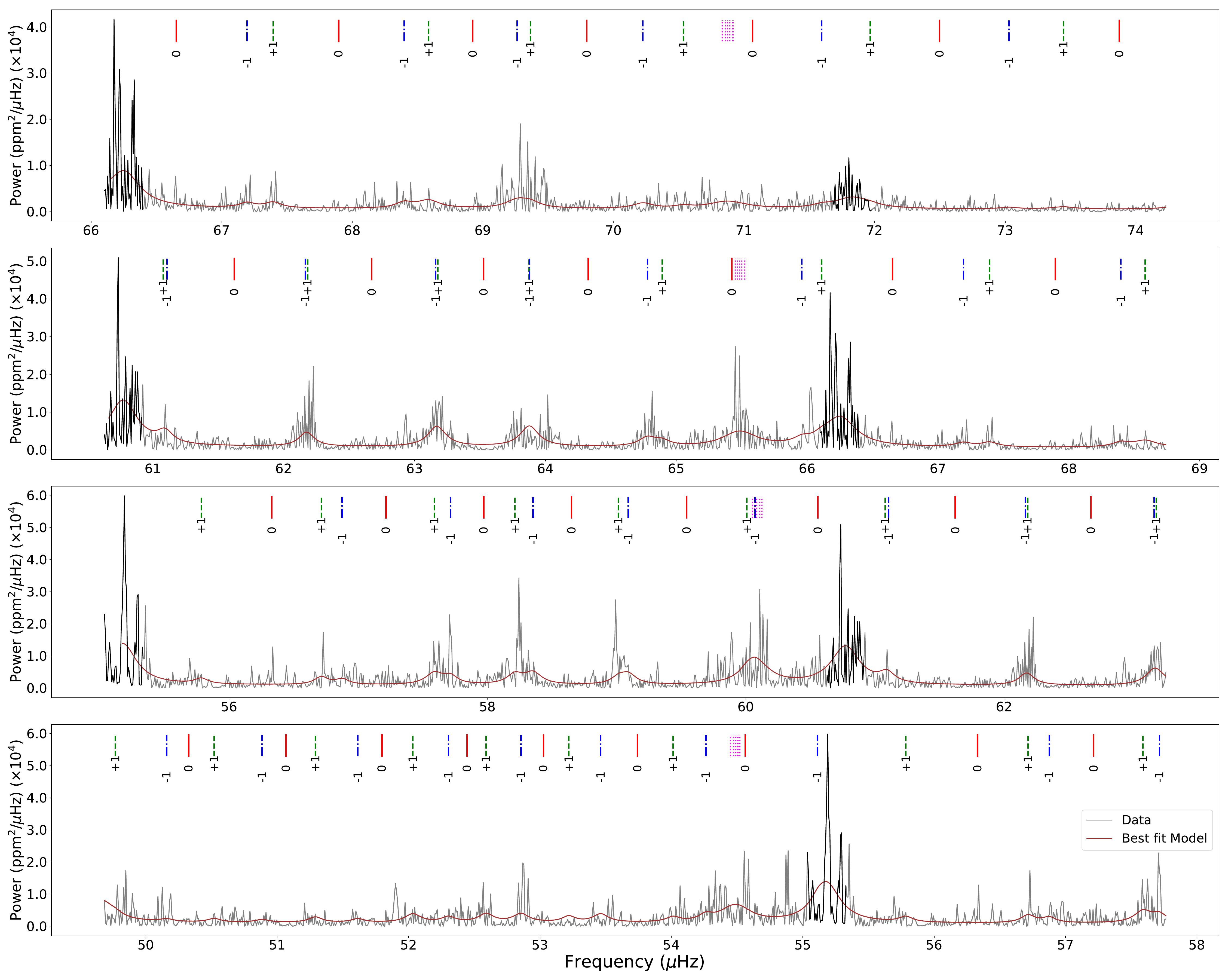}
\caption{Same as Figures \ref{fig:detailed_fit_11615944} and \ref{fig:detailed_fit_11546972}, but for KIC 6695665.}
\label{fig:detailed_fit_6695665}
\end{figure*}

\begin{figure*}[!ht]
\centering
\includegraphics[width=\linewidth]{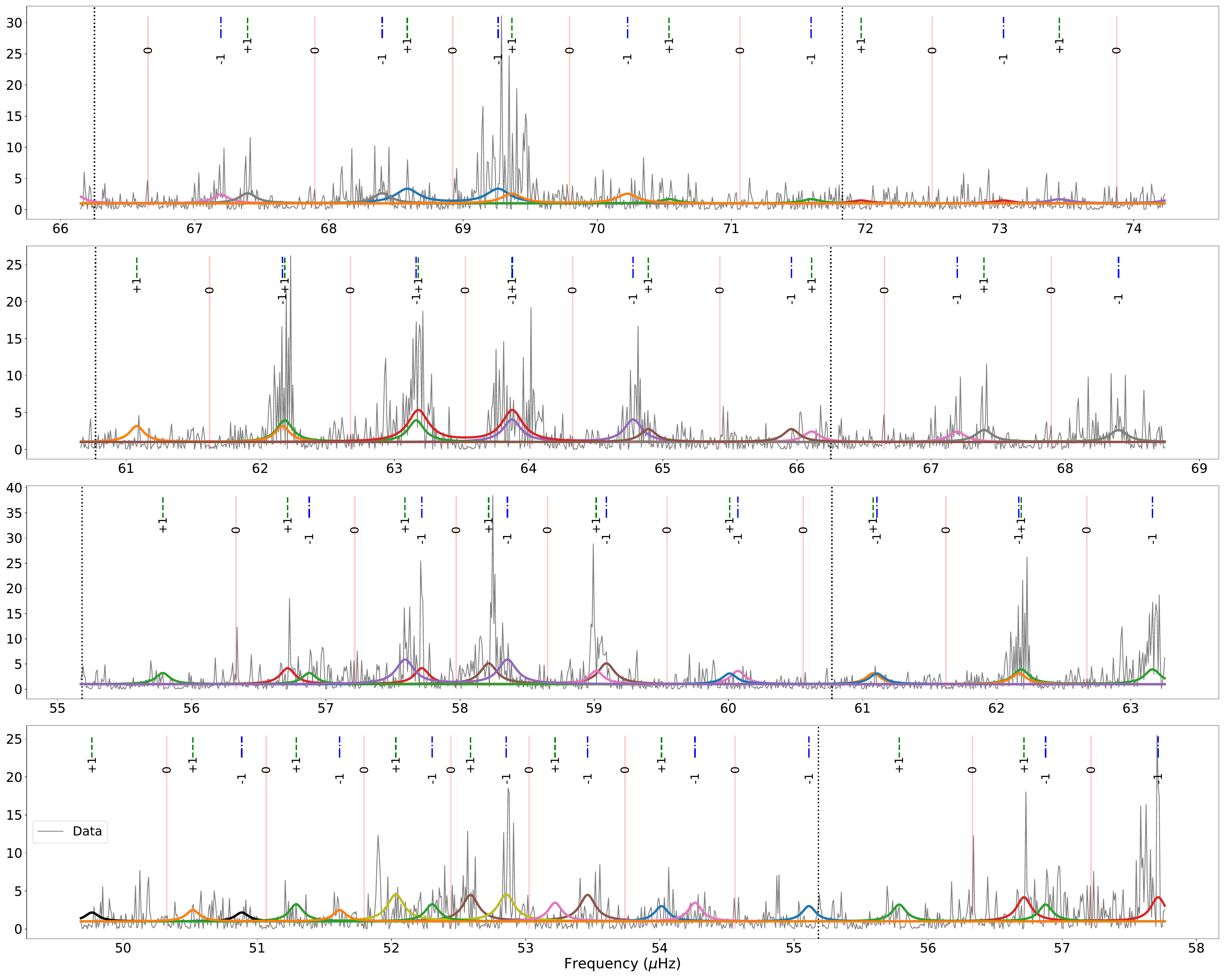}
\caption{This plot shows different $\ell=1$ modes obtained by best-fit model compared to the relative SNR of mixed modes for the star KIC 6695665. The power spectrum is divided by the sum of noise profile, $\ell=0$ and $\ell=2$ modes obtained by the best fit model to calculate relative SNR. This method removes the effect of $\ell=0$ and $\ell=2$ modes. In each panel, a solid colored line represents an $\ell=1$ mode and their respective azimuthal components. In all the mixed modes, the $m=0$ components have negligible amplitude. Nearly all peaks obtained by the model fall in the place of observed peaks.}
\label{fig:detailed_fit_6695665_l_1_modes}
\end{figure*}

\section{Analysis of best-fit models}\label{Analysis of best-fit models}

The detailed comparison of the best-fit model for the star KIC 11615944 is presented in Figure \ref{fig:detailed_fit_11615944}. The model aligns with the observations across all high-amplitude modes ($\gtrsim10^4$ ppm$^2/\mu$Hz). Additionally, all azimuthal components predicted by the best-fit model align with the observed data. We observed superimposition between three high-amplitude $\ell=2$ modes and the $\ell=1$ mixed modes. However, the $\ell=2$ mode at approximately $\sim\!65\,\mu$Hz shows discernible splittings and has significant amplitude compared to the background.  Figure \ref{fig:fig3_splitting} compares the model and the data around this $\ell=2$ mode, corroborating the high envelope rotation rate.

Figure \ref{fig:freq_mm_splitting_width} depicts the splittings and width variation with high-amplitude $\ell=1$ mode frequencies obtained from the best-fit model. The ratio of splitting to width consistently exceeds 0.5 for this model, meeting the criterion set by \cite{kamiaka:18} for reliable measurements for solar-like stars. Observations from Figures \ref{fig:fig3_splitting} and \ref{fig:detailed_fit_11615944} reveal that the g-dominated mixed mode at around 60.9$\,\mu$Hz does not exhibit any visibly split modes in the data that surpass the splitting suggested by the model. These observations support the reliability of the best-fit model and confirm the low core rotation rate.

A detailed comparison of the best-fit model for the star KIC 11546972 is shown in Figure \ref{fig:detailed_fit_11546972}. The model demonstrates alignment with the data across all modes with high amplitudes (exceeding $2\times10^4$ ppm$^2/\mu$Hz). Furthermore, the azimuthal components predicted by the best-fit model closely correspond to the observed data. In particular, we observed superimposition between two high-amplitude $\ell=2$ modes and $\ell=1$ mixed modes. However, the $\ell=2$ mode at approximately $143\,\mu$Hz remains distinguishable, exhibiting clear splittings and significant amplitude compared to the background. Figure \ref{fig:extended_fig_1_splitting} portrays the comparison between the model and the data around this $\ell=2$ mode, providing additional evidence supporting a high envelope rotation rate.

It has been established that $p$-dominated $\ell=1$ mixed modes exhibit high amplitudes and are situated near the center of two $\ell=0$ modes, while $g$-dominated mixed modes show lower amplitudes and are located closer to the $\ell=0$ modes. In the star KIC 11615944, $p$-dominated mixed modes at $\sim$63$\,\mu$Hz and $\sim$69$\,\mu$Hz display a higher splitting compared to the $g$-dominated mixed modes at $\sim$61$\,\mu$Hz and $\sim$64.2$\,\mu$Hz. Similarly, in the star KIC 11546972, $p$-dominated mixed modes at $\sim$118$\,\mu$Hz and $\sim$129$\,\mu$Hz exhibit a higher splitting compared to the $g$-dominated mixed modes at $\sim$119.8$\,\mu$Hz and $\sim$131$\,\mu$Hz. Through these observations, and the application of equation \ref{eq:rot_mm}, it may be inferred that the envelope rotation rate exceeds the core rotation rate.

The detailed comparison of the best-fit model to the data for the star KIC 6695665 is presented in Figure \ref{fig:detailed_fit_6695665}. While the individual $\ell=1$ azimuthal components may not exhibit a precise alignment with the data, it is evident that the $\ell=0$ modes and $\ell=2$ modes show good agreement with the data, and the overall best-fit model aligns well. For further analysis, we have divided the data by the sum of background noise, $\ell=0$ modes, and $\ell=2$ modes to eliminate the influence of these modes from the existing data. The residual data, which has been obtained by this division is compared to individual $\ell=1$ modes present in the best-fit model obtained by MCMC in Figure \ref{fig:detailed_fit_6695665_l_1_modes}. Given that the best-fit inclination angle is close to 90$^{\circ}$ (as shown in Figure \ref{fig:extended_fig_2_fit_6695665}), it is observed that the amplitudes of $m=0$ modes are close to zero. The peaks of most $\ell=1$ modes align well with the existing peaks in the data.

\begin{figure*}[!ht]
\centering
\includegraphics[width=\linewidth]{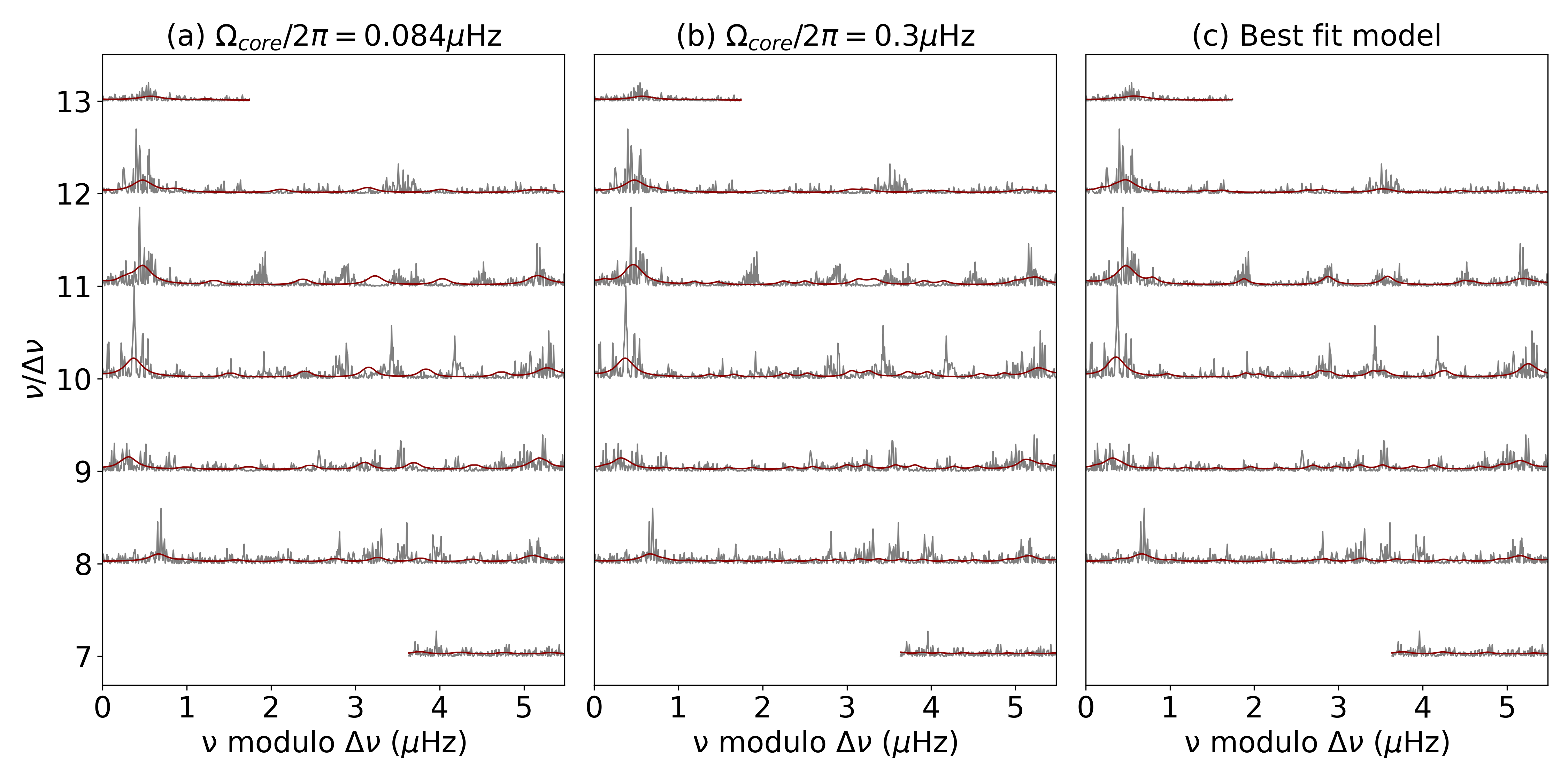}
\caption{Comparison of three models with different rotation rates to the data for the star KIC 6695665. Models (a) and (b) show core rotation rates of 0.084 and 0.3$\,\mu$Hz respectively whereas panel (c) compares best-fit model for which the rotation rate is $1.16\,\mu$Hz. All the parameters except $\Omega_{\rm core}/2\pi$ remain same. The observed peaks align well with the best-fit model compared to other two models.}
\label{fig:model_comparison_rot_rates}
\end{figure*}

\begin{figure*}[!ht]
\centering
\includegraphics[width=\linewidth]{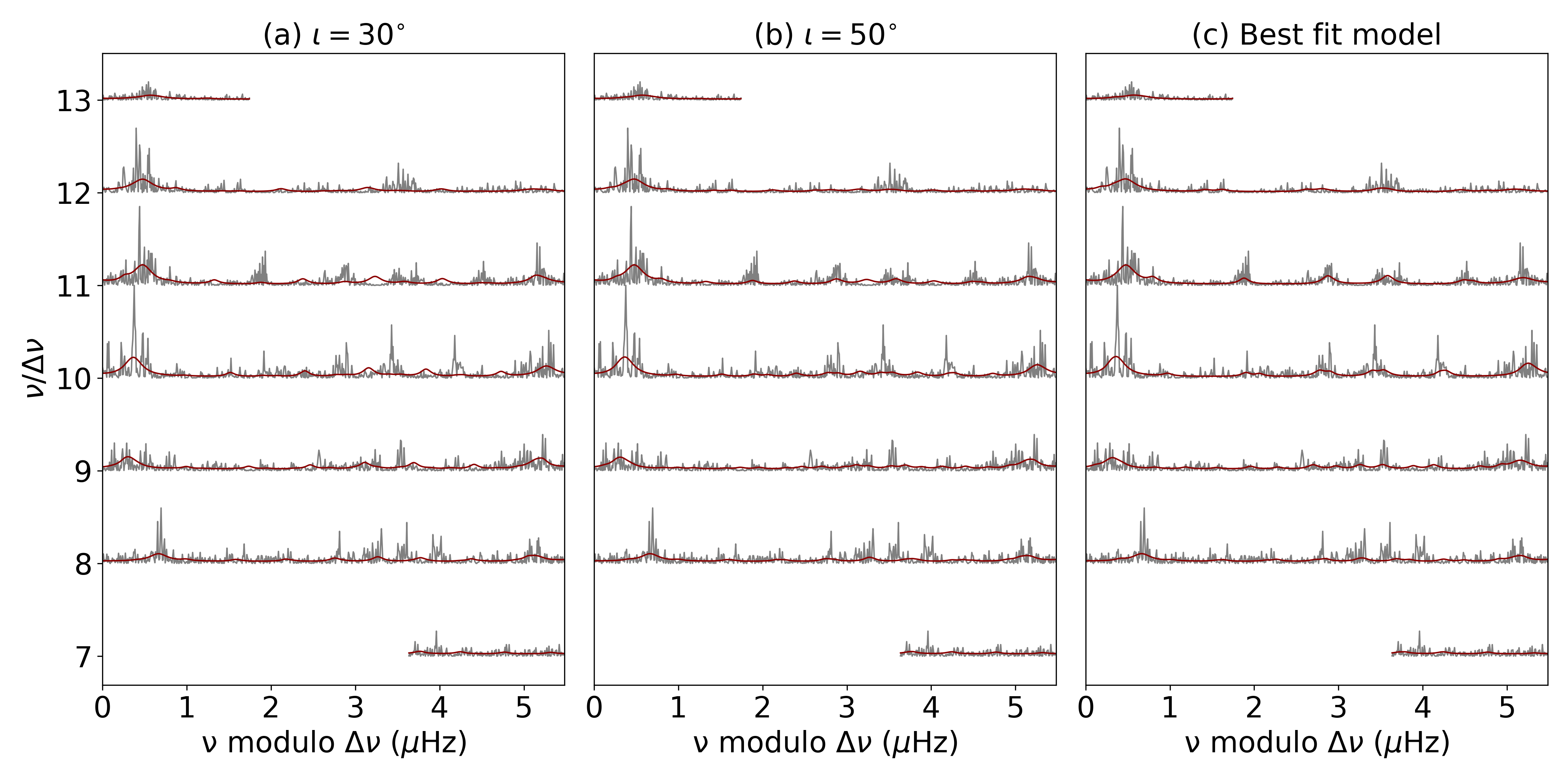}
\caption{Comparison of three models with different inclination angles to the data for the star KIC 6695665. Models (a) and (b) show core rotation rates of 30$^{\circ}$ and 50$^{\circ}$ respectively whereas panel (c) compares best-fit model for which the inclination angle is $85.4^{\circ}$. All the parameters except $\iota$ remain same. The observed peaks align well with the best-fit model compared to other two models.}
\label{fig:model_comparison_inc}
\end{figure*}

\begin{figure*}[!ht]
\centering
\includegraphics[width=\linewidth,height=\textheight]{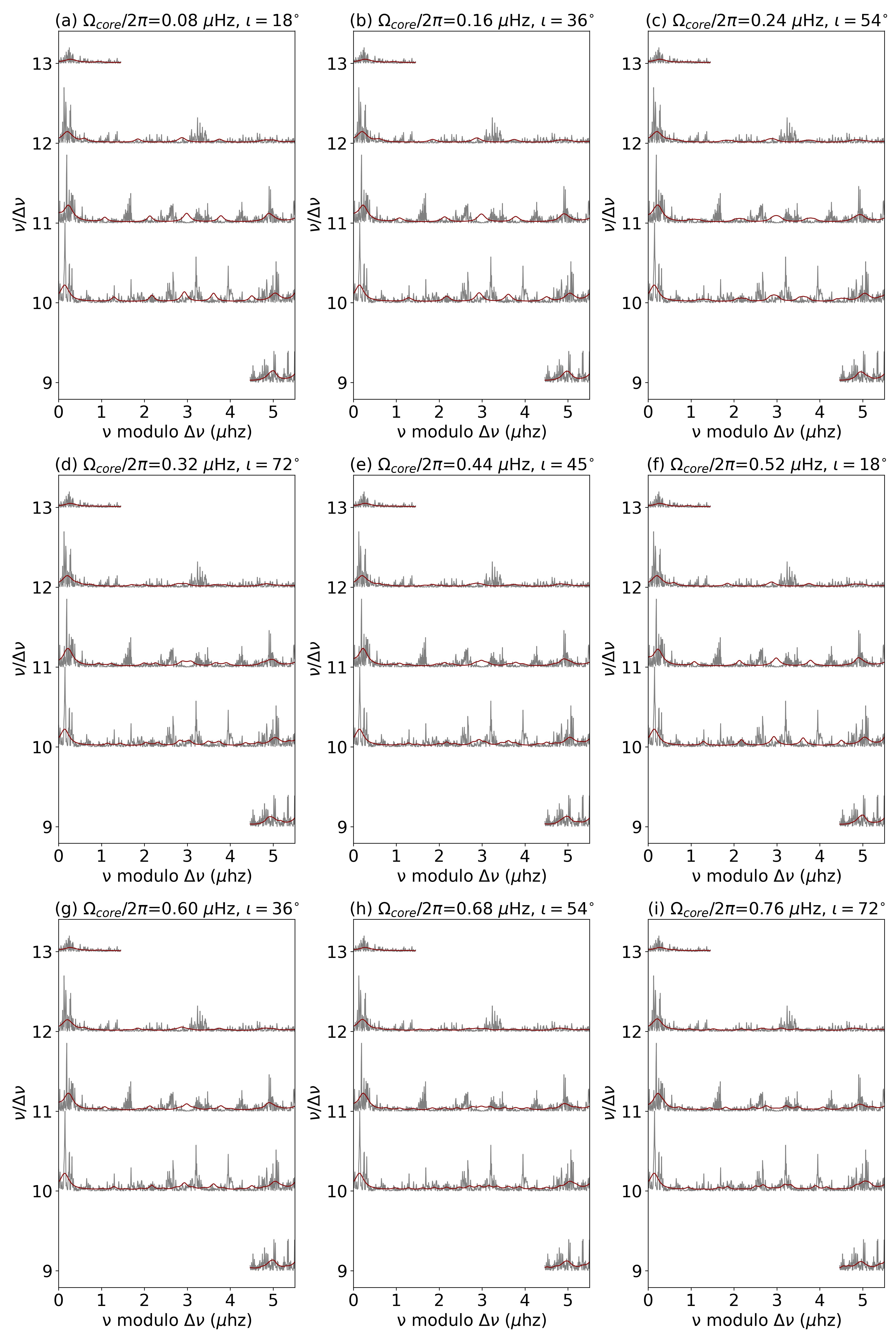}
\caption{Comparison of nine models with lower inclination angles and core rotation rates with observations for star KIC 6695665. The data is shown in grey, while models are marked red. These models do not align as closely with the data as the best-fit model shown in figures \ref{fig:detailed_fit_6695665}, \ref{fig:model_comparison_inc} and \ref{fig:model_comparison_rot_rates}.}
\label{fig:comparison_low_mid_acr_inc_best_ref_v3}
\end{figure*}

While there is overlap among these $\ell=1$ modes, distinct azimuthal components may be identified to account for the rapid core rotation. Specifically, two peaks are observed in the range of 67-67.5$\,\mu$Hz, and two additional peaks are noted at 68.5-69.5$\,\mu$Hz. The mismatch between synthetics produced by low core rotation rates and the data is illustrated in Figure \ref{fig:model_comparison_rot_rates}. We also show the disagreement between simulations produced by lower inclination angles and the data in Figure \ref{fig:model_comparison_inc}. We illustrate the mismatch between synthetics produced by the combination of low core-rotation rates and low inclination angles as compared to the observations in Figure \ref{fig:comparison_low_mid_acr_inc_best_ref_v3}. These provide additional indications supporting a high core rotation rate.

MCMC analysis in Figure \ref{fig:extended_fig_2_fit_6695665} also reveals that $\Delta \nu=5.48\,\mu$Hz, $\Delta \Pi = 304.0^{+0.5}_{-0.5}\,$s, and $q=0.31^{+0.01}_{-0.01}$ for the star KIC 6695665. The observed parameter values are firmly situated within the established distributions characteristic of clump stars \citep{vrard:16,mosser:17,dhanpal:23}. These values represent the average characteristics of this stellar population, excluding the possibility of peculiar structural anomalies.

\begin{figure*}[!ht]
\centering
\includegraphics[width=\linewidth]{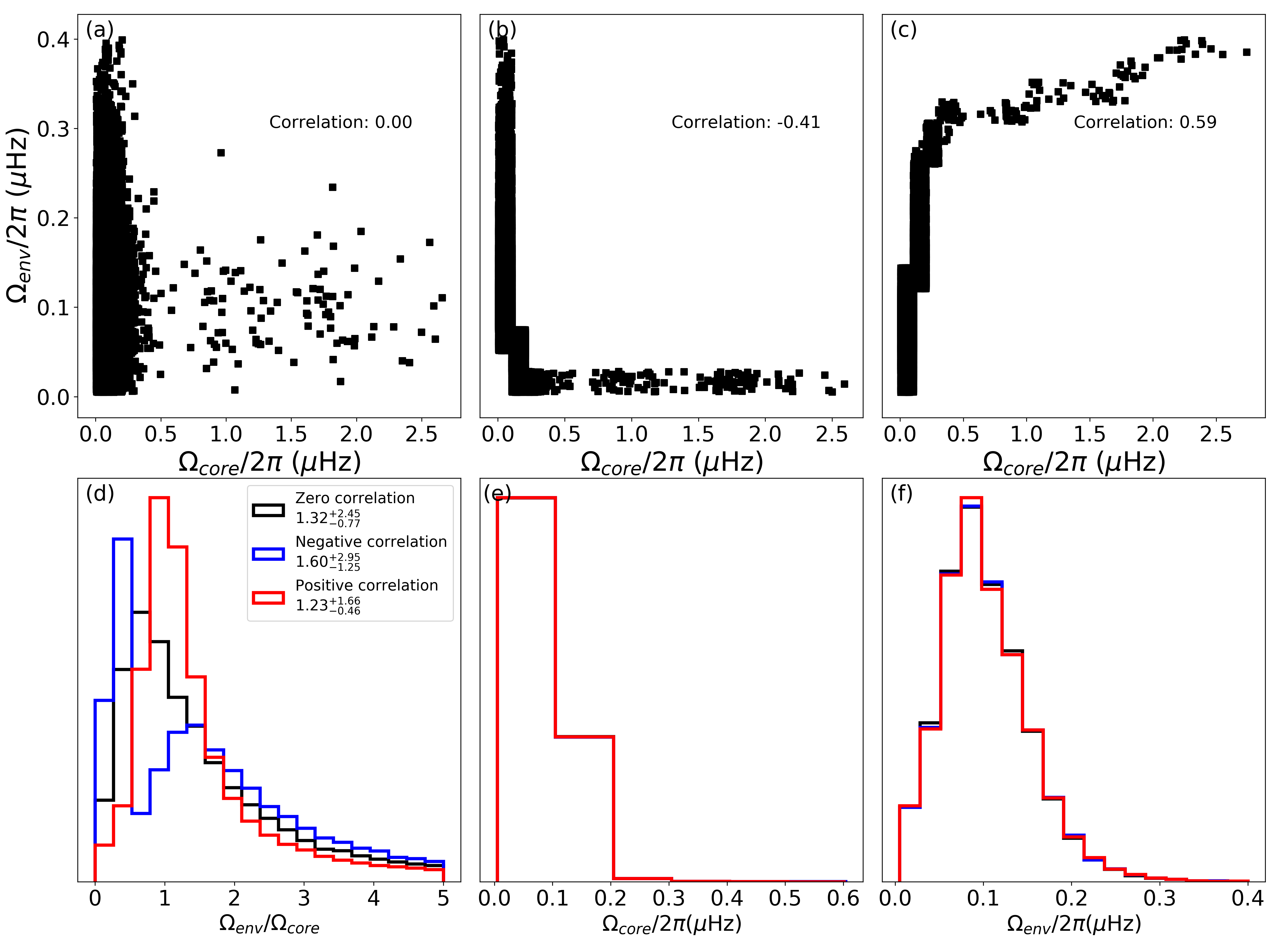}
\caption{Distributions of the ratios $\Omega_{\rm env}/\Omega_{\rm core}$ for KIC 11600442 in three different cases of correlation (a), (b), and (c). The distributions of $\Omega_{\rm env}$ and $\Omega_{\rm core}$  are similar to each other, as inferred by the machine, as shown in plots (e) and (f). However, the distributions of the ratio change with correlation, as shown in plot (d). The largest 1-$\sigma$ interval is associated negative correlation; we therefore use these 1-$\sigma$ values as our uncertainties for the rotation-rate ratios.}
\label{fig:results_correlation_acr_aer}
\end{figure*}

\begin{figure*}[!ht]
    \centering
    \includegraphics[width=0.7\linewidth]{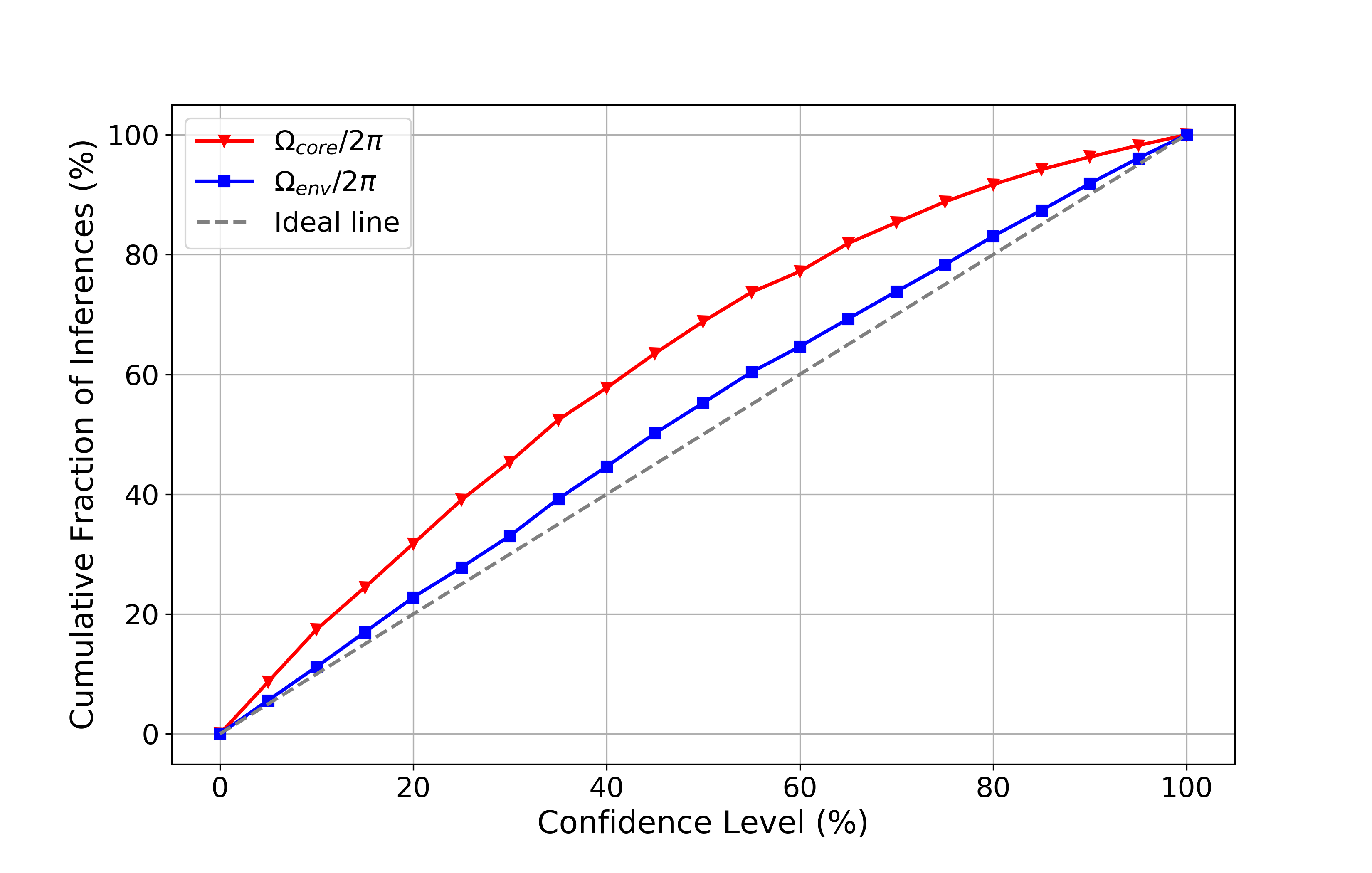}
    \caption{This plot presents results from an uncertainty calibration test, illustrating the frequency with which the ground truth lies within the x\% confidence interval of inferred values for both core and envelope rotation rates, where x ranges from 0 to 100. The plot includes only confident inferences based on synthetic data, as shown in Figure \ref{fig:synthetics_results}.}
    \label{fig:pp_frac_conf}
\end{figure*}

\begin{figure*}[!ht]
    \centering
    \includegraphics[width=\linewidth]{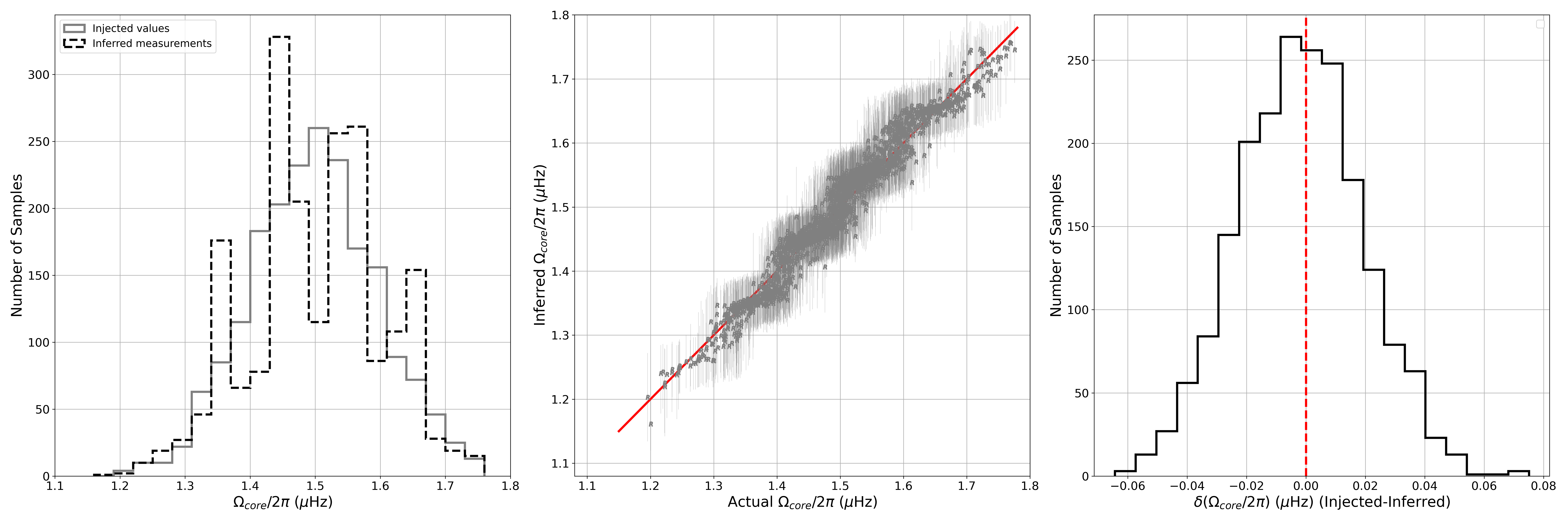}
    \caption{(a)  The true distribution of core rotation rates across 2000 simulated samples, alongside the distribution of inferred rotation rates for these same samples.(b) Inferred core rotation rates plotted against the actual core rotation rates for all 2000 samples. (c) Distribution of the difference between the inferred and actual core rotation rates. These plots demonstrate that the inferences are both reliable and unbiased.}
    \label{fig:gaussian_dist_result}
\end{figure*}

\begin{figure*}[!ht]
\centering
\includegraphics[width=\linewidth]{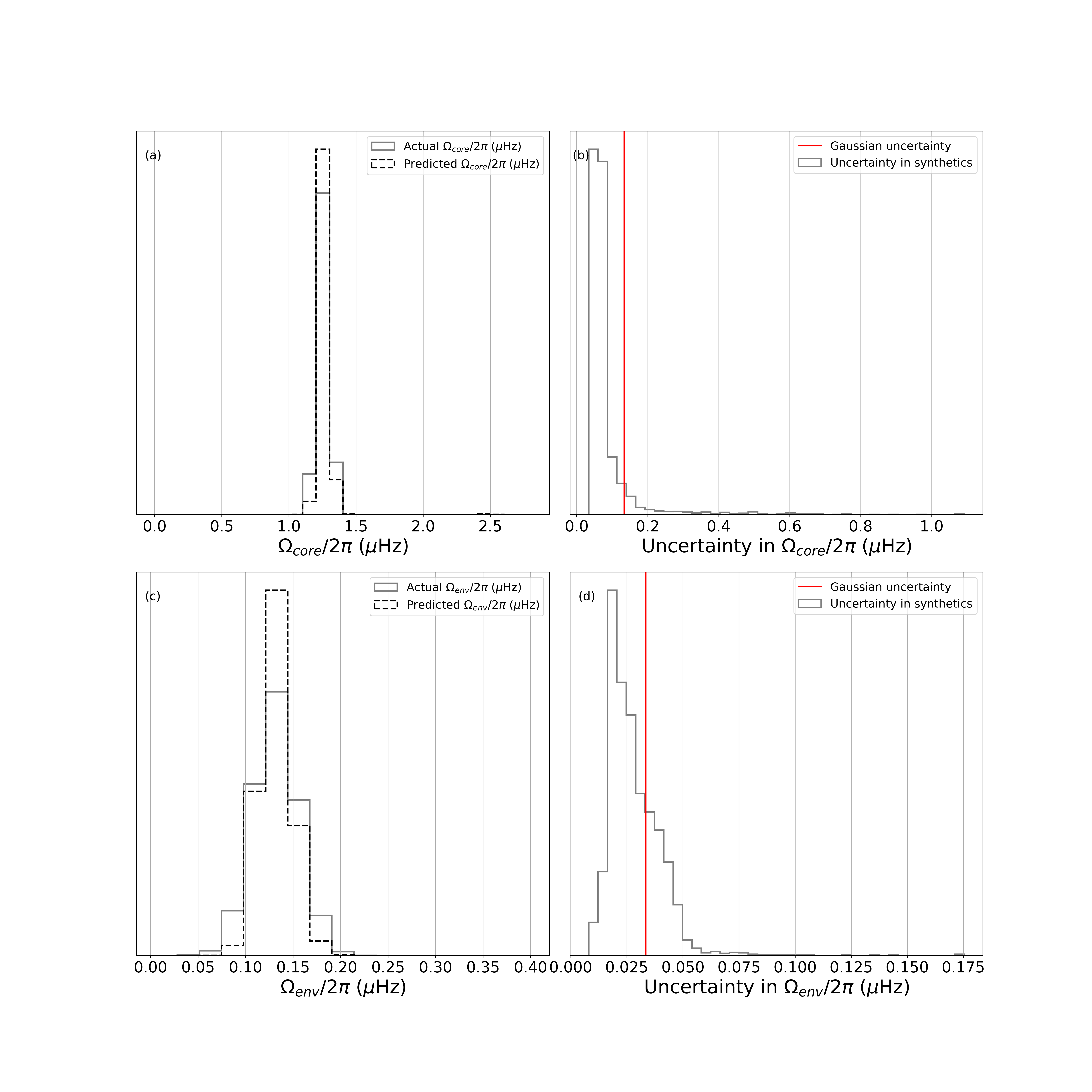}
\caption{Plots (a) and (c) display neural network inferences on a synthetic example, compared to a Gaussian distribution centered around ground truth. The widths of these Gaussian distributions are derived from the standard distribution of errors shown in Figure \ref{fig:synthetics_results}. Plots (b) and (d) illustrate the distributions of uncertainties for the synthetic data. Over 80\% of the inferences exhibit lower uncertainties than the corresponding Gaussian values.}
\label{fig:uncertainty_synthetics_actual}
\end{figure*}

\begin{figure*}[!ht]
\centering
\includegraphics[width=\linewidth]{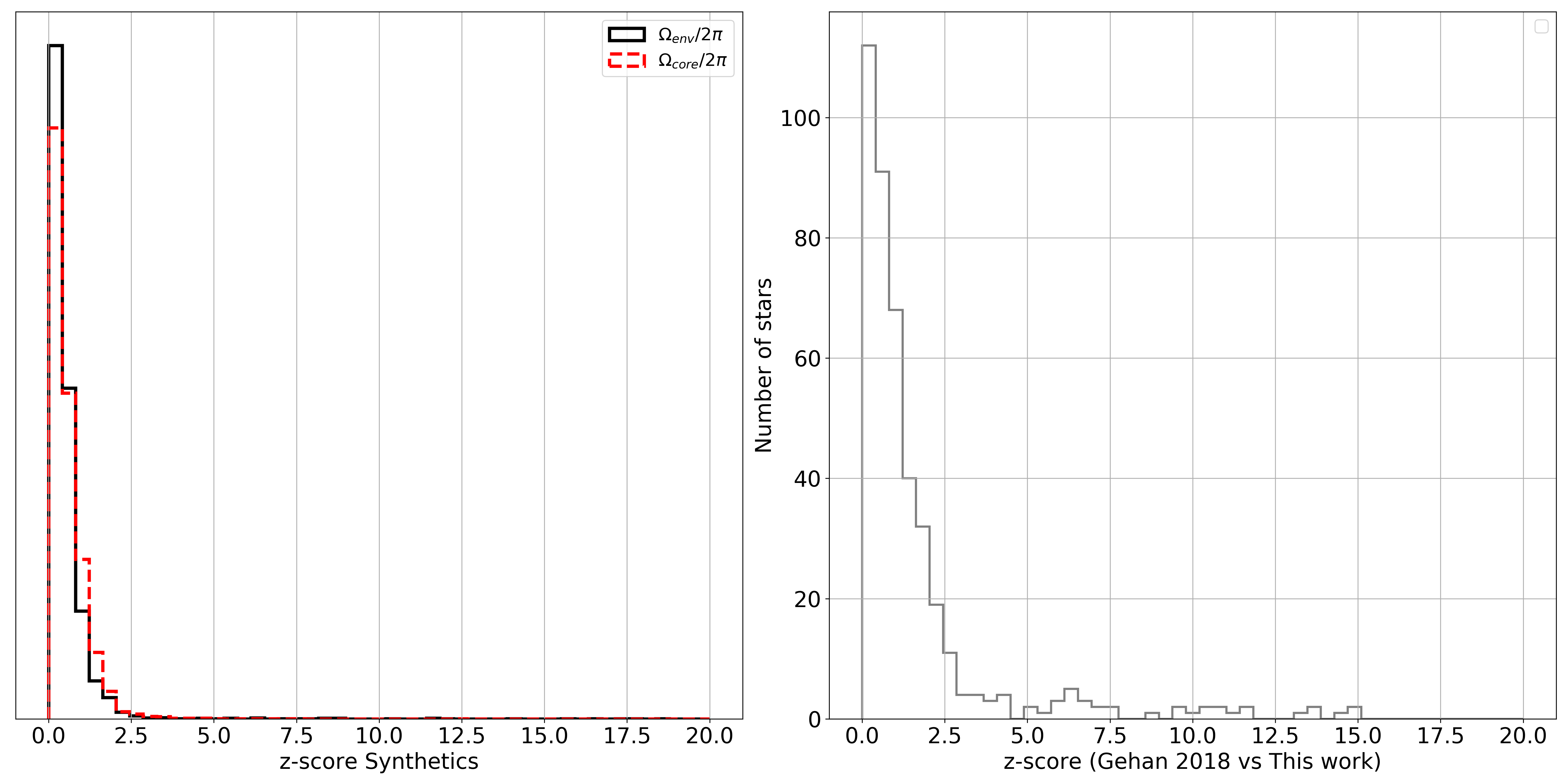}
\caption{(a) The distributions of z-scores for core and envelope rotation rate inferences relative to Gaussian distributions centered on the injected simulation values. Over 80\% of the inferences have z-scores less than 1. (b) Distribution of z-scores for Kepler red giants, based on the network's inferences and \cite{gehan:18}. Approximately 55\% of the confident inferences show z-scores less than 1.}
\label{fig:z_score_comparison}
\end{figure*}
\section{Computation of ratio}

The neural network is trained to output only the distributions of core and envelope rotation rates. However, we may compute the rotation-rate ratios using these two distributions. By introducing a correlation, we may draw samples from these distributions and calculate the ratios of these samples. As shown in Figure \ref{fig:results_correlation_acr_aer}, the distribution of the ratios obtained by introducing a negative correlation results in an increased width. Consequently, we minimized the correlation between the core and envelope rotation rates to obtain the largest intervals of rotation ratio. We show these measurements in Table \ref{tab:tab_ml_catalog}. In addition to these measurements, we also calculate the probability of the core-to-envelope rotation ratio exceeding 1 based on the obtained distributions of the ratio, and present these values in Table \ref{tab:tab_ml_catalog}. These probabilities reflect the strengths of each anomalous-rotator inference.

\section{Analysis of uncertainties}\label{Analysis of uncertainties}

We present three distinct tests to demonstrate the quality of the estimated uncertainties. In the first test, we evaluated the fraction of inferences that contain the ground truth within specified confidence intervals, covering a range from 0\% to 100\%. Results from the Figure \ref{fig:pp_frac_conf} indicate that for 84\% of inferences, the ground truth in $\Omega_{core}/2\pi$ falls within the 1-$\sigma$ (68\%) confidence interval. Similarly, for 71\% of inferences, the ground truth of $\Omega_{env}/2\pi$ is within the 1-$\sigma$ (68\%) confidence interval. Overall, we observe that the fraction of inferences containing the ground truth generally exceeds the nominal confidence interval, indicating that uncertainties are slightly overestimated on synthetic data and suggesting that the neural network tends to provide conservative uncertainty estimates.

In the second test, we conducted the following steps:
\begin{itemize}
    \item We generated 2000 synthetic spectra using asymptotic theory and added random noise drawn from a $\chi^2$ distribution with 2 degrees of freedom. For these simulations, we kept all seismic parameters constant except for the core rotation rate. Specifically, we set the parameters as follow: $\Delta \nu = 16$$\,\mu$Hz, $\Delta \Pi = 85\,$s, and $q = 0.16$. The core rotation rates were sampled from a Gaussian distribution centered at 1.5$\,\mu$Hz with a standard deviation of 0.1$\,\mu$Hz, as shown in Figure \ref{fig:gaussian_dist_result}(a).
    \item We then applied our neural network model to these synthetic spectra, obtaining an inferred rotation rate for each sample. When we compared these inferred values to the true core rotation rates for each sample, we observed a strong correlation, as illustrated in Figure \ref{fig:gaussian_dist_result}(b). This high correlation indicates that the neural network has been effectively trained go infer the rotation-rate parameter.
    \item We examined the difference between the median of each inference and the corresponding ground-truth rotation rate. This difference distribution was approximately Gaussian, centered around zero, as shown in Figure \ref{fig:gaussian_dist_result}(c). This result suggests that the neural network’s predictions are unbiased relative to the true values.
    \item Additionally, we observed that the distribution of the inferred medians closely matches the Gaussian distribution of the true core rotation rates. This similarity confirms that the neural network’s inference error is within 0.1$\,\mu$Hz, indicating that the model’s uncertainties are not inflated beyond the expected range.
\end{itemize}

In order to create realistic Kepler-like synthetics, the samples that we generate here contain intrinsic random-noise realization. Due to noise, the inferred parameters deviate slightly from the ground truth. In figure \ref{fig:synthetics_results}, we observe that the errors for core rotation rate are less than 0.041$\,\mu$Hz for 68\% of samples. In figure \ref{fig:synthetics_results}, we observe that the errors for envelope rotation rate are less than 0.023$\,\mu$Hz for 68\% of samples.

In the following test, we compared the inferences computed by the network with the distributions of ground truth. However, the ground truth is a single value. Thus, in order to create a distribution of ground truths, we modeled it as a Gaussian distribution with the mean coinciding with the ground truth and width to be 0.041$\,\mu$Hz for $\Omega_{core}/2\pi$ and 0.023$\,\mu$Hz for $\Omega_{env}/2\pi$. These Gaussian distributions are compared to the inferences of core and envelope rotation as shown in Figures \ref{fig:uncertainty_synthetics_actual}(a) and \ref{fig:uncertainty_synthetics_actual}(c). Apart from the agreement in medians of inferences and ground-truth values shown in the figure \ref{fig:uncertainty_synthetics_actual}, more than 80\% of uncertainties on these samples are smaller than the assumed Gaussian widths.   

In figures \ref{fig:uncertainty_synthetics_actual}(a) and (c), we  compute z-scores between the respective two distributions of core and envelope rotation. When the z-score is less than 1, it is indicative of 1-$\sigma$ agreement. More than 89\% (84\%) of the z-scores for comparisons between the Gaussian distributions and the neural network inferences are less than 1 for core (envelope) rotation rate as presented in Figure \ref{fig:z_score_comparison} (a). These results show that more than 89\% (84\%) of inferences have better than 1-$\sigma$ agreement for core (envelope) rotation-rate measurements. Also, more than 95\% of these inferences have better than 1.5-$\sigma$ agreement for the inferences of both core and envelope rotation rates. Furthermore, some 55\% of confident measurements on Kepler red giants have z-scores less than 1 for core rotation when compared to the published distributions \citep{gehan:18}, as shown in Figure \ref{fig:z_score_comparison}(b). These results indicate reliability in the inferences for core and envelope rotation rates, including both median values and associated uncertainties.

\begin{figure*}[!ht]
    \centering
    \includegraphics[width=\linewidth]{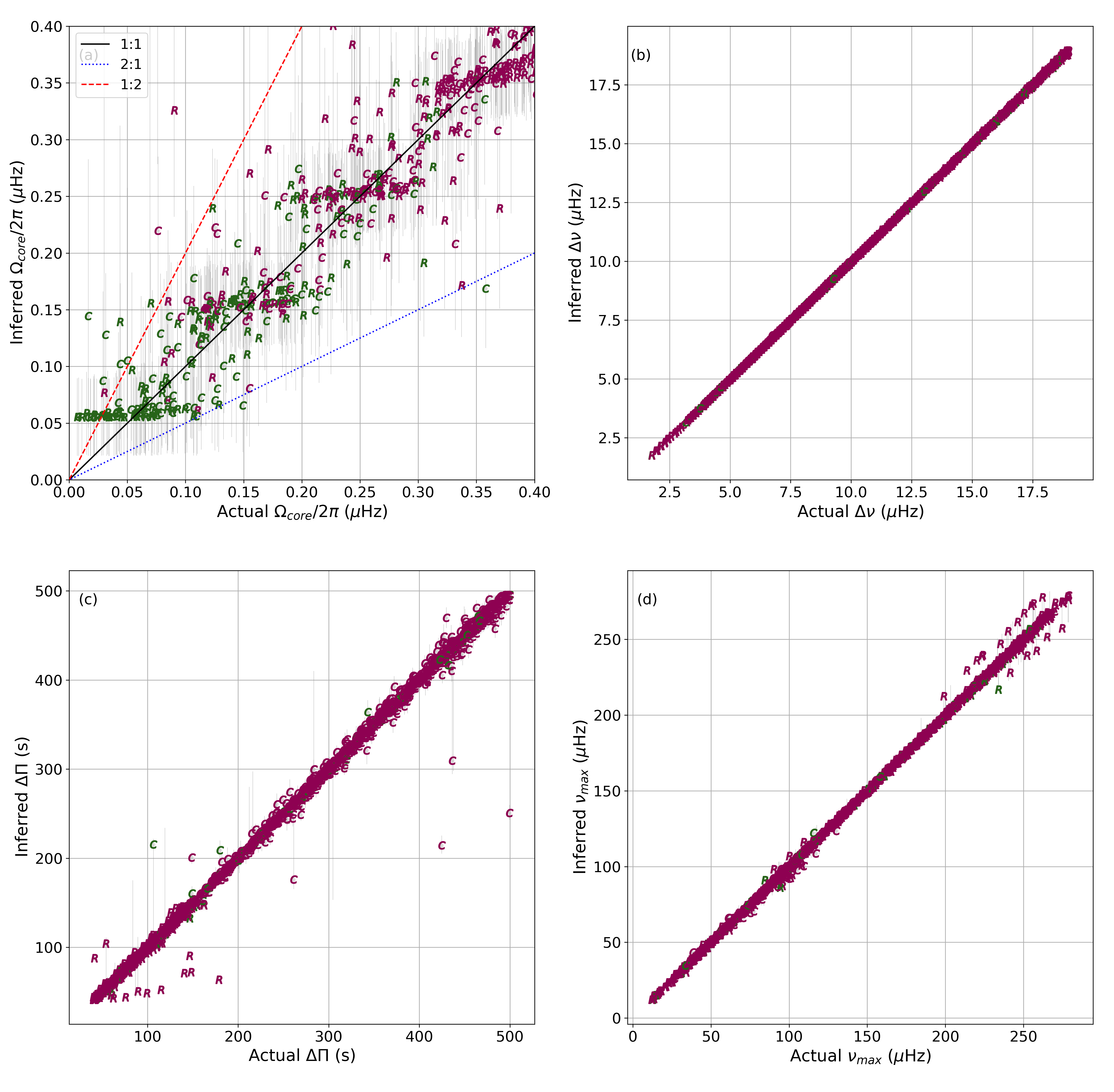}
    \caption{Inferred seismic parameters against the ground truth values for the simulations presented in Figure \ref{fig:synthetics_results}. Stars with $\Omega_{\rm env} > \Omega_{\rm core}$ are shown in green, while those with $\Omega_{\rm core} > \Omega_{\rm env}$ are shown in purple. Panel (a) highlights cases with low core rotation rates to examine anomalous rotators where the envelope rotates faster than the core. Panels (b), (c), and (d) demonstrate that the inferred values of $\Delta \nu$, $\Delta \Pi$, and $\nu_{\max}$ exhibit correlations exceeding 99.9\% with ground truth values. These results indicate that seismic parameters are accurately computed across both normal and anomalous cases.}
    \label{fig:strange_stars_synthetics_actual}
\end{figure*}

\begin{figure*}[!ht]
    \centering
    \includegraphics[width=0.7\linewidth]{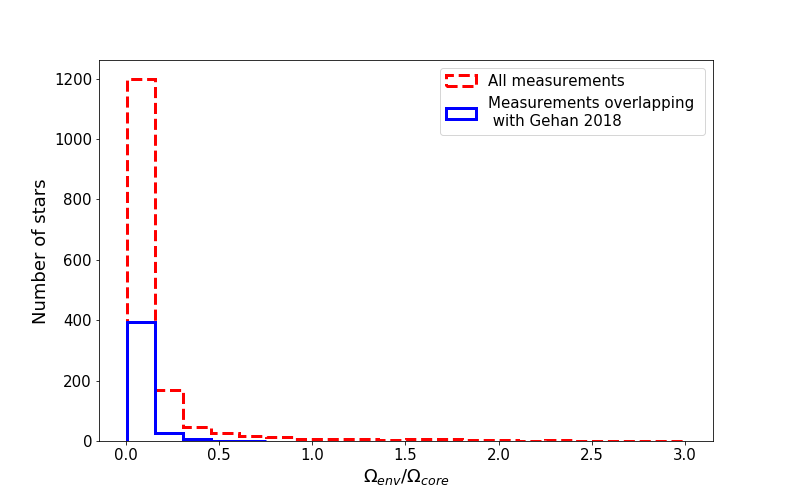}
    \caption{Distribution of rotation rate ratios for all stars listed in Table \ref{tab:tab_ml_catalog}, along with those overlapping with the sample from \cite{gehan:18}, as indicated in the legend. Among the stars with rotation rate ratios overlapping with those in \cite{gehan:18}, no anomalous cases are observed i.e., no stars exhibit an envelope rotation rate exceeding the core rotation rate.}
    \label{fig:env_core_all_gehan}
\end{figure*}

\begin{figure*}[!ht]
\centering
\includegraphics[width=0.95\linewidth,height=0.95\textheight]{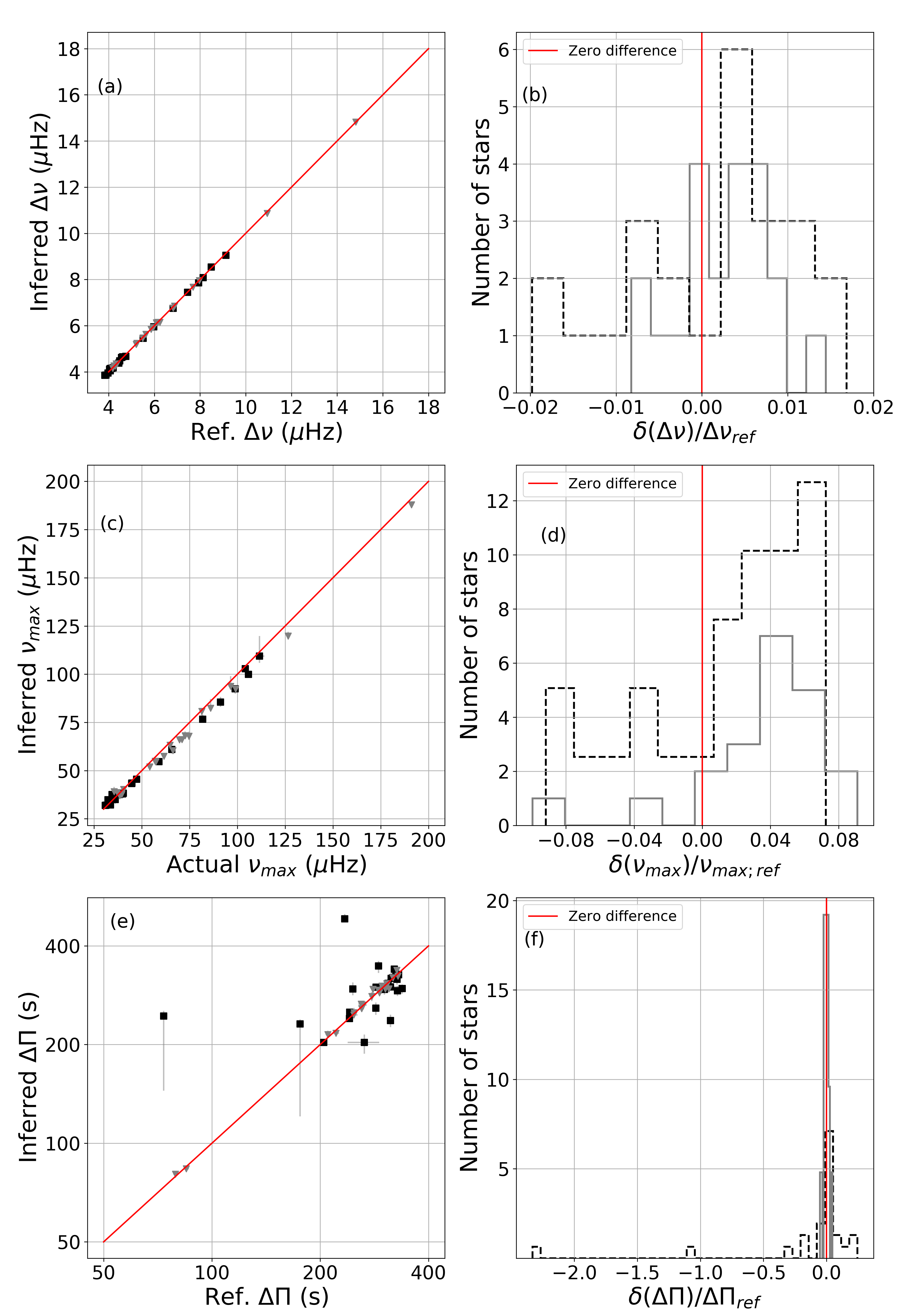}
\caption{Comparison of the inferences derived from the neural network with published measurements \cite{vrard:16} for global seismic parameters $\Delta \nu$, $\Delta \Pi$, and $\nu_{max}$ in the anomalous rotators. The black squares in panels (a), (c), and (e), along with the black-dashed histograms in panels (b), (d), and (f), represent anomalous red-clump rotators with core rotation rates exceeding 0.3$\,\mu$Hz. The grey squares in panels (a), (c), and (e), and the grey histograms in panels (b), (d), and (f), correspond to anomalous red-clump rotators with rotation rate ratios ($\Omega_{env}/\Omega_{core}$) greater than 1. \cite{vrard:16}.}
\label{fig:refInferredComparisonFastRotators}
\end{figure*}

\section{Parameter set of Anomalous stars}\label{Parameter set of Anomalous stars}

In this section, we analyze the parameters $\Delta \nu$, $\Delta \Pi$, and $\nu_{max}$ in synthetic data, with an emphasis on anomalous envelope rotators. Figures \ref{fig:strange_stars_synthetics_actual}(b), \ref{fig:strange_stars_synthetics_actual}(c), and \ref{fig:strange_stars_synthetics_actual}(d) demonstrate that $\Delta \nu$, $\Delta \Pi$, and $\nu_{max}$ are computed with high accuracy across all confident rotation rate estimates shown in Figure \ref{fig:synthetics_results}. Even in cases where the envelope rotates faster than the core, no bias is observed in the core rotation rate, as illustrated in Figure \ref{fig:strange_stars_synthetics_actual}(a).

 As shown in Figure \ref{fig:env_core_all_gehan}, none of the anomalous rotators where the envelope rotation rate exceeds that of the core are included in the Kepler catalog of \cite{gehan:18}, precluding a direct bias comparison with prior estimates. However, we find that the inferred values of $\Delta \nu$, $\nu_{max}$, and $\Delta \Pi$ align well with those from previous studies, as shown in Figure \ref{fig:refInferredComparisonFastRotators} across various anomalous rotators. Over 68\% of the $\Delta \nu$ estimates fall within 1\% of the reference values, while 68\% of $\nu_{\rm max}$ and $\Delta \Pi$ estimates fall within 4\% and 5\%, respectively. These results show strong agreement with published values, supporting the reliability and lack of bias in the rotation-rate estimates presented here.

\clearpage

\begin{longtable*}{|l|l|l|l|l|l|l|l|l|l|}
\caption{Measurements of the seismic parameters and rotation rates obtained using neural network. The table presents median and 1-$\sigma$ values derived from $16^{th}$ and $84^{th}$ percentile. In this table, we present the measurements of $\Delta \nu$, $\Delta \Pi$, $q$, $\Omega_{core}/2\pi$, $\Omega_{env}/2\pi$ and $\iota$ obtained from the neural network. Additionally, we include the measurement of $\Omega_{env}/\Omega_{core}$, derived from the distribution of core and envelope rotation. We also provide the probability of $\Omega_{env}>\Omega_{core}$. The inferences marked with $\dagger$ should be interpreted with care, as the power spectra are in proximity to the Nyquist frequency. \label{tab:tab_ml_catalog}}\\
\hline
{} & KIC & $\Delta \nu$ ($\mu$Hz) & $\Delta \Pi$ (s) & $q$ & $\Omega_{core}/2\pi$ ($\mu$Hz) & $\Omega_{env}/2\pi$ ($\mu$Hz) & $\iota$ (in deg.) & $\Omega_{env}/\Omega_{core}$ & P($\frac{\Omega_{env}}{\Omega_{core}}>$1)\\
\hline
\endfirsthead

\multicolumn{10}{c}{series Table \thetable{} -- continued from previous page} \\
\hline
{} & KIC & $\Delta \nu$ ($\mu$Hz) & $\Delta \Pi$ (s) & $q$ & $\Omega_{core}/2\pi$ ($\mu$Hz) & $\Omega_{env}/2\pi$ ($\mu$Hz) & $\iota$ (in deg.) & $\Omega_{env}/\Omega_{core}$ & P($\frac{\Omega_{env}}{\Omega_{core}}>$1)\\
\hline
\endhead
1 & 10000151 & $18.39^{+0.09}_{-0.06}$ & $91.44^{+1.32}_{-1.12}$ & $0.16^{+0.02}_{-0.02}$ & $0.86^{+0.09}_{-0.06}$ & $0.16^{+0.04}_{-0.03}$ & $48.33^{+5.00}_{-3.96}$ & $0.19^{+0.05}_{-0.06}$ & 0.00 \\
2 & 10000198 & $10.94^{+0.04}_{-0.06}$ & $76.15^{+1.09}_{-1.09}$ & $0.18^{+0.02}_{-0.02}$ & $0.68^{+0.08}_{-0.06}$ & $0.04^{+0.02}_{-0.01}$ & $86.28^{+2.53}_{-4.26}$ & $0.06^{+0.02}_{-0.03}$ & 0.00 \\
3 & 10000228 & $13.76^{+0.07}_{-0.05}$ & $95.07^{+1.92}_{-2.04}$ & $0.13^{+0.02}_{-0.02}$ & $0.69^{+0.11}_{-0.07}$ & $0.03^{+0.02}_{-0.01}$ & $74.09^{+5.40}_{-3.83}$ & $0.04^{+0.02}_{-0.04}$ & 0.00 \\
4 & 10000547 & $13.26^{+0.06}_{-0.05}$ & $80.30^{+1.59}_{-1.91}$ & $0.11^{+0.02}_{-0.01}$ & $0.52^{+0.06}_{-0.08}$ & $0.04^{+0.02}_{-0.02}$ & $83.11^{+4.45}_{-4.51}$ & $0.08^{+0.04}_{-0.06}$ & 0.00 \\
5 & 10001994 & $14.15^{+0.04}_{-0.04}$ & $81.63^{+1.71}_{-1.14}$ & $0.15^{+0.02}_{-0.01}$ & $0.34^{+0.04}_{-0.05}$ & $0.04^{+0.03}_{-0.02}$ & $82.46^{+5.25}_{-6.42}$ & $0.13^{+0.07}_{-0.11}$ & 0.00 \\
6 & 10003039 & $16.83^{+0.05}_{-0.08}$ & $88.96^{+1.42}_{-1.13}$ & $0.11^{+0.02}_{-0.02}$ & $0.84^{+0.04}_{-0.05}$ & $0.03^{+0.02}_{-0.02}$ & $83.45^{+4.29}_{-3.92}$ & $0.04^{+0.02}_{-0.02}$ & 0.00 \\
7 & 10004898 & $17.74^{+0.04}_{-0.04}$ & $88.70^{+1.95}_{-2.03}$ & $0.22^{+0.02}_{-0.02}$ & $0.76^{+0.04}_{-0.04}$ & $0.08^{+0.02}_{-0.02}$ & $52.21^{+4.82}_{-4.94}$ & $0.10^{+0.03}_{-0.03}$ & 0.00 \\
8 & 10005428 & $17.46^{+0.04}_{-0.04}$ & $88.62^{+1.09}_{-1.09}$ & $0.18^{+0.02}_{-0.02}$ & $0.65^{+0.04}_{-0.04}$ & $0.03^{+0.02}_{-0.02}$ & $85.83^{+2.83}_{-4.32}$ & $0.04^{+0.02}_{-0.03}$ & 0.00 \\
\hline
\end{longtable*}

\bibliography{apssamp}
\bibliographystyle{aasjournal}

\end{document}